\documentclass{article}
\usepackage{emulateapj,pstricks,apjfonts}
\def\exosat     {{\em EXOSAT}\/}
\def\axaf       {{\em AXAF}\/}
\def\ginga      {{\em Ginga}\/}
\def\asca       {{\em ASCA}\/}
\def\heao       {{\em HEAO}\/-1}
\def\einstein   {{\em Einstein}\/}
\def\spacelab   {{\em Spacelab}\/-2}
\def\rosat      {{\em ROSAT}\/}
\def\am         {$^\prime$}
\def\as         {$^{\prime\prime}$}
\def\deg        {$^{\circ}$}
\def\kmsmpc     {~km$\;$s$^{-1}\,$Mpc$^{-1}$}
\def\ergcms     {~erg$\;$s$^{-1}\,$cm$^{-2}$}
\def\cmsq       {~cm$^{-2}$}
\def\gax        {\gtrsim}
\begin{document}

\submitted{ApJ in press; astro-ph/9711289}

\lefthead{TEMPERATURE STRUCTURE OF CLUSTERS}
\righthead{MARKEVITCH ET AL.}

\title{THE TEMPERATURE STRUCTURE OF 30 NEARBY CLUSTERS OBSERVED WITH \asca.
SIMILARITY OF TEMPERATURE PROFILES}

\author{Maxim Markevitch\altaffilmark{1,3}, William R. Forman\altaffilmark{1},
Craig L. Sarazin\altaffilmark{2}, and Alexey Vikhlinin\altaffilmark{1,3}}

\altaffiltext{1}{Harvard-Smithsonian Center for Astrophysics, 60 Garden St.,
Cambridge, MA 02138; maxim, wrf, alexey @head-cfa.harvard.edu}

\altaffiltext{2}{Department of Astronomy, University of Virginia,
Charlottesville, VA 22903; cls7i@virginia.edu}

\altaffiltext{3}{Also Space Research Institute, Russian Academy of Sciences}

\begin{abstract}

We present an analysis of \asca\ spatially resolved spectroscopic data for a
nearly complete sample of bright clusters with redshifts between 0.04 and
0.09. Together with several clusters analyzed elsewhere using the same
method, this sample consists of 30 objects with $T_e\gax 3.5$ keV for which
we obtained projected temperature profiles and, when possible, crude
two-dimensional temperature maps. The clusters are A85, A119, A399, A401,
A478, A644, A754, A780, A1650, A1651, A1795, A2029, A2065, A2142, A2256,
A2319, A2597, A2657, A3112, A3266, A3376, A3391, A3395, A3558, A3571, A3667,
A4059, Cygnus A, MKW3S, and Triangulum Australis. All clusters, with the
possible exception of a few with insufficiently accurate data, are found to
be nonisothermal with spatial temperature variations (apart from cooling
flows) by a factor of 1.3--2. \asca\ temperature maps for many clusters
reveal merger shocks. The most notable of these are A754, A2065, A3558,
A3667, and Cygnus A; merging can also be inferred with lower confidence from
the A85, A119, and A2657 temperature maps and from the A3395 and Triangulum
Australis entropy maps. About half of the sample shows signs of merging; in
about 60\% of the sample, we detect cooling flows. Nearly all clusters show
a significant radial temperature decline at large radii. For a typical 7 keV
cluster, the observed temperature decline between 1 and 6 X-ray core radii
(0.15 and 0.9\,$h^{-1}$ Mpc) can be approximately quantified by a polytropic
index of 1.2--1.3.  Assuming such a polytropic temperature profile and
hydrostatic equilibrium, the gravitating mass within 1 and within 6 core
radii is approximately 1.35 and 0.7 times the isothermal $\beta$-model
estimates, respectively.

Most interestingly, we find that temperature profiles, excluding those for
the most asymmetric clusters, appear remarkably similar when the temperature
is plotted against radius in units of the estimated virial radius. We
compare the composite temperature profile to a host of published
hydrodynamic simulations. The observed profiles appear steeper than
predictions of most Lagrangian simulations (Evrard, Metzler, \& Navarro
1996; Eke, Navarro, \& Frenk 1997). The predictions for $\Omega=1$
cosmological models are most discrepant, while models with low $\Omega$ are
closer to our data. We note, however, that at least one $\Omega=1$
Lagrangian simulation (Katz \& White 1993) and the recent high-resolution
Eulerian simulation (Bryan \& Norman 1997) produced clusters with
temperature profiles similar to or steeper than those observed. Our results
thus provide a new constraint for adjusting numerical simulations and,
potentially, discriminating among models of cluster formation.

\end{abstract}

\keywords{Cosmology --- galaxies: clusters: individual ---
intergalactic medium --- X-rays: galaxies}

\section{INTRODUCTION}

It has been twenty years since the realization that the extended X-ray
emission from clusters (e.g., Kellogg et al.\ 1972) is thermal and arises
from optically thin plasma filling the clusters (e.g., Mitchell et al.\
1976). This plasma is in hydrostatic equilibrium in most clusters and
delineates the distribution of the cluster gravitational potential. If one
could measure the spatial distributions of the density and temperature of
this plasma, it is possible to calculate the distribution of total cluster
mass, including its dominant dark matter component (e.g., Bahcall \& Sarazin
1977; Mathews 1978). Such measurements have cosmological implications. Most
importantly, the observed high fraction of hot gas in the total cluster mass
is a strong argument for a low value of the cosmological density parameter
$\Omega_0$ (e.g., White et al.\ 1993). The spatial distribution of the
cluster plasma temperature is interesting in itself, because it is an
indicator of the cluster dynamical state.  Hydrodynamic simulations show
that clusters which recently formed via mergers of smaller subunits should
have a complex temperature structure which becomes more regular with time
(e.g., Schindler \& M\"uller 1993; Roettiger, Burns, \& Loken 1993).
Determination of the dynamical state of present-day clusters can constrain
cosmological models, because rich clusters should be dynamically older in an
open universe than in a high-density universe (White \& Rees 1978;
Richstone, Loeb, \& Turner 1992). Although recent simulations (e.g., Thomas
et al.\ 1997) suggest that the degree of irregularity of the cluster images
is not as strongly dependent on $\Omega_0$ as it was thought, temperature
maps contain additional information on the cluster dynamics that may enable
more sensitive tests.  The spatial distribution of cluster temperatures also
may provide clues on the significance of nongravitational sources of gas
thermal energy, such as supernovae-driven galactic winds (e.g., Metzler \&
Evrard 1997).

The most promising quantitative approach to these interesting problems is to
compare the observed cluster temperature and density structure with
predictions of cosmological hydrodynamic simulations.  Presently,
independent simulation techniques seem to converge on the same qualitative
results for cluster dark matter and gas distributions within the framework
of their assumed physical models (e.g., Frenk et al.\ 1998), although their
resolution is still insufficient to model the fine details of the gas
density distribution. In some sense, the opposite holds for present-day
observations --- while the X-ray imaging instruments such as \einstein\ and
\rosat\ have obtained high-resolution gas density maps for a large number of
clusters, until recently, detailed spatially resolved temperature data were
unavailable. The early X-ray spectroscopic instruments lacked imaging
capability, while imaging instruments had limited spectral resolution or
bandwidth. Coarse spatial temperature distributions were obtained only for
the nearest clusters such as Coma, Perseus, and Virgo with \exosat\ (Hughes,
Gorenstein, \& Fabricant 1988), \spacelab\ (Eyles et al.\ 1991; Watt et al.\
1992), and \ginga\ (Koyama, Takano, \& Tawara 1991). The \rosat\ PSPC was
the first instrument to combine good spatial and adequate spectral
resolution for energies below 2 keV. It obtained temperature distributions
for several galaxy groups and cool clusters (e.g., Ponman \& Bertram 1993;
David, Jones, \& Forman 1995). The \rosat\ temperature maps for several hot
clusters were published as well (e.g., Briel \& Henry 1994, 1996; Henry \&
Briel 1995, 1996), but see Markevitch \& Vikhlinin (1997a, hereafter MV97a)
for a more realistic estimate of their uncertainties.

The \asca\ X-ray observatory (Tanaka, Inoue, \& Holt 1994) is capable of
measuring the temperature distributions in nearby clusters. The \asca\
energy band (0.5--11 keV) is well-suited for clusters, and the 3\am\
half-power diameter angular resolution is adequate.  Although analysis of
the \asca\ data is complicated by the mirror effects such as the energy
dependence of the point spread function (Takahashi et al.\ 1995) and
sometimes stray light (Ishisaki 1996), there are published results on
cluster temperature structure that take these effects into account.
Markevitch et al.\ (1994, 1996a) presented results for A2163. The analysis
method presented in the latter paper was used for several other published
cluster temperature maps (see references in Table 1 below) and is used in
this paper. Applying independent techniques, Ikebe et al.\ (1996, 1997)
analyzed the Fornax and Hydra~A clusters. Loewenstein (1997) reported
preliminary results on A2218; Honda et al.\ (1996) presented a temperature
map of Coma, while Ezawa et al.\ (1997) reported on the temperature and
abundance profile of AWM7. A temperature map of A1367 has been obtained by
Donnelly et al.\ (1997), and a more detailed map of the central part of Coma
by Jones et al.\ (1997)

In this paper, we report on the first systematic study of a representative
sample of 30 nearby hot clusters, which we have undertaken to determine the
common properties of their spatial temperature distributions. Because of the
statistical purpose of this investigation, we give only a brief description
of individual clusters, some of which will be discussed in more detail in
subsequent papers. We concentrate on the cluster-scale temperature structure
and only perform the necessary minimum of modeling of such details as
cooling flows, point sources in the field, etc. For these components, we
allow maximum parameter freedom to obtain model-independent
measurements for the main cluster gas. We use
$H_0=100\,h$ \kmsmpc\ and $q_0=0.5$.

\begin{table*}[tbh]
\small
\renewcommand{\arraystretch}{1.3}
\renewcommand{\tabcolsep}{2mm}
\begin{center}
TABLE 1
\vspace{1mm}

{\sc Cluster Sample}
\vspace{2mm}

\begin{tabular}{lllcccccl}
\hline \hline
Name  & \multicolumn{2}{c}{Exposure, ks$^{\rm e}$} & $z$ & $T_e$ single-comp.$^{\rm a}$ & $T_X$ weighted$^{\rm \;b}$ & $r_{180}$ &
Ref. & Notes$^{\rm d}$\\ \cline{2-3} 
  & GIS & SIS & & keV & keV & arcmin$^{\rm c}$ & & \\
\hline
A85	   &27+14&21+10&0.052& 6.1 $\pm0.2$ & 6.9 $\pm0.4	$   & 38.9&  & m, CF\\
A119	   & 32 &  18  &0.044& 5.8 $\pm0.6$ & 5.6 $\pm0.3	$   & 40.6&  & m  \\
A399	   &29+38&22+31&0.072& 7.4 $\pm0.7$ & 7.0 $\pm0.4	$   & 29.2&  &    \\
A401	   &32+38&27+31&0.074& 8.3 $\pm0.5$ & 8.0 $\pm0.4	$   & 30.6&  &    \\
A478	   & 33 &  15  &0.088& 7.1 $\pm0.4$ & 8.4 ~~~$^{+0.8}_{-1.4}$&26.9&  & CF \\
A644	   & 52 &  45  &0.071& 7.1 $\pm0.6$ & 7.9 $\pm0.8	$   & 31.5& 1& m, CF?\\
A754	   & 21 &  16  &0.054& 9.0 $\pm0.5$ & 9.5 ~~~$^{+0.7}_{-0.4}$&44.0& 2& M  \\
A780 (Hydra A)& 26& 23 &0.057& 3.8 $\pm0.2$ & 4.3 $\pm0.4	$   & 28.5&  & CF \\
A1650	   & 49 &  42  &0.085& 5.6 $\pm0.6$ & 6.7 $\pm0.8	$   & 24.9&  & CF \\
A1651	   & 32 &  25  &0.085& 6.3 $\pm0.5$ & 6.1 $\pm0.4	$   & 23.7&  &    \\
A1736	   & 17 &  ... &0.046& 3.5 $\pm0.4$ & ... 		    & ... &  &    \\
A1795	   & 36 &  28  &0.062& 6.0 $\pm0.3$ & 7.8 $\pm1.0	$   & 35.2&  & CF \\
A2029	   & 34 &  30  &0.077& 8.7 $\pm0.3$ & 9.1 $\pm1.0	$   & 31.6& 3& CF \\
A2065	   &23+23&20+21&0.072& 5.4 $\pm0.3$ & 5.5 $\pm0.4	$   & 25.9& 4& M, CF\\
A2142	   &14+17& 9+14&0.089& 8.8 $\pm0.6$ & 9.7 ~~~$^{+1.5}_{-1.1}$&28.6&  & CF \\
A2256	   &28+35&21+24&0.058& 7.5 $\pm0.4$ & 6.6 $\pm0.4^{\rm f}$  & 34.4& 5&    \\
A2319	   &13+15&10+12&0.056& 9.2 $\pm0.7$ & 8.8 $\pm0.5	$   & 41.4& 6&    \\
A2597	   & 39 &  30  &0.085& 3.6 $\pm0.2$ & 4.4 ~~~$^{+0.4}_{-0.7}$&20.0&  & CF \\
A2657	   & 45 &  32  &0.040& 3.7 $\pm0.3$ & 3.7 $\pm0.3	$   & 36.3&  & m, CF\\
A3112	   & 34 &  15  &0.070& 4.7 $\pm0.4$ & 5.3 ~~~$^{+0.7}_{-1.0}$&26.0&  & CF \\
A3266	   & 33 &  24  &0.055& 7.7 $\pm0.8$ & 8.0 $\pm0.5	$   & 40.2&  &    \\
A3376	   & 21 &  ... &0.046& 4.3 $\pm0.6$ & 4.0 $\pm0.4	$   & 32.9&  &    \\
A3391	   & 21 &  14  &0.054& 5.7 $\pm0.7$ & 5.4 $\pm0.6	$   & 33.5&  & $\gamma$\\
A3395	   & 31 &  22  &0.050& 4.8 $\pm0.4$ & 5.0 $\pm0.3	$   & 34.6&  & m  \\
A3558	   & 17 &  12  &0.048& 5.5 $\pm0.3$ & 5.5 $\pm0.4	$   & 37.6& 7& M, CF, $\gamma$?\\
A3571&23+19+22&15+15+19&0.040& 6.9 $\pm0.3$ & 6.9 $\pm0.2	$   & 50.0&  & CF \\
A3667	   & 16 &  13  &0.053& 7.0 $\pm0.6$ & 7.0 $\pm0.6	$   & 38.5& 4& M, CF\\
A4059	   & 36 &  26  &0.048& 4.1 $\pm0.3$ & 4.4 $\pm0.3	$   & 33.5&  & CF \\
Cygnus A   &29+33&24+25&0.057& 6.5 $\pm0.6$ & 6.1 $\pm0.4	$   & 33.7& 4& M, $\gamma$  \\
MKW3S      & 30 &  24  &0.045& 3.5 $\pm0.2$ & 3.7 $\pm0.2	$   & 32.6&  & CF \\
Triangulum &11+7 & 9+4 &0.051& 9.5 $\pm0.7$ & 9.6 $\pm0.6	$   & 46.8& 8& m; $\gamma$ or CF?\\
\hline
\end{tabular}
\vspace{2mm}

\begin{minipage}{16.5cm}
$^{\rm a}$ Single-temperature fit to the spectrum of the whole cluster.
Errors are 90\%\\
$^{\rm b}$ Emission-weighted gas temperature excluding
cooling flow and other contaminating components, see \S\ref{tempsec}\\
$^{\rm c}$ Estimated from $T_X$, see \S\ref{profsec}\\
$^{\rm d}$ Detected in our temperature data: CF --- significant central
cool component; $\gamma$ --- central power-law component; M --- major
merger; m --- some indication of merging\\ 
$^{\rm e}$ Entries with ``+'' are exposures for multiple pointings\\
$^{\rm f}$ Including secondary cluster. Excluding secondary, $7.3\pm
0.5$ keV.\\
References: 1---Bauer \& Sarazin (1998); 2---map presented in Henriksen \&
Markevitch (1996); 3---Sarazin et al.\ (1997); 4---maps in M98; 5---M96,
MV97b; 6---map in M96; 7---MV97a; 8---map in MSI.
\end{minipage}
\end{center}
\vspace*{-7mm}
\end{table*}

\section{THE SAMPLE}
\label{samplesec}

We selected clusters in the redshift interval $0.04\leq z\leq 0.09$ with
0.1--2.4 keV fluxes greater than about $2\times 10^{-11}$\ergcms\ from the
\rosat\ All Sky Survey X-ray-bright Abell cluster sample (Ebeling et al.\
1996), and included Cygnus A, MKW3S and Triangulum Australis missing from
the Abell catalog due to Galactic obscuration. The redshift range is chosen
so that the most distant clusters of the sample are still well resolved by
\asca, while for the most nearby clusters, radii of interest
($0.5-1\,h^{-1}$ Mpc) and the bright cluster cores are covered by the single
\asca\ field of view, so that stray light contamination from outside the
field of view of the observation (Ishisaki 1996) does not complicate the
analysis. The redshift and flux selection results in 35 clusters; we analyze
30 of them that currently have accessible \asca\ data (see Table 1). For our
present purposes, it is unimportant that the sample is not a complete
flux-limited one.  Rather, we aim at a representative sample of clusters
with a range of temperature and having different evolutionary stages. Our
list includes such diverse clusters as A754 which is undergoing a major
merger (e.g., Henry \& Briel 1995) as well as A1795 and A2029 which are
among the most regular X-ray and strongest cooling flow clusters (e.g.,
Buote \& Tsai 1996). The sample includes clusters with average temperatures
from 4 to 10 keV, with mergers and strong cooling flows at both ends of the
temperature interval.  In addition to the variety of X-ray morphologies,
A780 (Hydra A) and Cygnus A are examples of clusters with powerful central
radio sources, while A3558 is located in the dense environment of the
Shapley Supercluster.

\section{METHOD}
\label{methodsec}

The \asca\ mirrors have an energy-dependent point spread function (PSF) with
an average half-power diameter of 3\am\ (Serlemitsos et al.\ 1995). The PSF
effects should be taken into account when deriving spatially resolved
temperature distributions (e.g., Takahashi et al.\ 1995). The method for
doing this, which we use here, is described in Markevitch et al.\ (1996a)
and Markevitch (1996, hereafter M96). Here we present its brief outline and
relevant additional details. In this paper, we study spatial distributions
of projected cluster gas temperatures. For this, we divide each cluster
image into regions, usually selected to coincide with interesting surface
brightness features, and assume that the spectrum is uniform over each
region. Because \asca\ has insufficient angular resolution to obtain,
simultaneously, both surface brightness and temperature distributions with
sufficient accuracy for clusters in our sample, we use higher-resolution
images from the \rosat\ PSPC for all clusters except A1650 and A2065, for
which \einstein\ IPC and IPC combined with \rosat\ HRI were used,
respectively. These images are used as models of the surface brightness
distributions which determine relative normalizations between the projected
emission measures in different cluster regions.

\asca\ spectral data are collected from the regions in the detector plane that
correspond to the sky regions, applying the SIS and GIS field of view
boundaries. Direct and scattered flux contributions from each sky region to
each detector region are calculated by multiplying the model image by the
\asca\ effective area and convolving it with the \asca\ PSF (both dependent
on position and energy) for each energy bin, then integrating the result
over the respective detector regions. Then cluster temperatures in all model
regions are fitted simultaneously to the spectra from all regions, all
detectors, and all pointings if there is more than one.

\subsection{Brightness Model from ROSAT PSPC}

We used the Snowden et al.\ (1994) code to generate \rosat\ PSPC images in
the 0.5--2 keV energy band (or in the 0.9--2 keV band for clusters with the
highest Galactic absorption to maximize the signal to noise ratio). An X-ray
background was determined for each cluster individually by fitting the outer
radii of the PSPC field of view by a constant plus a power law cluster
radial brightness profile. All clusters in our sample are sufficiently small
so that separating the outer cluster halo and the background did not pose a
problem.  The \rosat\ band X-ray background averaged over areas similar in
size to the regions of our temperature maps ($\sim 10'$ in diameter) has an
rms scatter of about 10\%, while the background averaged over larger areas
similar to the outer bins of our radial temperature profiles ($\sim 30-40'$)
has a scatter of 5\% (Vikhlinin \& Forman 1995; Soltan et al.\ 1996). These
estimates of the PSPC background uncertainty are included in our confidence
intervals.

There is an ambiguity in defining a background when a significant fraction
of it is resolved into point sources. Hard point sources that are bright in
the \asca\ band were fitted individually along with the surrounding cluster
temperatures, to account for their scattered flux. Softer sources that are
bright only in the \rosat\ band were excluded from the fit by masking them
from the model and detector regions. Otherwise, they may lead to an
overestimate of the emission measure in the outer, low surface brightness
regions of the cluster, resulting in a slightly lower best-fit \asca\
temperature. Given the limited \asca\ resolution, it is not practical to
excise all point sources detectable in the \rosat\ image, therefore the
integral contribution of the remaining numerous weak sources should be
subtracted as part of the background. We have calculated the \rosat\
background consistently with this requirement, by excluding from the
background calculation region only sources as bright as those excluded from
the \asca\ fitting region.

\subsection{ASCA PSF Model and Choice of Energy Band}

Above 2 keV, the \asca\ PSF is modeled by interpolating between the Cyg X-1
GIS images at different focal plane positions (Takahashi et al.\ 1995; Ueno
1996). At lower energies, Cyg X-1 appears intrinsically extended and is
inadequate for PSF modeling. In this work, we use \asca\ data in the 1.5--11
keV band, excluding the 2--2.5 keV interval with a poorly calibrated
effective area, and extrapolate the PSF model from higher energies to the
1.5--2 keV interval (assuming a higher PSF uncertainty of 10\%). We have
chosen to include the latter interval despite the increasing PSF uncertainty
in order to take advantage of the high statistical precision data below the
2.2 keV mirror reflectivity edge. Another reason for including it is that
for many clusters, we found that discarding lower energy data resulted in
slightly higher (by about 0.5 keV) temperatures. This can be expected given
that the effective area calibration is more accurate over the whole \asca\
energy band than in any smaller interval. Even though this difference is
smaller than most individual temperature uncertainties, we tried to avoid
such a bias. For a check, we analyzed all observations with the energy
cutoff at 1.5 and 2.5 keV and found no significant differences in best-fit
temperature values except for that mentioned above.

For modeling of the SIS data, the GIS Cyg X-1 calibration images were
corrected for the energy dependence of the intrinsic GIS detector blurring
by additional smoothing resulting in a final constant resolution (Gaussian
$\sigma=0.5'$). The cluster SIS data also were smoothed to the same
resolution.

\subsection{Treatment of Cooling Flows}
\label{cfsec}

In this work, cluster cooling flows are considered ``contaminants'' whose
effect on the measured temperatures of the main cluster gas should be
removed. Usually we can confidently detect the presence of an additional
spectral component at the center of a cooling flow cluster, but in most
cases we cannot say whether it is a cooling flow, an additional lower
single-temperature, or a power law component without a detailed analysis
(e.g., of spectral lines and the surface brightness) which is beyond the
scope of this work. The best-fit temperatures in the outer cluster regions
are essentially independent of these details, as long as our chosen model
describes the central spectrum adequately and the relative normalization of
the central and the outer regions is correct.

Due to the spectral complexity of cooling flows, the relative emission
measure of the central cluster region calculated from the \rosat\ PSPC image
(as is done for other regions) is inadequate. For this reason and to allow
maximum model independence, we fitted, as free parameters, the normalization
of the central region relative to other regions, the fraction of a cooling
flow (or another spectral component) in the total emission from that model
region, and the temperature of its main thermal component, from which the
cooling was assumed to start. A cooling flow spectrum was modeled as
prescribed in Sarazin \& Graney (1991). Because of the large number of free
parameters for the central spectra and the PSF uncertainty associated with
the small regions ($r = $ 1.5\am\ or 2\am) used to model cooling flows, the
derived central temperatures for strong cooling flow clusters are weakly
constrained. Generally, the cooling flow model parameters are not usefully
constrained and therefore are not presented.

Bright point sources that are seen both in \rosat\ and \asca\ images were
fitted following the same procedure.

\subsection{Data and Model Coordinate Alignment}

It is important that the model image, derived from \rosat\ or \einstein, is
accurately aligned with the \asca\ image. A small offset can result in an
error of the model flux in the regions surrounding the brightness peak that
is greater than the statistical uncertainty, especially for clusters with
strong cooling flows. The standard \asca\ $1\sigma$ coordinate accuracy of
24\as\ (Gotthelf 1996) is insufficient for this purpose. Therefore, we
corrected the \asca-\rosat\ coordinate offset for each individual
observation by comparing the \rosat\ image (convolved with the \asca\ PSF)
to the actual \asca\ image. The uncertainty of such corrections is 6--13\as\
($1\sigma$), depending on how peaked the cluster is.

In the course of this work we also have noticed a systematic 0.4\am\ offset
between the GIS and SIS detector coordinates. For the same reason as above,
this offset can result in an inconsistency between the GIS and
SIS temperature maps of cooling flow clusters. We corrected this offset; our
previous papers did not consider strong cooling flows and therefore were not
significantly affected.

\subsection{ASCA Data Filtering and Background Calculation}
\label{bgsec}

We have applied a conservative version of the standard filtering criteria to
the \asca\ data (ABC Guide)
http://heasarc.gsfc.nasa.gov/docs/asca/abc/abc.html). In addition, SIS data
that showed telemetry saturation were excluded. The resulting exposures
(averages of two SISs and two GISs) are presented in Table 1. For the GIS,
we used only the data within 18\am\ of the detector centers, and corrected
the detector gain maps for their long-term time dependence (Makishima 1996).
For the SIS, we analyzed faint and bright mode data and different clocking
modes together.  Our results are insensitive to the time dependence of the
SIS spectral resolution, and a 1.5 keV energy cutoff ensures that they are
also insensitive to the uncertainty of the SIS efficiency at the energies
below $\sim 0.8$ keV (e.g., Sarazin, Wise, \& Markevitch 1997).

For the background calculation, we used \asca\ observations of blank fields.
For the SIS, the normalization was calculated using the total useful
exposure, applying a correction for the long-term mode-dependent degradation
of the SIS efficiency (Dotani et al.\ 1995) since the background fields were
observed early in the mission. A $1\sigma$ uncertainty of 20\% was assumed
for the SIS background normalization.

For the GIS, the blank field data were normalized by exposures in the
individual cut-off rigidity intervals as a first approximation. For the
smaller and fainter clusters in our sample, it was possible to model and
subtract the cluster emission from the same GIS image in the hard energy
band (where the cluster is faint and at the same time the background is most
important), and normalize the blank field background to the residual. An
uncertainty of such normalization was 5--7\% ($1\sigma$). The normalization
determined in such a way showed good agreement with that determined from the
exposure values, with a typical deviation of less than 10\%. We therefore
assumed a 10\% $1\sigma$ uncertainty for bright clusters for which such a
direct background estimate was impossible.

\subsection{Summary of Systematic Uncertainties}

To summarize, the following systematic uncertainties ($1\sigma$) were
included in our temperature confidence intervals in addition to the
statistical uncertainties:

\vspace{1mm}

GIS and SIS background normalization errors of 5--20\%;

An \asca\ effective area calibration uncertainty of 5\% of the model flux in
each energy interval (this incorporates the uncertainty of the mirror
optical axis positions, see Gendreau \& Yaqoob 1997);

A PSF error of 10\% for radial temperature profiles, of 10\% for temperature
maps in the PSF core for relatively small regions (e.g., cooling flows), and
of 15\% for temperature maps in the wings of the PSF;

A \rosat\ surface brightness model error, including the \rosat\ statistical
error and background uncertainties of 5\% for the temperature profiles
and 10\% for the temperature maps;

A 6--20\as\ relative \asca-\rosat\ image offset uncertainty.

\vspace{1mm}

Effects of the PSF, effective area and \asca\ background uncertainties on
the spatially-resolved temperature values obviously are reduced when the
cluster is observed in several offset pointings, ideally with different
telescope roll angles, because possible spatially dependent miscalibrations
then average out. Ten clusters in the sample have multiple pointings, of
them A2256, A3571 and Cygnus A have been observed with different roll
angles. Use of all four \asca\ detectors also reduces the effects of PSF and
effective area errors, since the mirror optical axes are offset by a few
arcminutes with respect to each other.

The PSF scattering seriously complicates our measurements. In the outermost
radial bins of our clusters, only 30--60\% of the flux at an energy of 5 keV
originates in the corresponding region in the sky, while the remaining flux
is that scattered from the bright inner cluster regions.  We will see below
that for the distant cooling flow clusters that have sharp central
brightness peaks and therefore the greatest scattered fraction, the PSF
uncertainty translates into large uncertainties in the outer temperatures.
This usually precludes the reconstruction of accurate two-dimensional
temperature maps.

\subsection{Fitting Procedure}
\label{fitsec}

The GIS and SIS spectra from each detector (SIS chips were treated as
independent detectors) were binned in several (5--13) intervals with
different width so that $\chi^2$ minimization could be performed. To avoid
finding false minima of $\chi^2$ as a function of many free parameters, we
used the annealing minimization technique (e.g., Press et al.\ 1992).  The
Raymond \& Smith (1977, 1992 version) model for thermal emission was used.
In most cases, abundances in different regions were not usefully constrained
and therefore were fixed at the cluster average values. Given the \asca\
spectral resolution, this practically does not affect the obtained best-fit
temperature values.

Generally, we used spectra from all detectors for each region to reduce
systematic uncertainties. We excluded from the fit the spectra with
insufficient statistics, for example, due to the partial coverage when a
given region is better covered by another pointing, or due to the SIS small
field of view when the GIS has sufficient data for this region.  When
fitting the temperature maps, we also excluded SIS spectra for regions with
small partial chip coverage to avoid problems due to the possible
temperature nonuniformity inside each region.  Normalizations between
different detectors were free parameters, while relative normalizations
between different model regions were fixed or freed as described above. Due
to our rather conservative inclusion of systematic uncertainties, values of
$\chi^2_{\rm min}$ per degree of freedom were less than 1 in all fits.
One-parameter confidence intervals for the fitted parameters were estimated
by Monte-Carlo simulations that included all statistical and systematic
uncertainties.

For a consistency check, we fitted GIS and SIS data separately, finding
reasonable agreement in all cases but two. For A3376, we were unable to
obtain consistent results from GIS and SIS, most likely due to the anomalous
SIS background in that observation, and chose to use only the GIS data. For
A1736, we obtained unacceptable $\chi^2$ values for all SIS fits and for GIS
spatially-resolved fits, again possibly for the background reasons. This
cluster only has a relatively low-quality IPC image which did not allow a
more accurate GIS background normalization as described in \S\ref{bgsec}.
Therefore, for A1736, we only present a wide-beam GIS temperature which
should not be significantly affected by this uncertainty.

For those clusters with more than one pointing, we also performed separate
fits for different pointings for another consistency check. The results were
consistent within their expected errors, confirming the adequacy of the
adopted systematic error estimates.

\section{RESULTS}
\label{resultsec}

\subsection{Wide-beam Temperatures}

To check the consistency between our results and earlier wide-beam
measurements, we have performed single-temperature fitting of the overall
cluster spectra (for $r<16-18'$) without exclusion of point sources (except
for Cygnus A which was fit by a thermal plus an absorbed power law
component) or use of cooling flow models. GIS and SIS data were fitted
simultaneously, except that SIS data were not used when it was observing in
1-CCD mode with a field of view of only $11'\times 11'$ (A644, A1650, A2657,
A3266, A3391, Triangulum Australis). Absorption columns were fixed at their
Galactic values taken from Dickey \& Lockman (1990). The resulting
temperature values are given in Table 1. Fig.\ 1 shows these values compared
to those from David et al.\ (1993), who presented wide-beam \einstein\ MPC
results and compiled the most accurate \exosat\ (Edge \& Stewart 1991) and
\ginga\ (Hatsukade 1989) measurements.  We added a later Cygnus A \ginga\
result from Ueno et al.\ (1994). Only those David et al.\ values with an
accuracy of better than 50\% are shown. For several clusters where MPC and
either \exosat\ or \ginga\ disagreed, both measurements are shown.

The figure shows that over the whole range of temperatures in the sample,
our temperatures are in good agreement with previous measurements. The most
notable exceptions, marked by circles, merit explanation. The
highest-temperature deviation is Triangulum Australis, whose \exosat\
temperature is lower than ours. Markevitch, Sarazin, \& Irwin (1996b,
hereafter MSI) showed that most of this discrepancy is caused by different
assumed $N_H$ values (Edge \& Stewart obtained a best-fit $N_H$ value higher
than Galactic; however, \rosat\ PSPC observations yielded a value very close
to the Galactic value which we use).  Also, an update of the GIS gain
calibration (Makishima 1996) has resulted in a slightly lower \asca\
temperature than the one reported in MSI.  Further, our 7.7 keV temperature
for A3266 is inconsistent with the lower MPC value. A soft point source seen
in the \rosat\ PSPC image just outside our fitting region (see Fig.\ 2) may
contribute to this discrepancy.  Our 6.0 keV value for A1795 is higher than
the \ginga\ value but agrees with the MPC value (the MPC and \ginga\ values
are themselves discrepant). This cluster has a strong cooling flow and
instruments with different energy coverage are likely to obtain different
single-temperature fits. As expected, our single-temperature fit for A1795
depends on the adopted low energy cutoff. The energy band difference between
\asca\ and the MPC may also be the reason of the A2065 discrepancy. For this
cluster, we obtain $T=5.4$ keV but also detect large spatial temperature
variations probably due to a merger (Markevitch et al.\ 1998, hereafter
M98).  The separate GIS and SIS single-temperature fits are similar, and it
is noteworthy that our A2065 temperature is in a better agreement with the
$L_X-T$ relation (David et al.\ 1993) than the higher MPC value.

\pspicture(0,-1.3)(8.8,9.8)

\rput[tl]{0}(-0.5,9.7){\epsfxsize=8.8cm
\epsffile{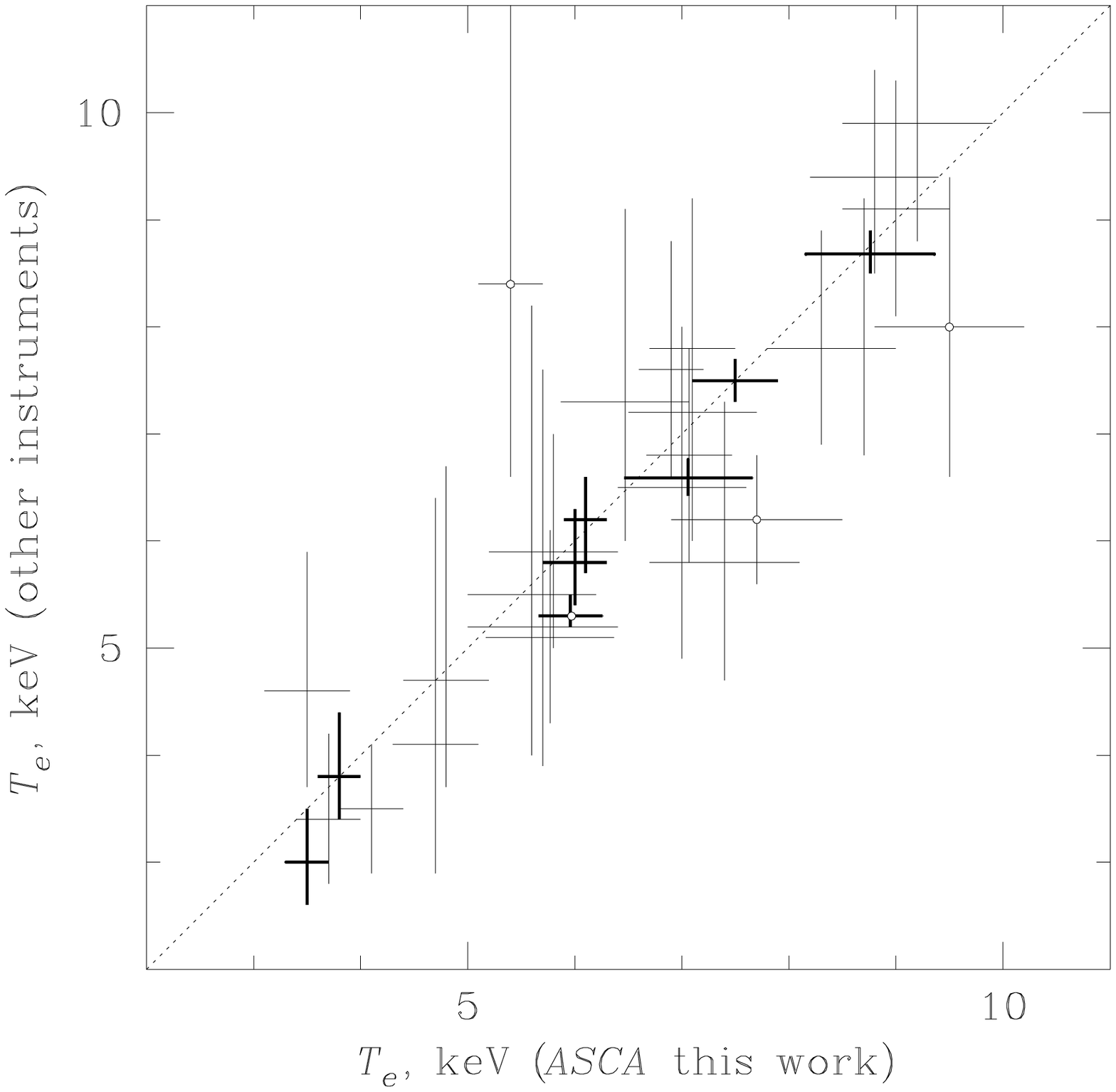}}

\rput[tl]{0}(-0.45,1){
\begin{minipage}{8.75cm}
\small\parindent=3.5mm
{\sc Fig.}~1.---Comparison of our temperatures with those from \einstein\
MPC, \exosat\ and \ginga\ (David et al.\ 1993; Ueno et al.\ 1994). Errors
are 90\%. Values from more than one earlier instrument are shown if they
disagreed. Bold crosses denote the most accurate of the David et al.\
values.  Circles denote the most notable discrepancies, discussed in text.
\end{minipage}
}
\endpspicture

Having performed this consistency check, we show below that
single-temperature values often have little meaning, because clusters are
nonisothermal and many have strong cooling flow contributions.

\begin{figure*}[thbp]
\pspicture(0,-1)(18.5,23.3)

\rput[tl]{0}(0.5,23.2){\epsfxsize=8.5cm
\epsffile{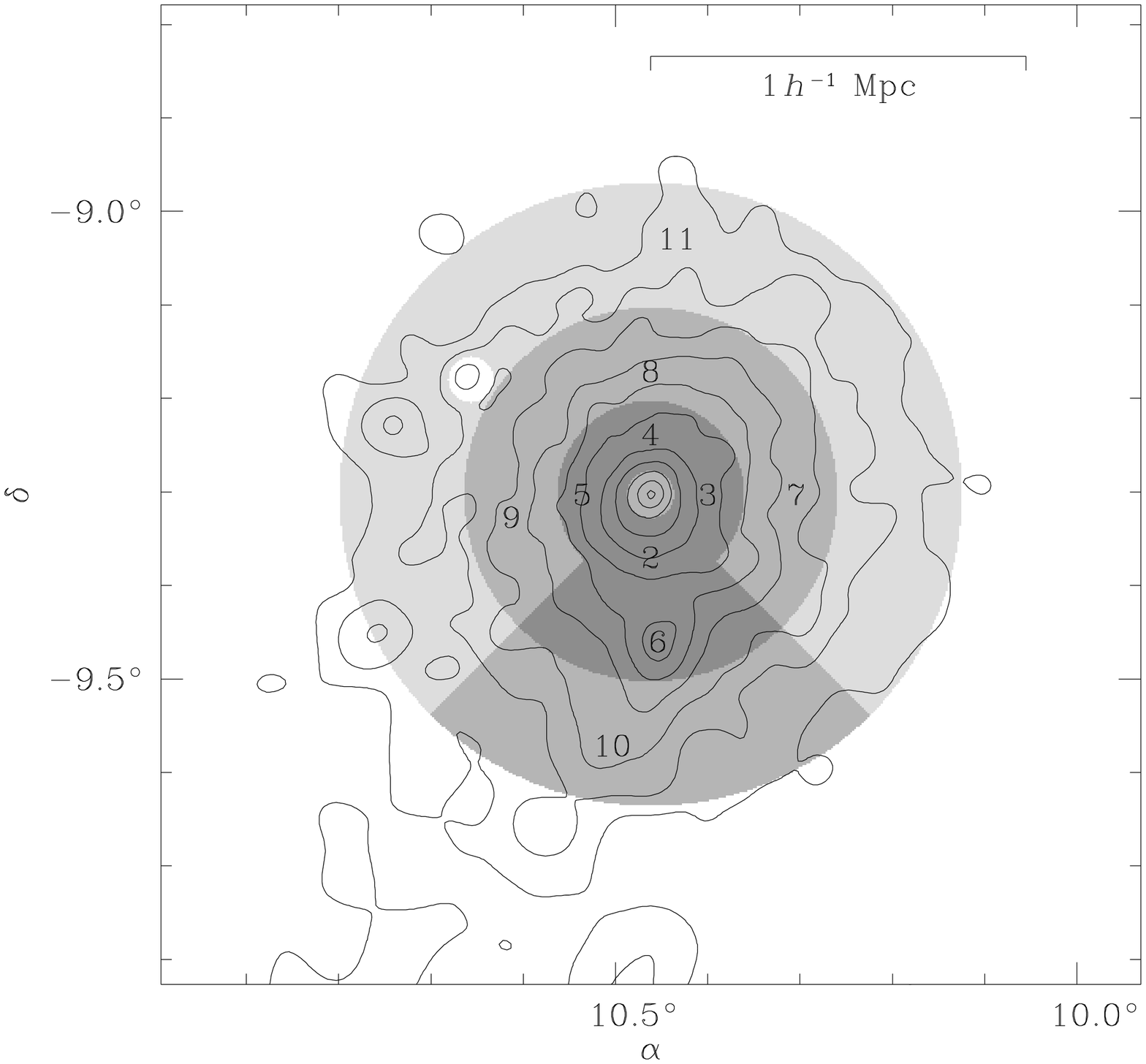}}
\rput[tl]{0}(0.75,16.2){\epsfxsize=7.75cm
\epsffile[30 428 530 678]{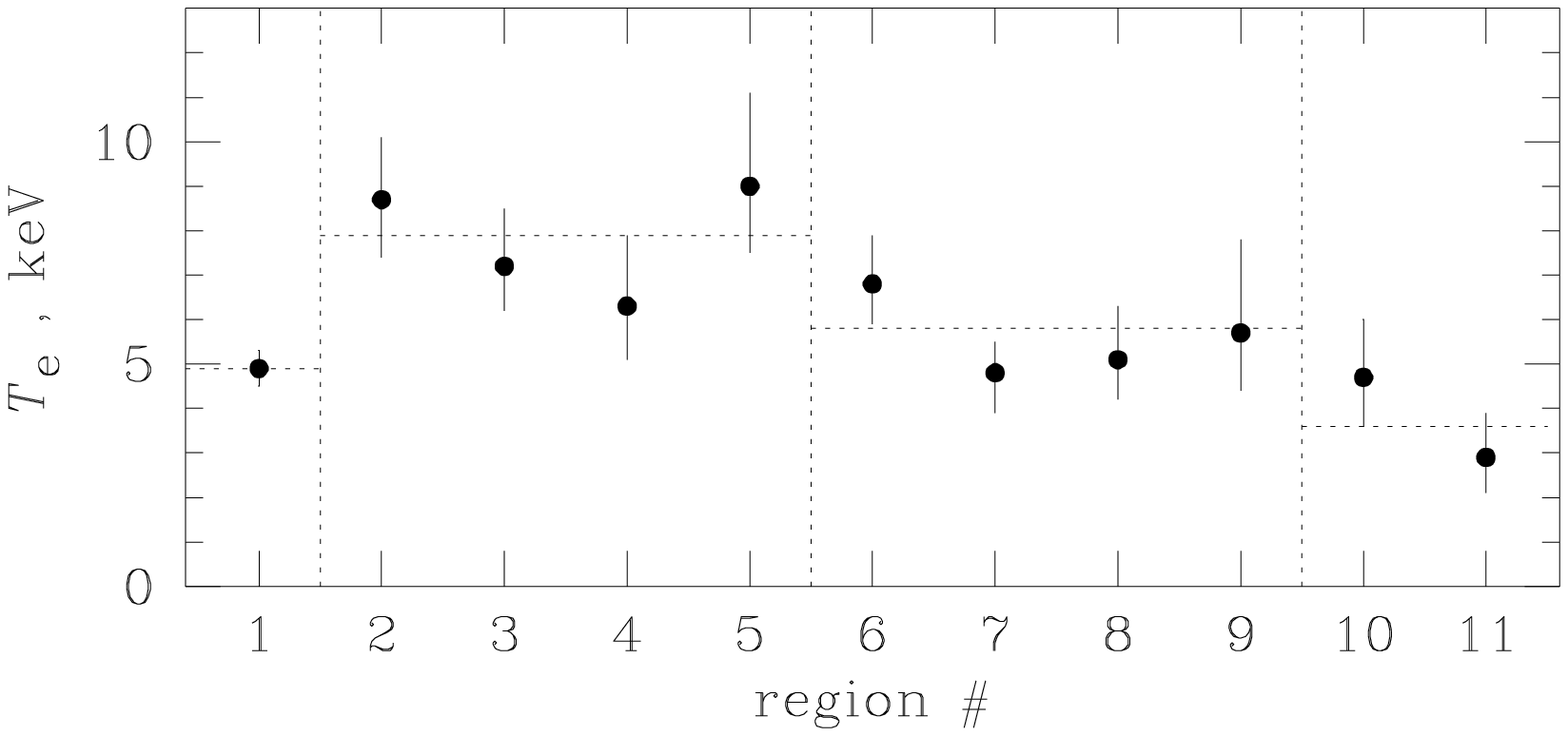}}

\rput[tl]{0}(9.5,23.2){\epsfxsize=8.5cm
\epsffile{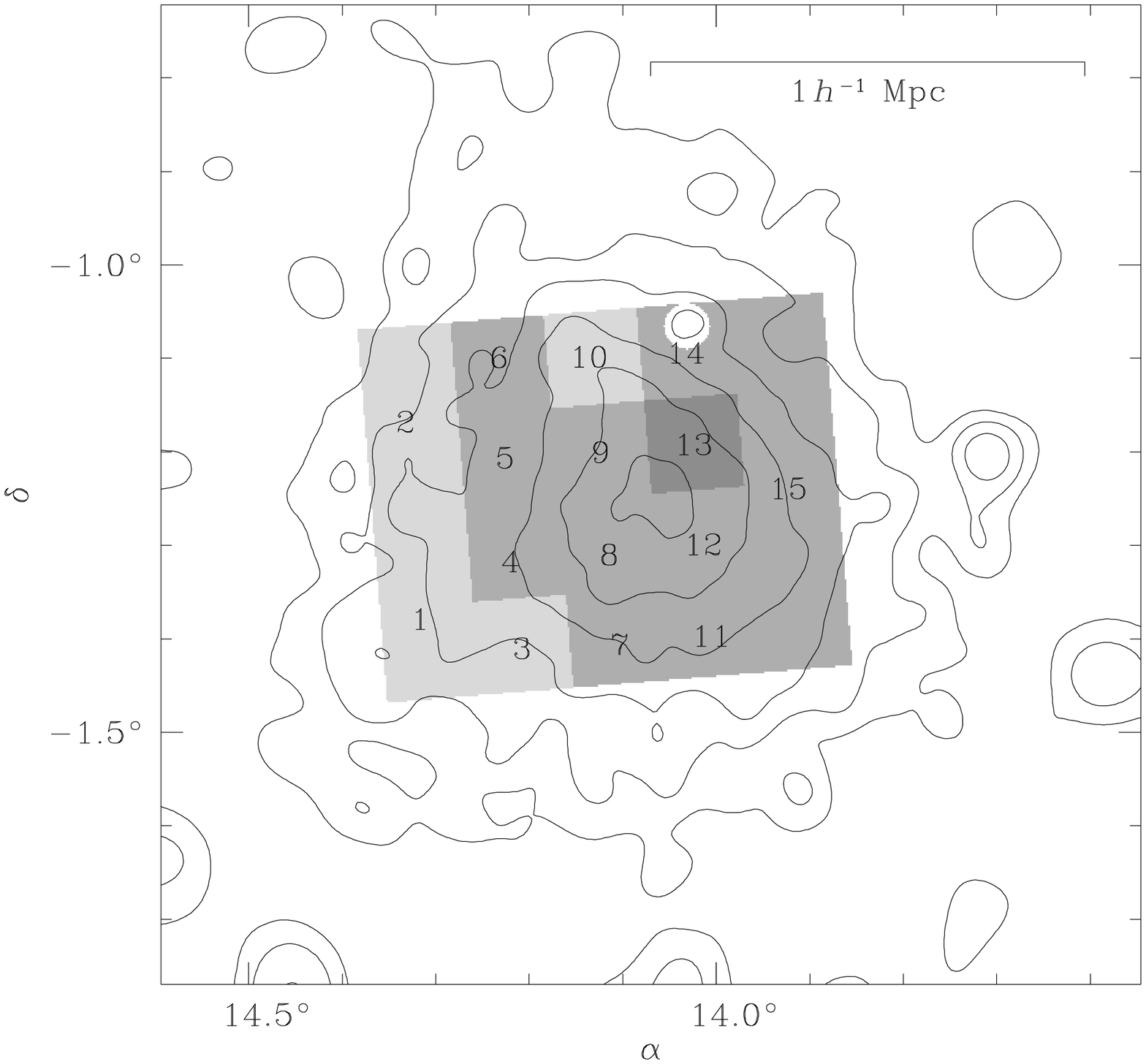}}
\rput[tl]{0}(9.75,16.2){\epsfxsize=7.75cm
\epsffile[30 428 530 678]{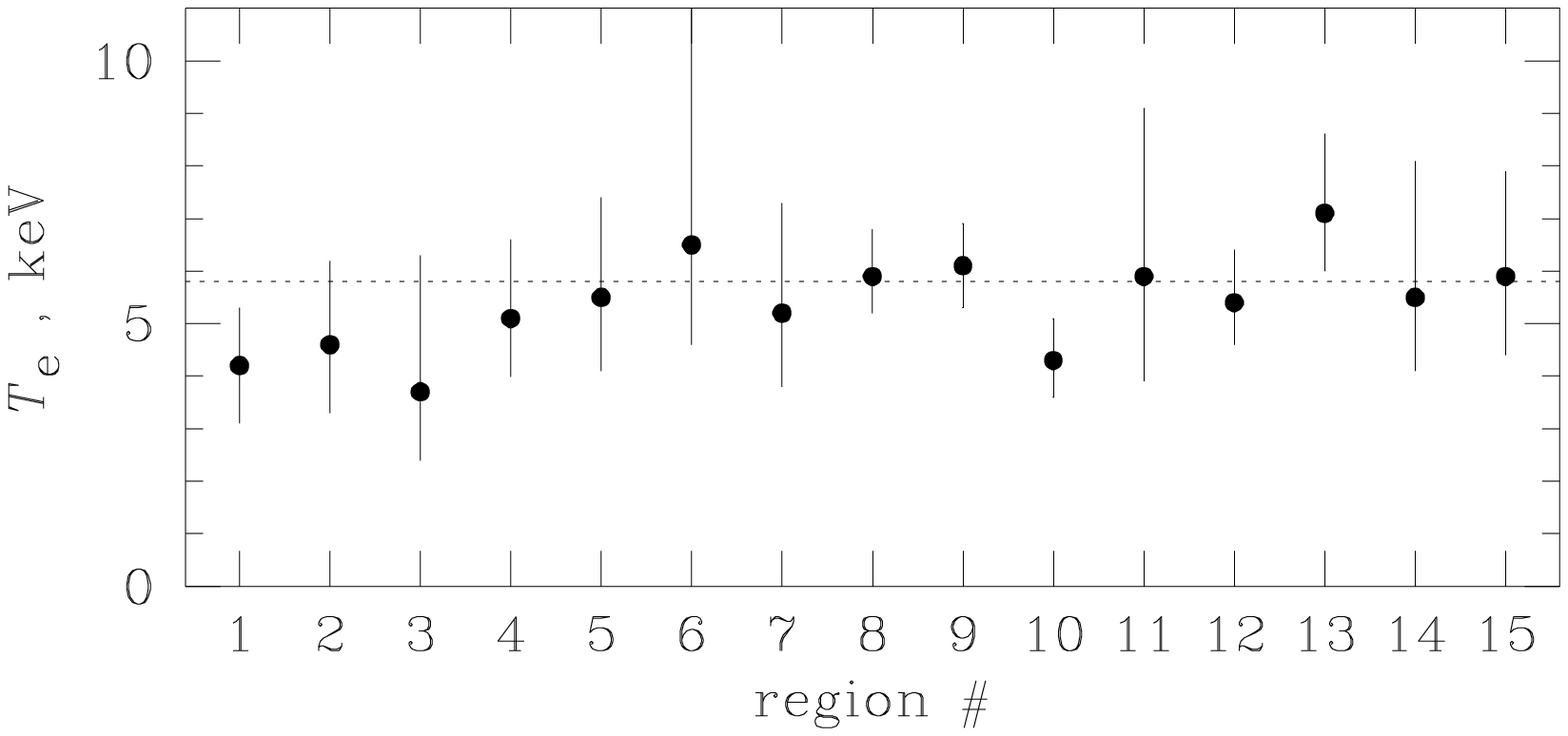}}

\rput[tl]{0}(0.5,11.9){\epsfxsize=8.5cm
\epsffile{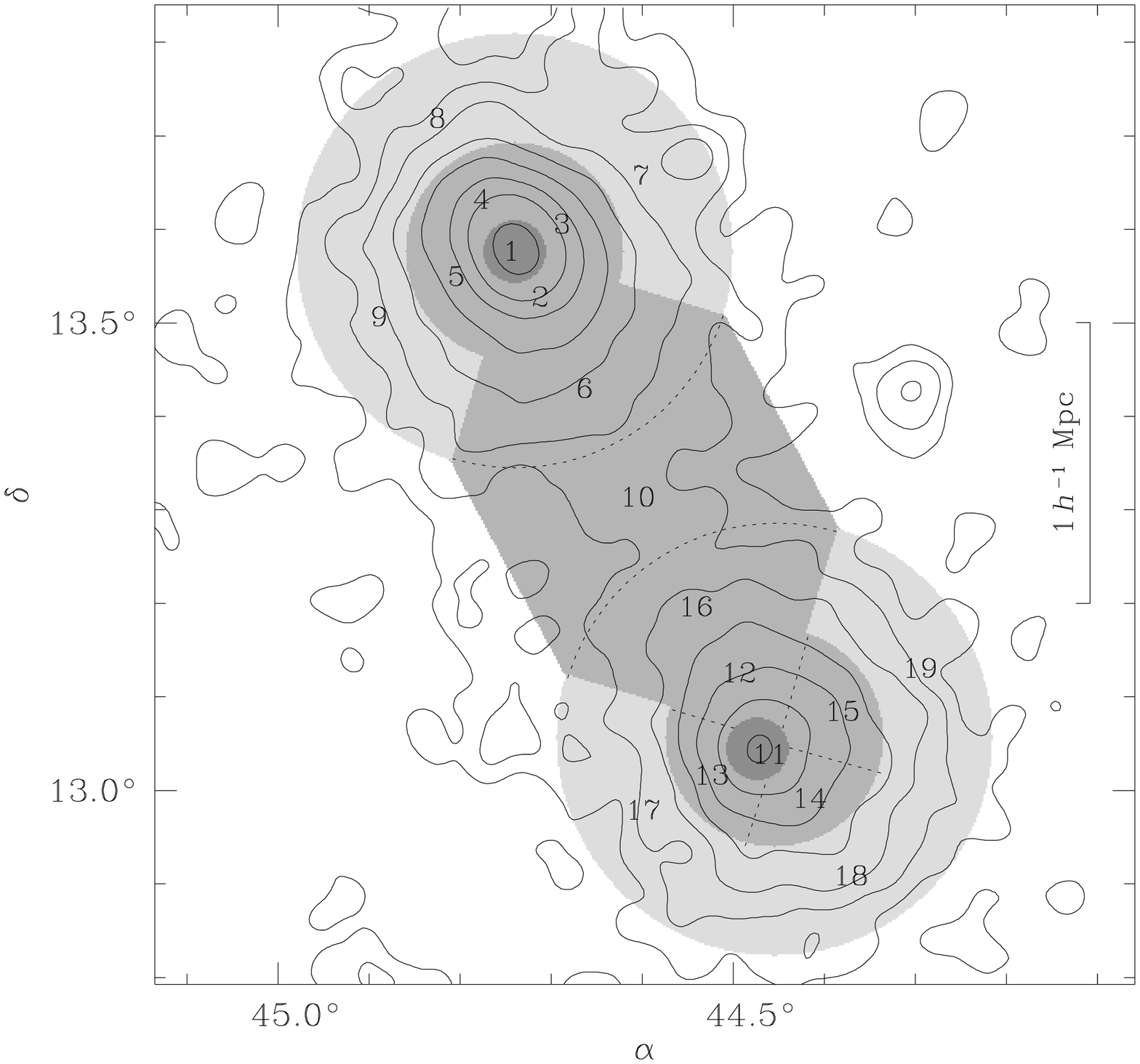}}
\rput[tl]{0}(0.75,4.9){\epsfxsize=7.75cm
\epsffile[30 428 530 678]{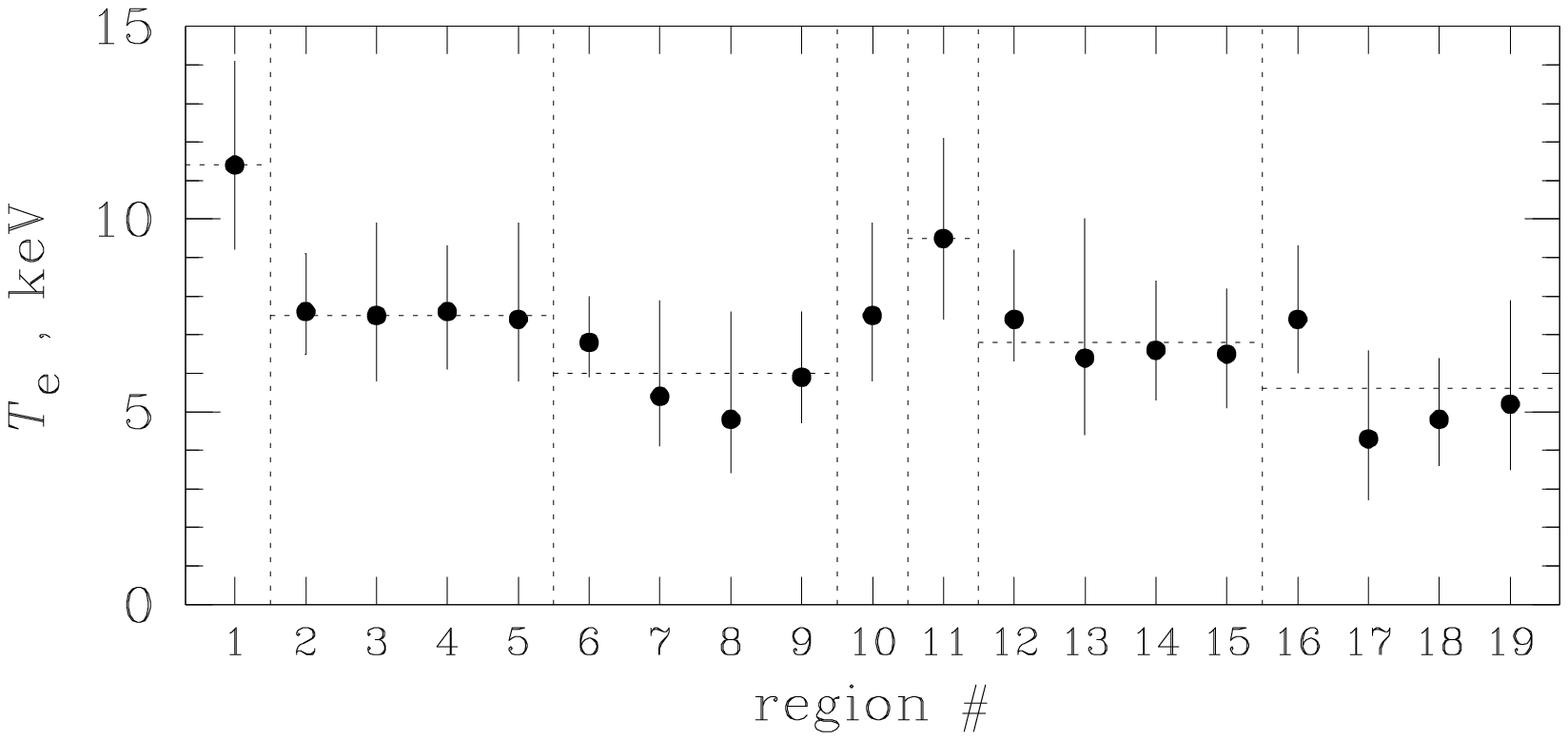}}

\rput[tl]{0}(9.5,11.9){\epsfxsize=8.5cm
\epsffile{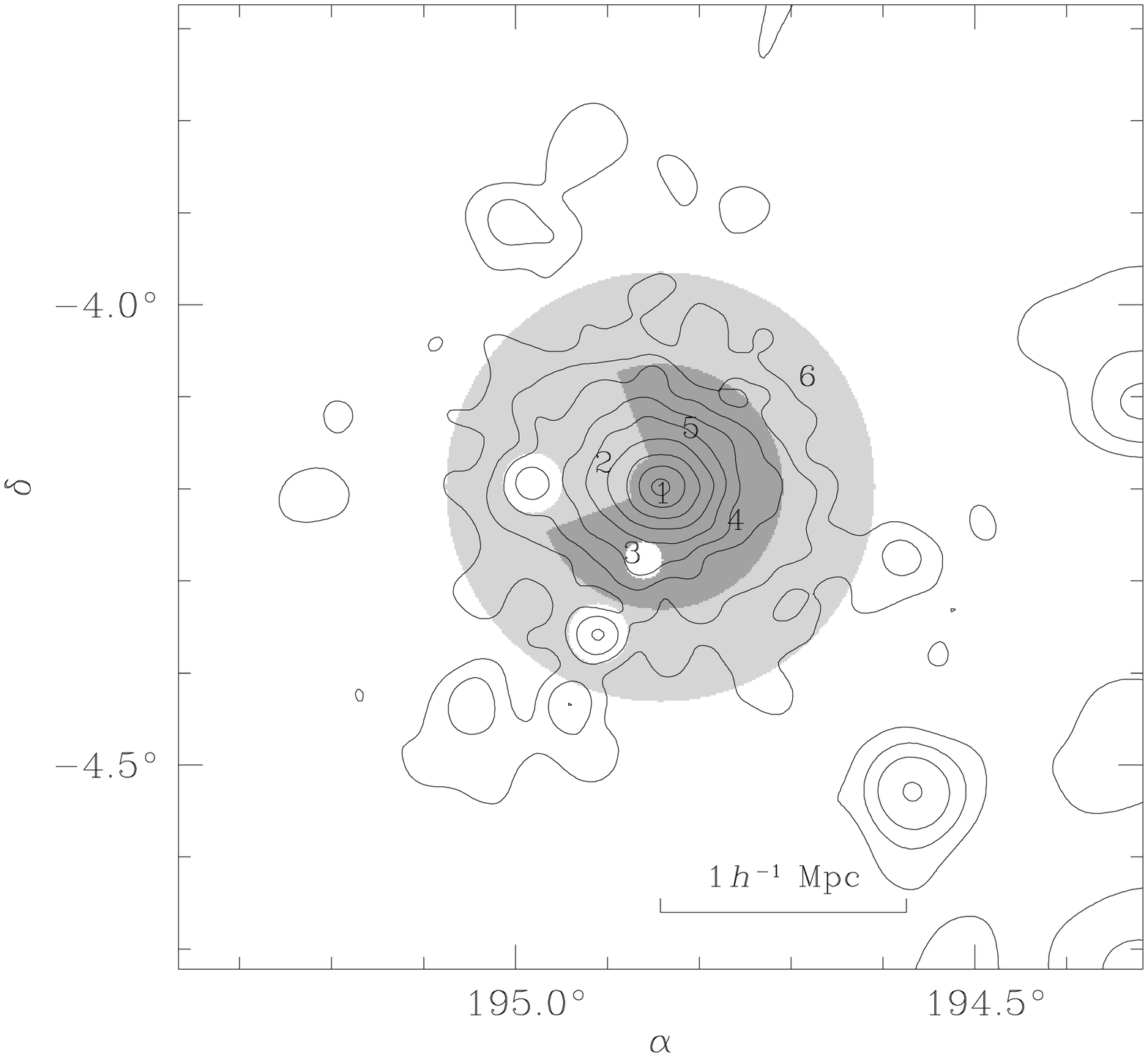}}
\rput[tl]{0}(9.75,4.9){\epsfxsize=7.75cm
\epsffile[30 428 530 678]{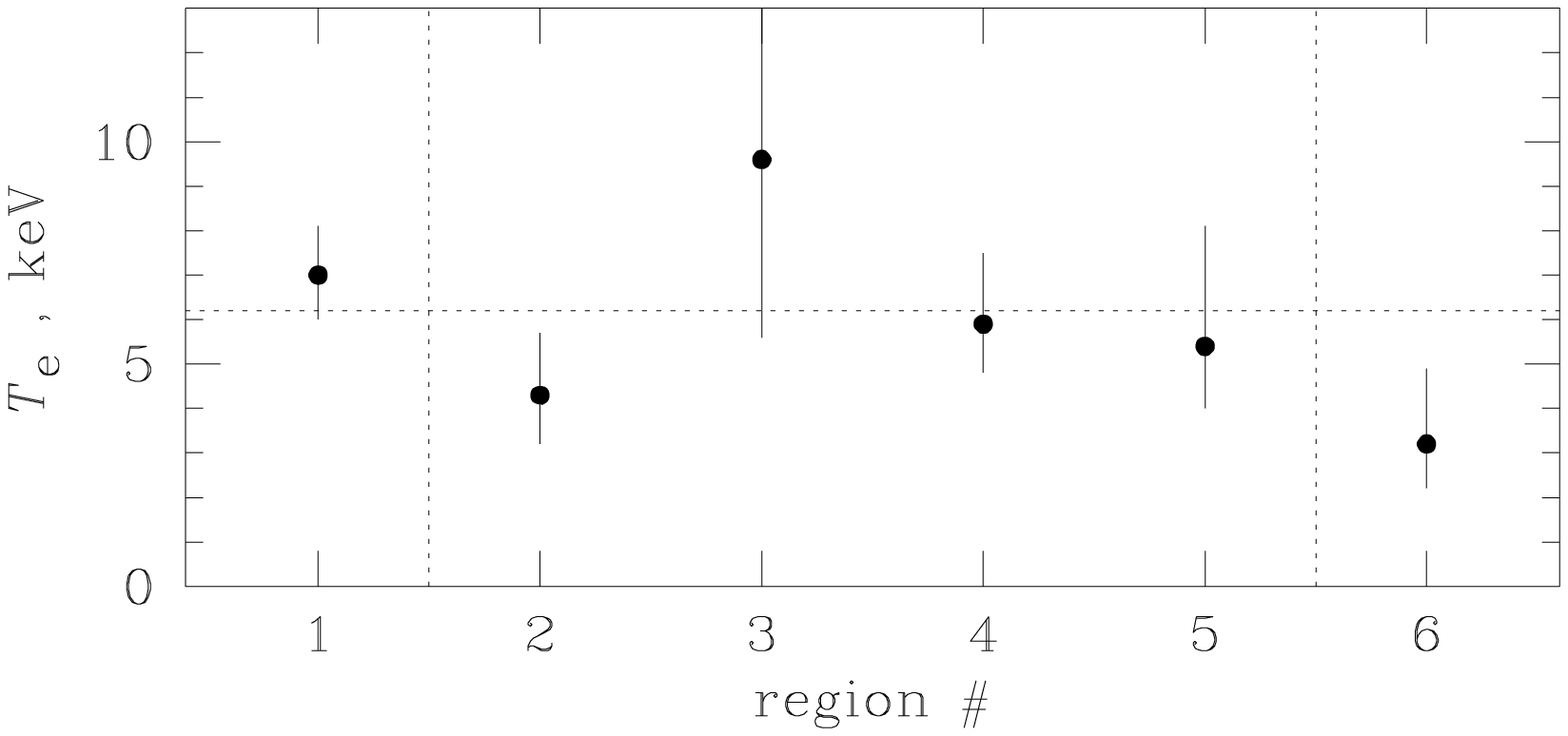}}

\rput[lc]{0}( 2.1,22.5){\small A85}
\rput[lc]{0}(11.1,22.5){\small A119}
\rput[lc]{0}( 2.1,11.2){\small A401}
\rput[lc]{0}( 2.1, 6.0){\small A399}
\rput[lc]{0}(11.1,11.2){\small A1651}

\rput[tl]{0}(0,0.5){
\begin{minipage}{18cm}
\small\parindent=3.5mm
{\sc Fig.}~2.---Cluster temperature maps. Contours show \rosat\ PSPC
brightness, grayscale shows \asca\ temperatures. Regions (sectors, boxes, or
their combination, delineated by dotted lines in the nonobvious places) are
numbered and their temperatures with 90\% confidence intervals (including
systematics) are shown in the lower panels. Dotted horizontal lines in lower
panels show emission-weighted average temperatures either in the annulus or
over the whole cluster; dotted vertical lines separate annuli. Point sources
excluded or fitted separately are shown as blank circles, but some of them
are not shown for clarity.
\end{minipage}
}
\endpspicture
\end{figure*}

\begin{figure*}[htbp]
\pspicture(0,0.7)(18.5,23.3)

\rput[tl]{0}(0.5,23.2){\epsfxsize=8.5cm
\epsffile{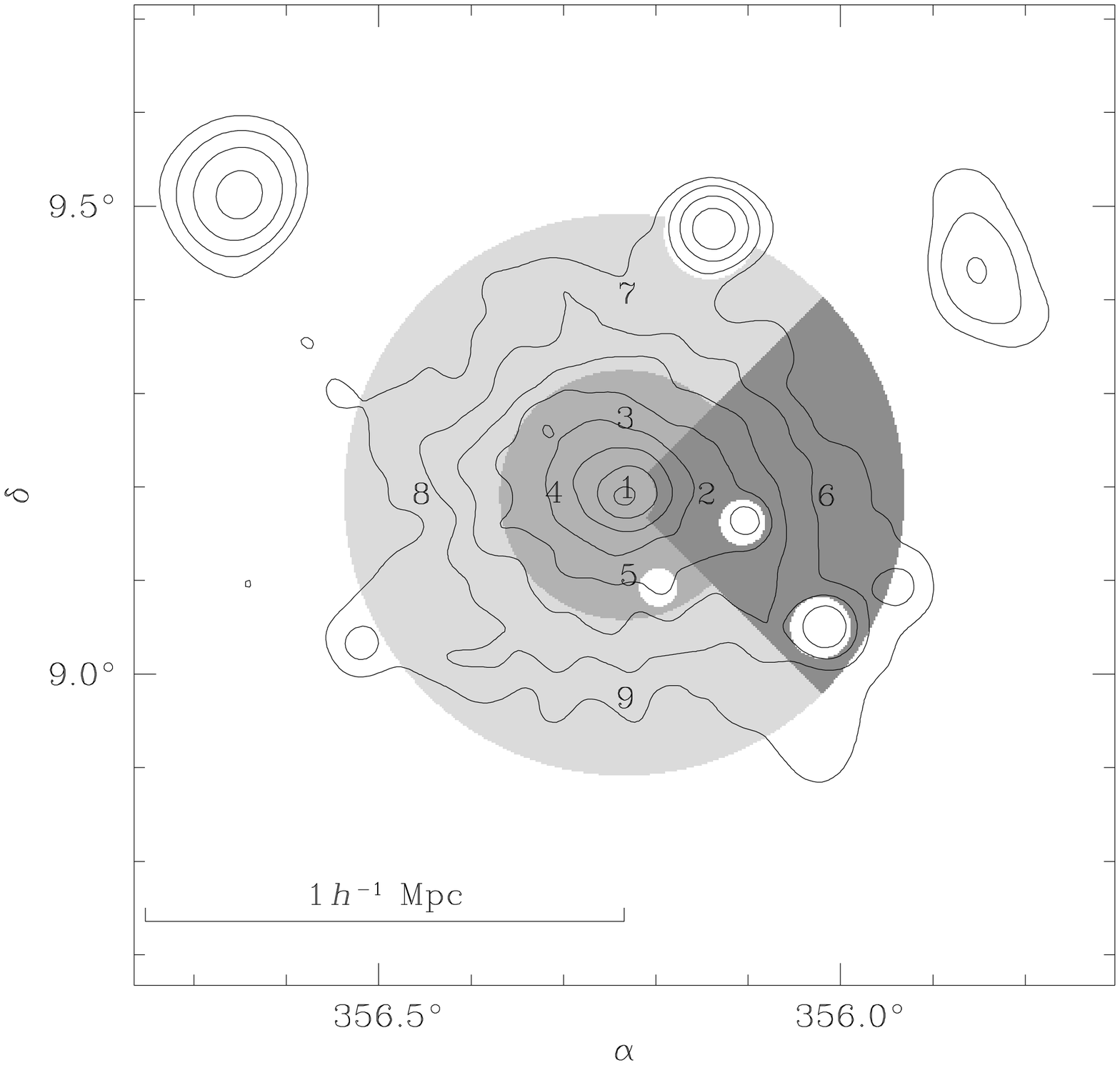}}
\rput[tl]{0}(0.75,16.2){\epsfxsize=7.75cm
\epsffile[30 428 530 678]{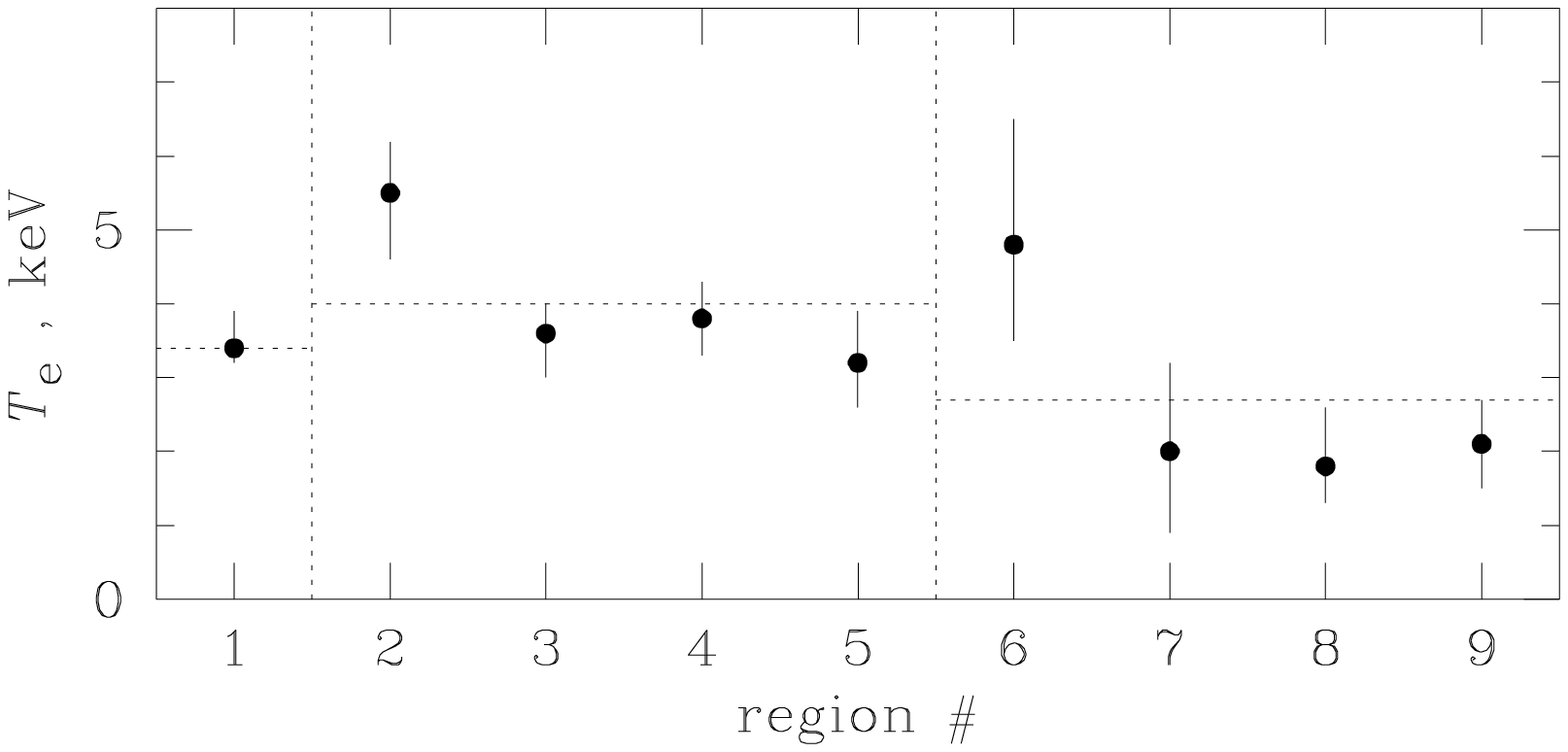}}

\rput[tl]{0}(9.5,23.2){\epsfxsize=8.5cm
\epsffile{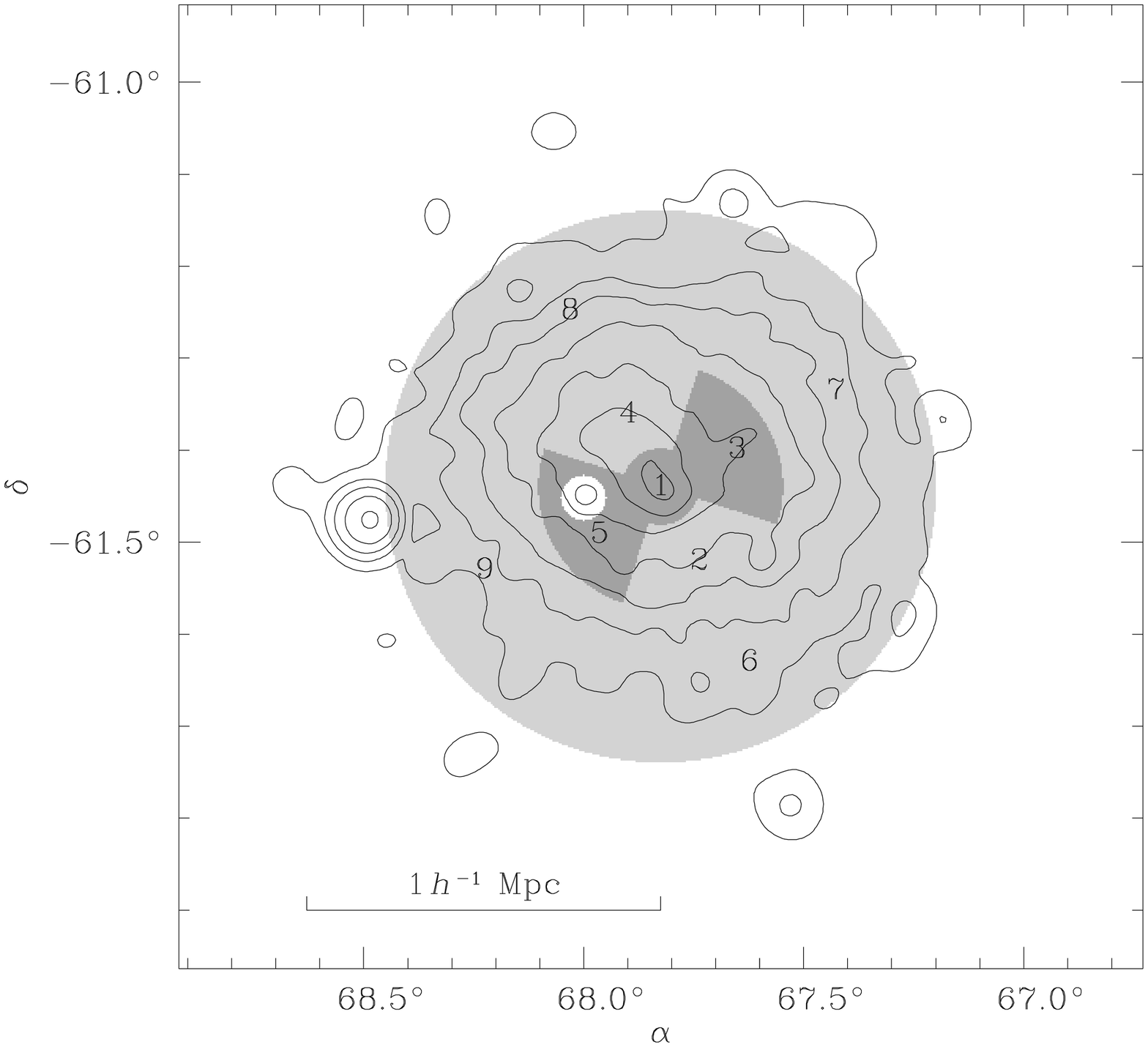}}
\rput[tl]{0}(9.75,16.2){\epsfxsize=7.75cm
\epsffile[30 428 530 678]{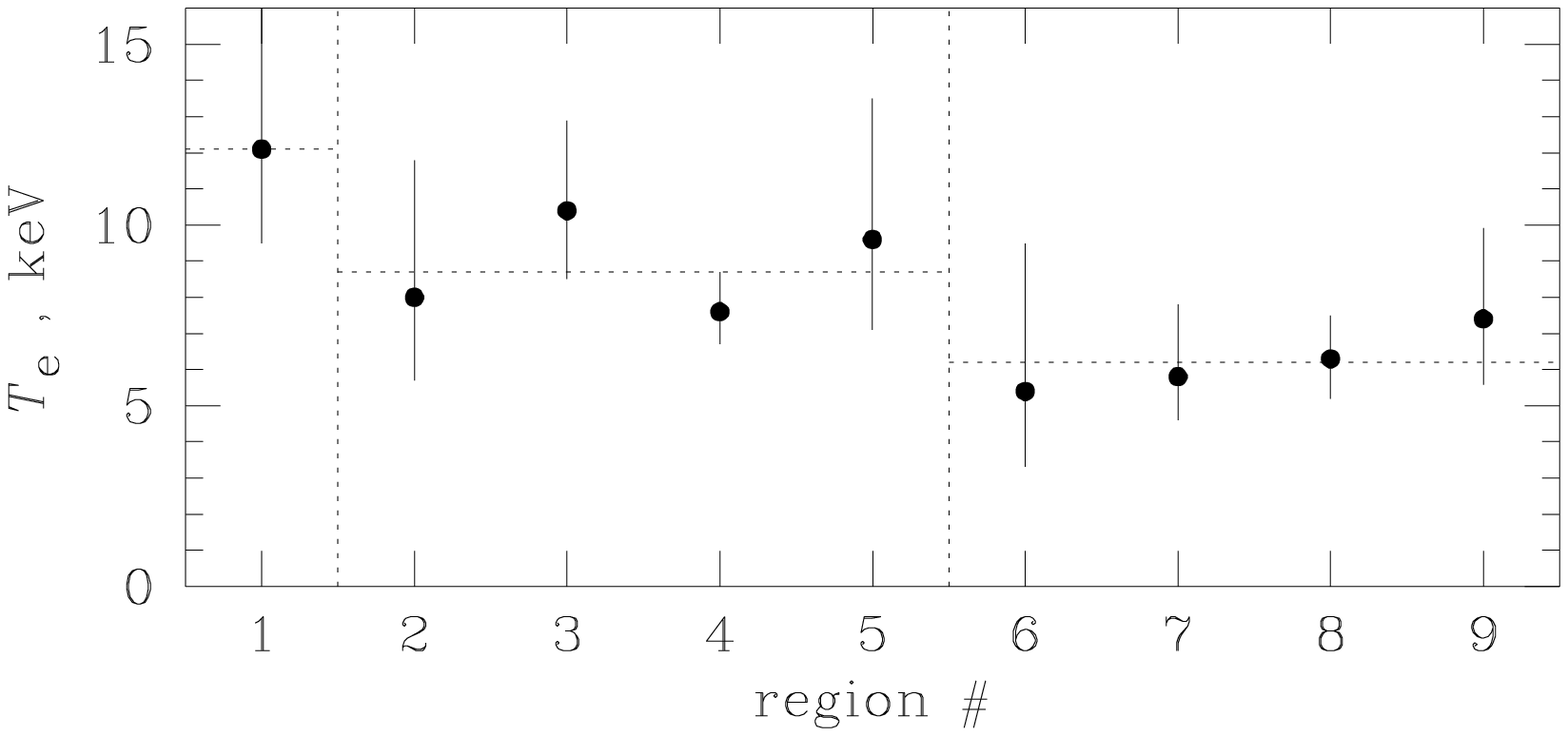}}

\rput[tl]{0}(0.5,11.9){\epsfxsize=8.5cm
\epsffile{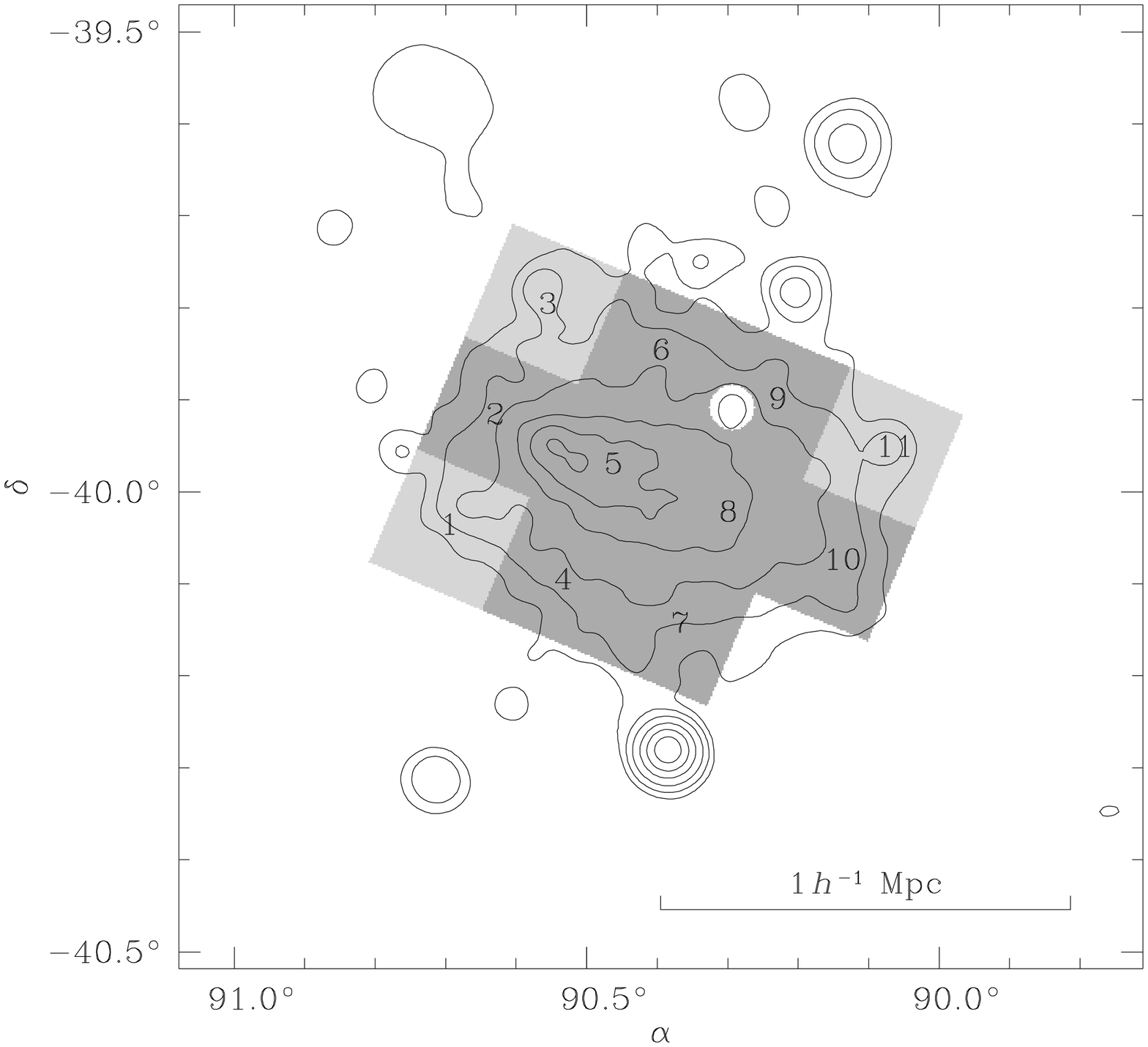}}
\rput[tl]{0}(0.75,4.9){\epsfxsize=7.75cm
\epsffile[30 428 530 678]{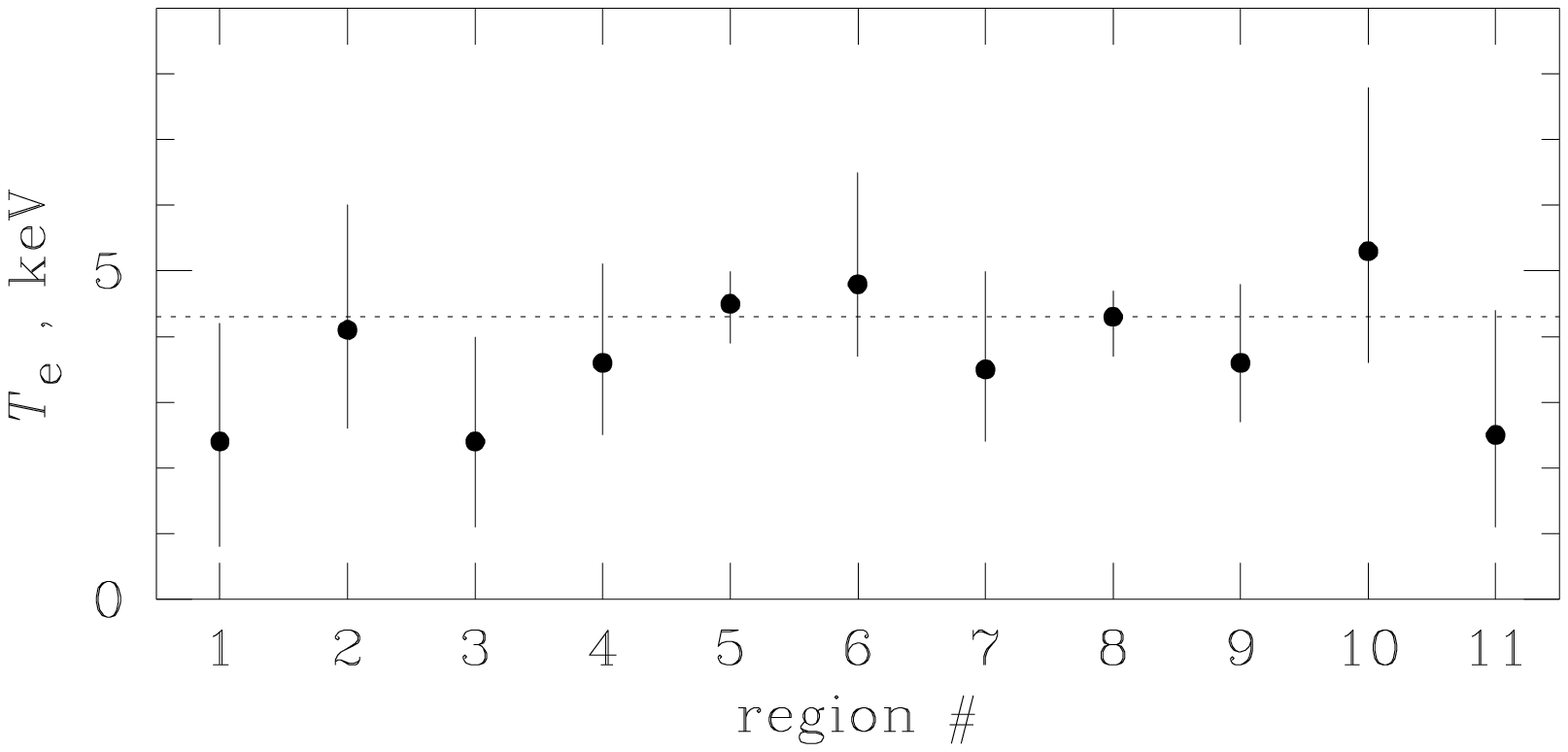}}

\rput[tl]{0}(9.5,11.9){\epsfxsize=8.5cm
\epsffile{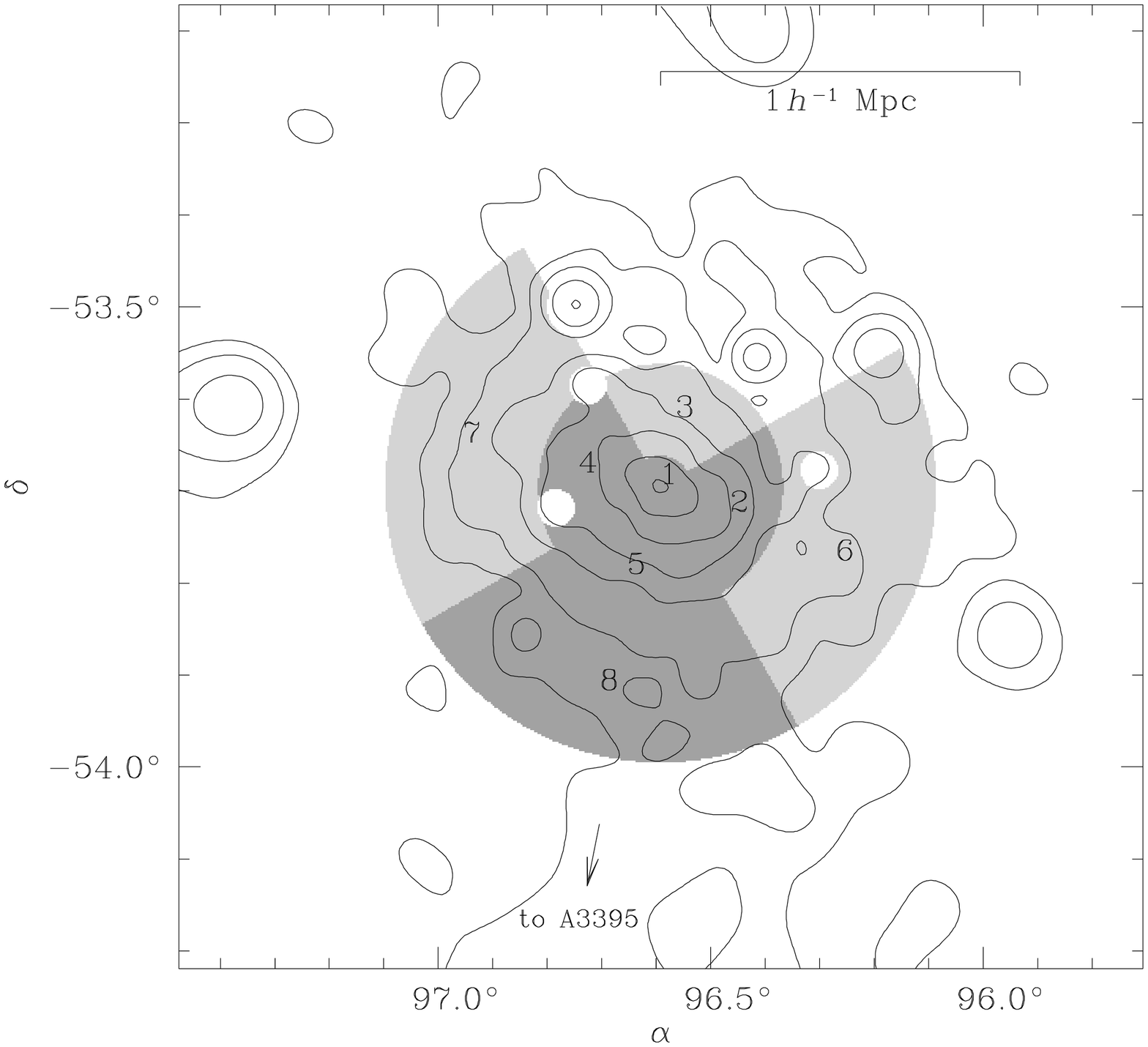}}
\rput[tl]{0}(9.75,4.9){\epsfxsize=7.75cm
\epsffile[30 428 530 678]{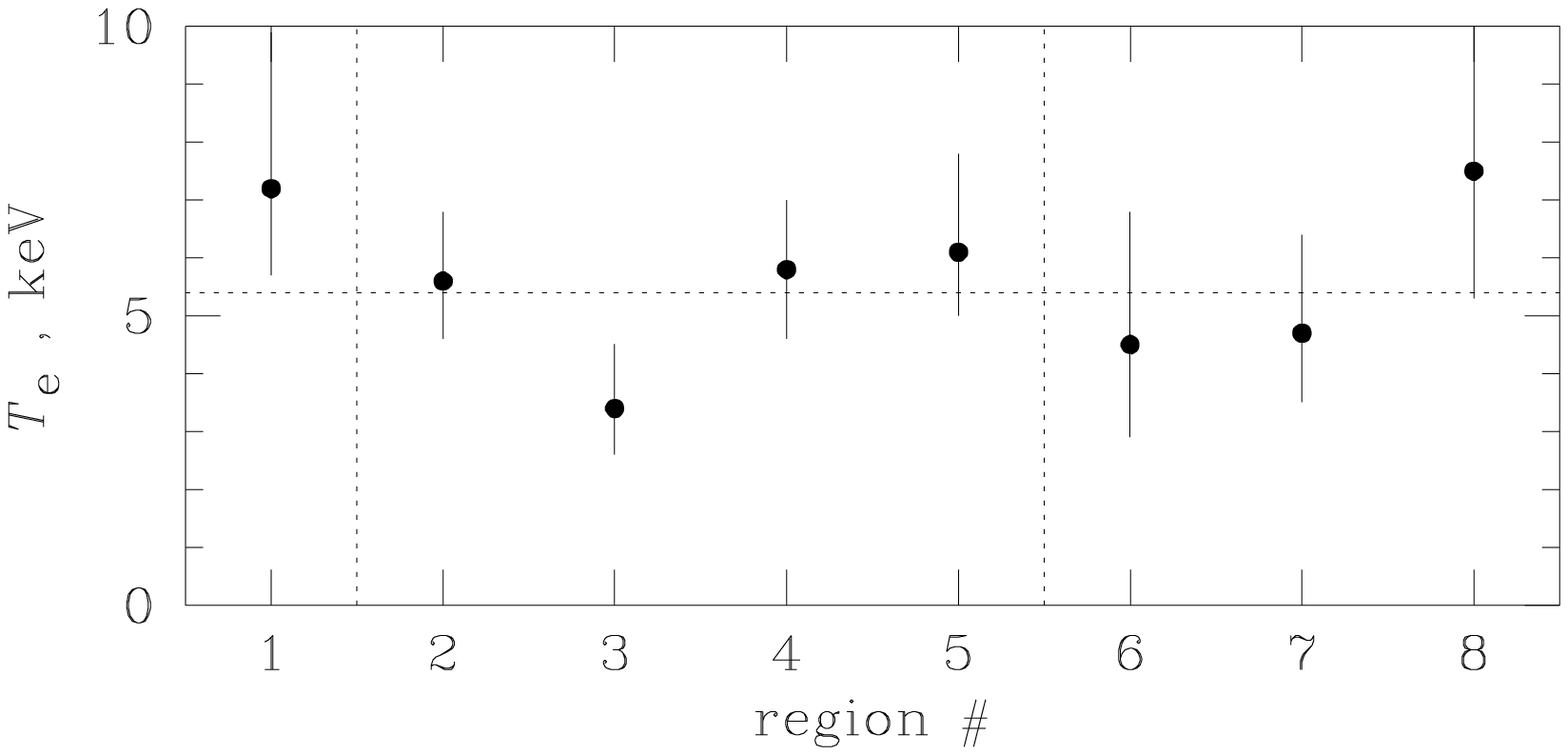}}

\rput[lc]{0}( 2.1,22.5){\small A2657}
\rput[lc]{0}(11.1,22.5){\small A3266}
\rput[lc]{0}( 2.1,11.2){\small A3376}
\rput[lc]{0}(11.1,11.2){\small A3391}

\rput[tl]{0}(0,0.5){
\begin{minipage}{18cm}
\small\parindent=3.5mm
{\sc Fig.}~2.---Continued
\end{minipage}
}
\endpspicture
\end{figure*}

\begin{figure*}[ht]
\pspicture(0,11.7)(18.5,23.3)

\rput[tl]{0}(0.5,23.2){\epsfxsize=8.5cm
\epsffile{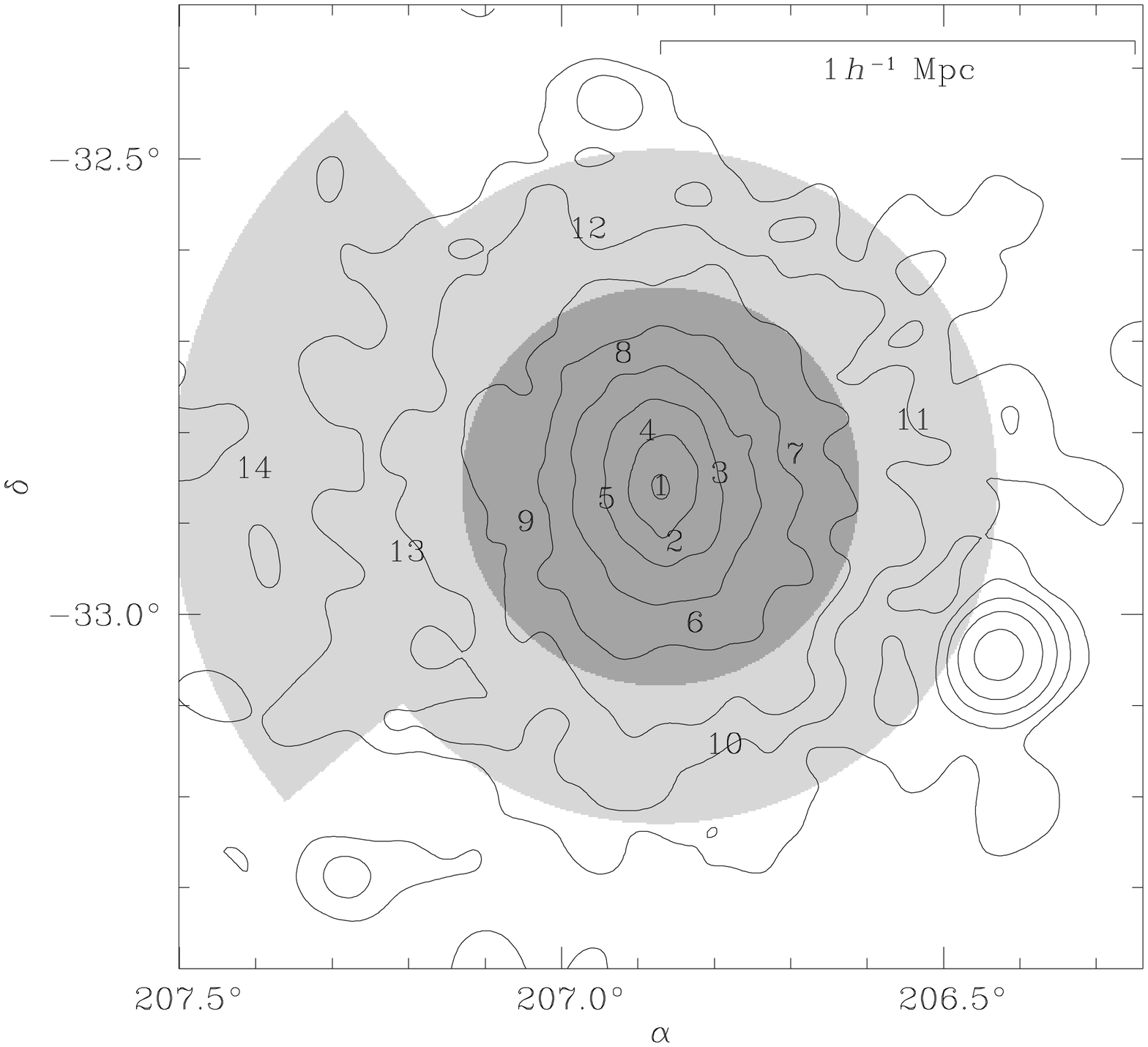}}
\rput[tl]{0}(0.75,16.2){\epsfxsize=7.75cm
\epsffile[30 428 530 678]{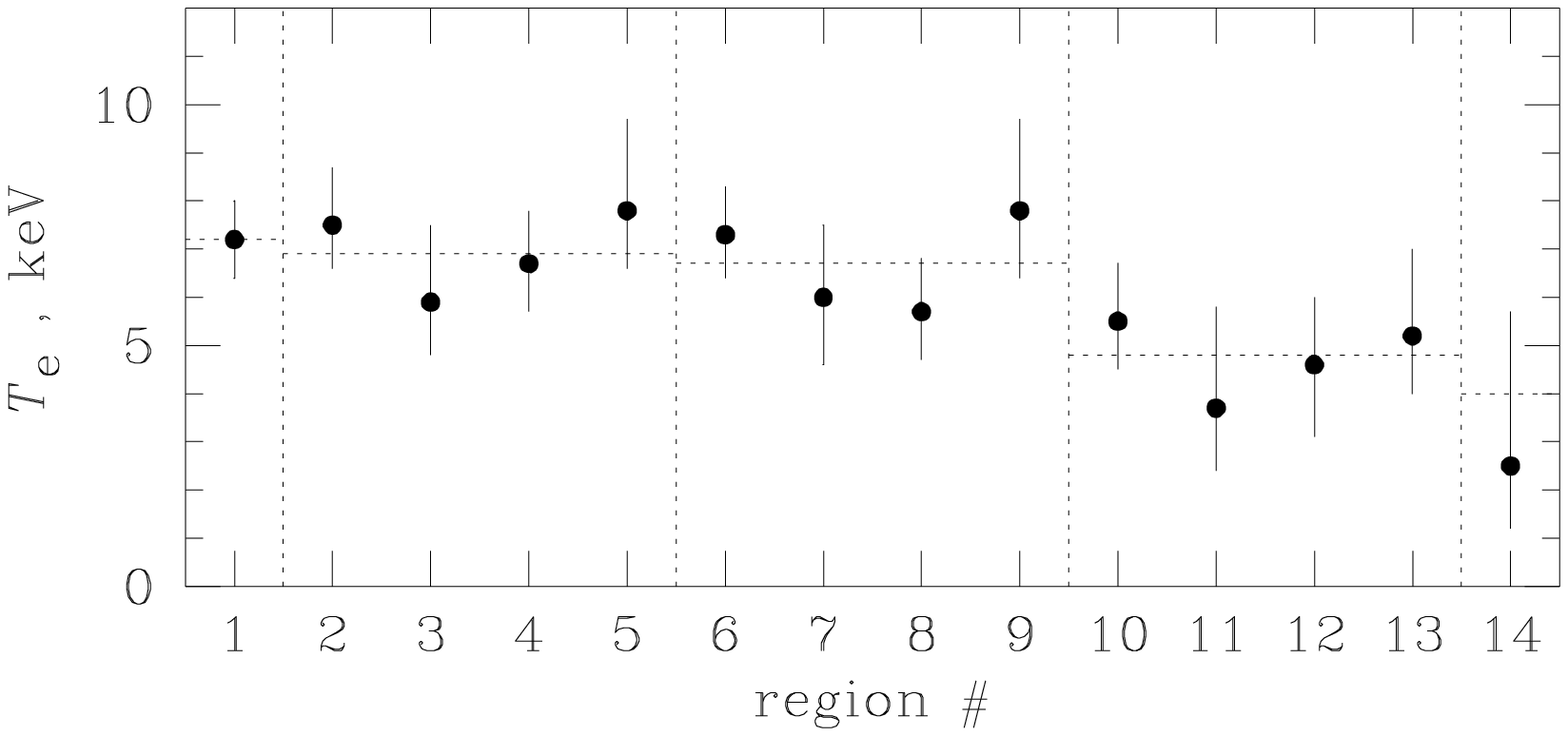}}

\rput[tl]{0}(9.5,23.2){\epsfxsize=8.5cm
\epsffile{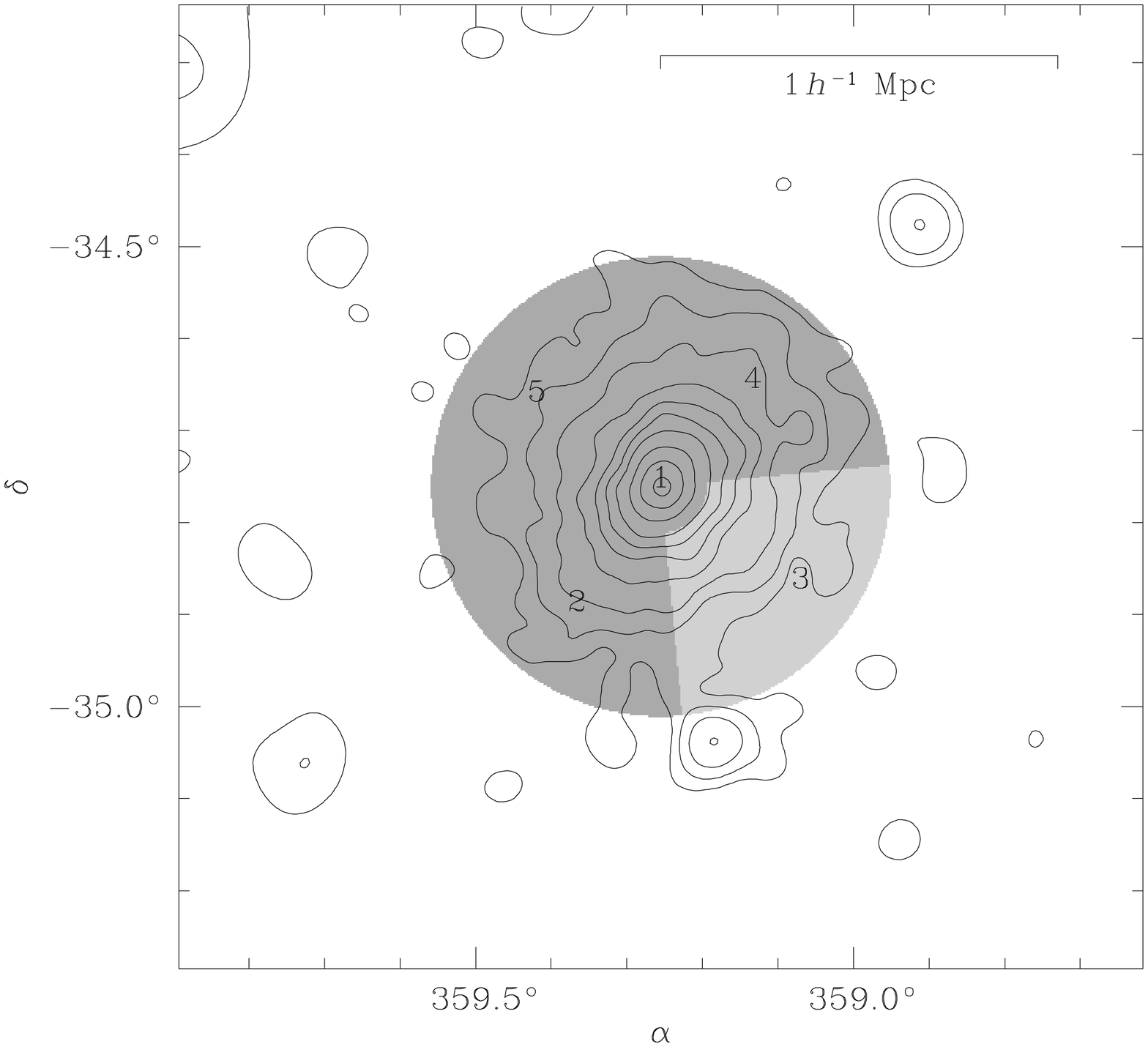}}
\rput[tl]{0}(9.75,16.2){\epsfxsize=7.75cm
\epsffile[30 428 530 678]{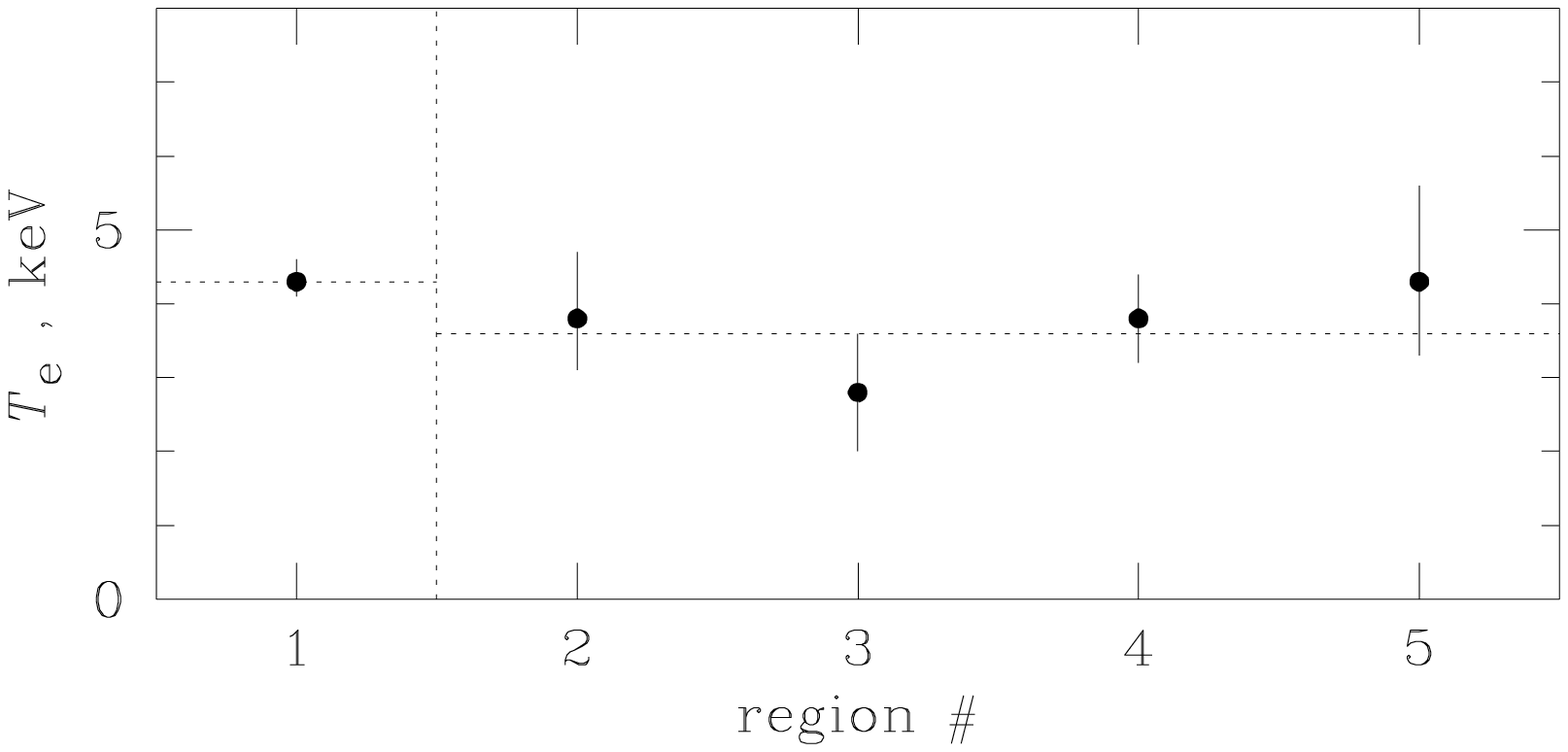}}

\rput[lc]{0}( 2.1,22.5){\small A3571}
\rput[lc]{0}(11.1,22.5){\small A4059}

\rput[tl]{0}(0,11.8){
\begin{minipage}{18cm}
\small\parindent=3.5mm
{\sc Fig.}~2.---Continued
\end{minipage}
}
\endpspicture
\end{figure*}

\begin{figure*}[hbt]
\pspicture(0,10.9)(18.5,23.3)

\rput[tl]{0}(0.5,23.2){\epsfxsize=8.5cm
\epsffile{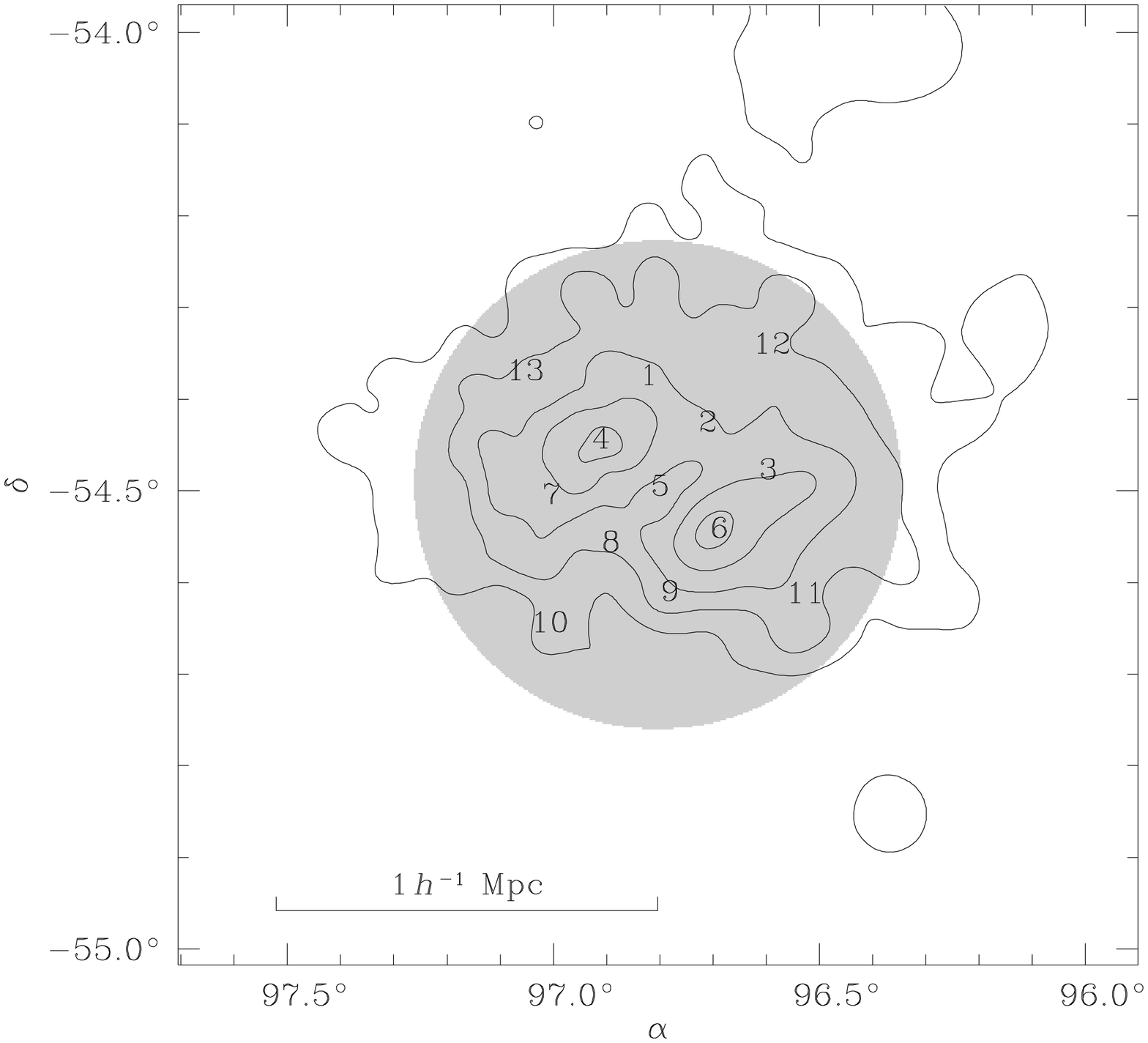}}
\rput[tl]{0}(0.75,16.2){\epsfxsize=7.75cm
\epsffile[30 428 530 678]{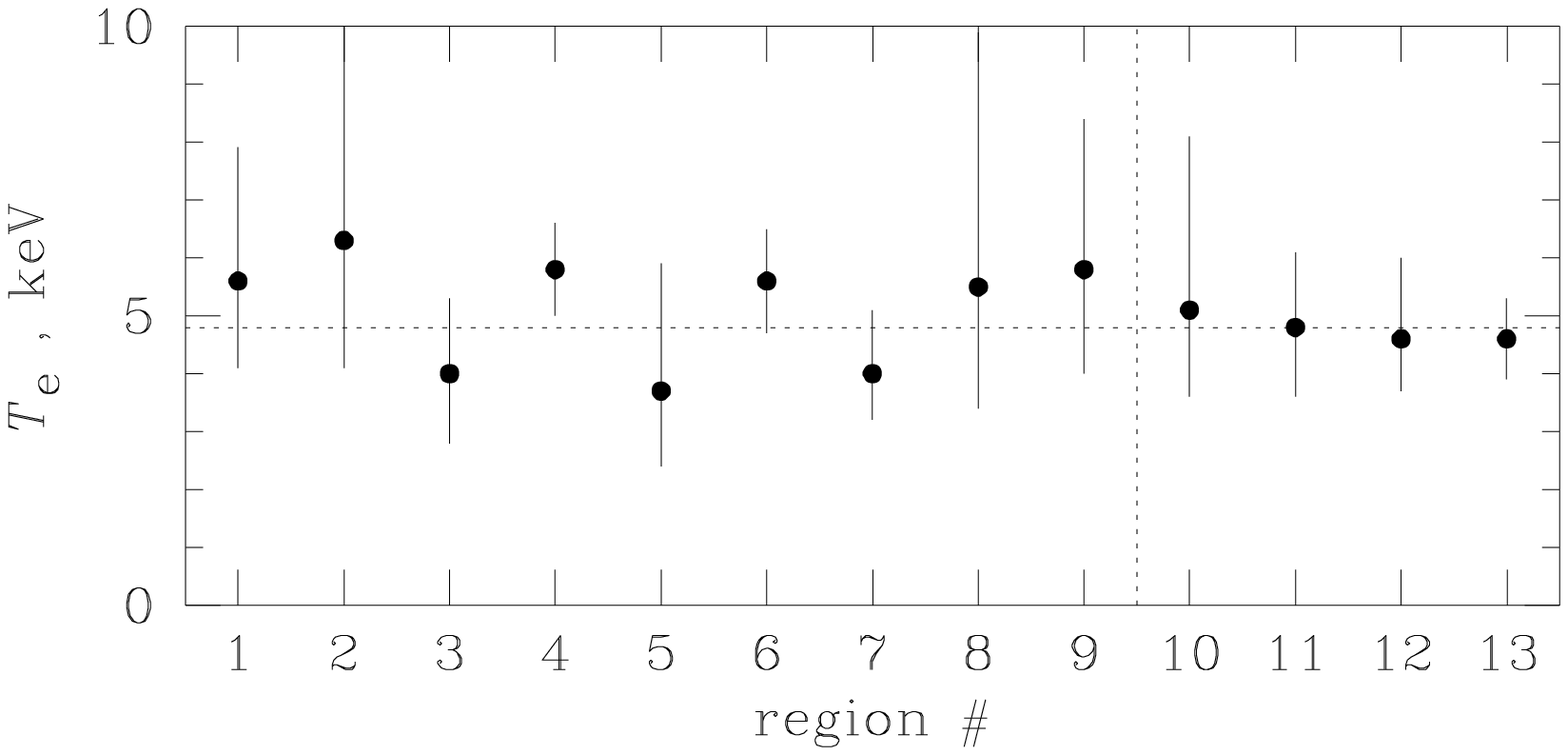}}

\rput[tl]{0}(9.5,23.2){\epsfxsize=8.5cm
\epsffile{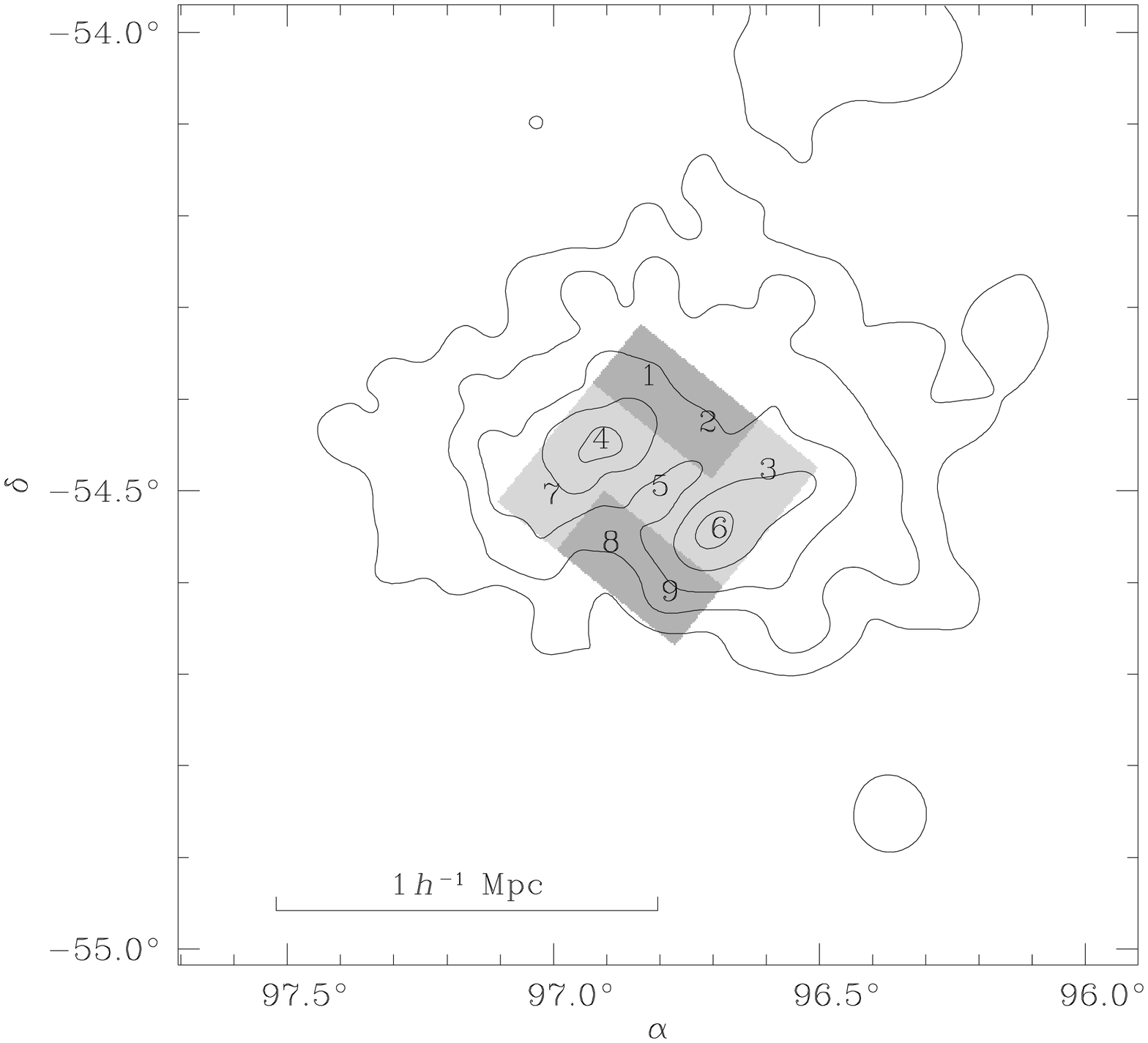}}
\rput[tl]{0}(10.75,16.2){\epsfxsize=7.75cm
\epsffile[30 428 530 678]{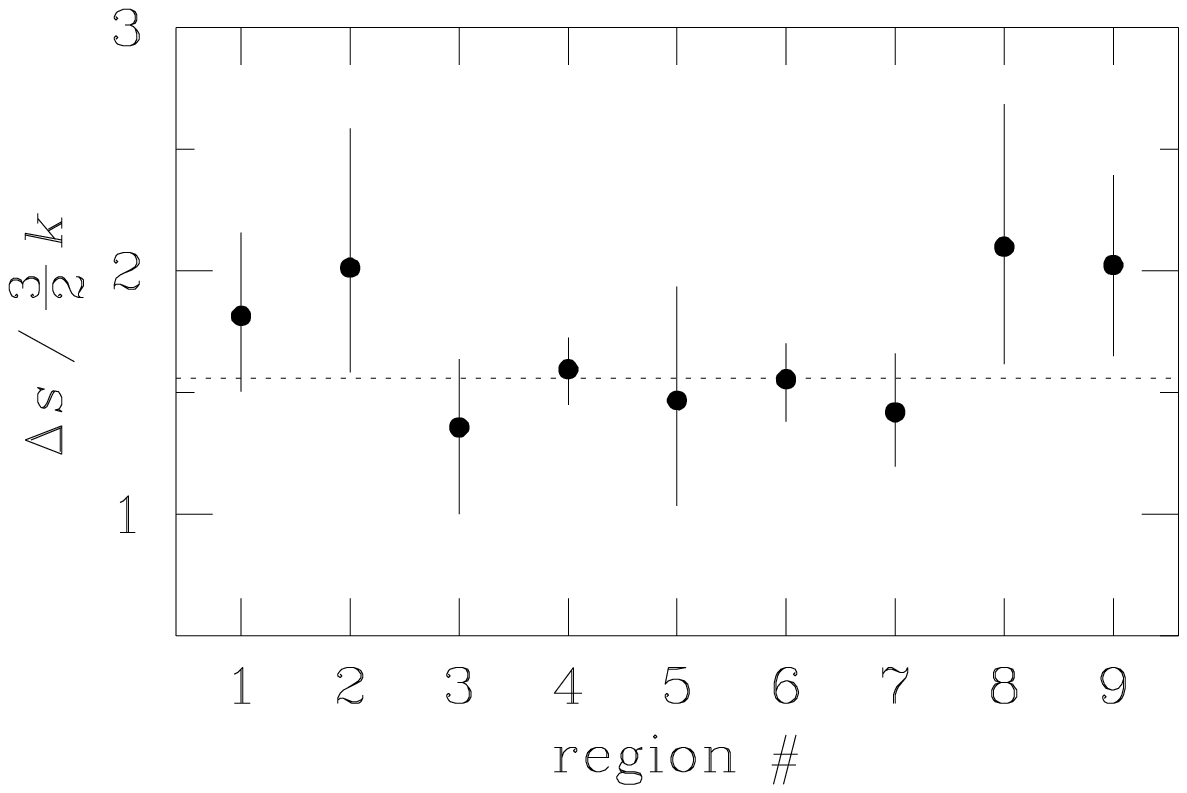}}

\rput[lc]{0}( 2.1,22.5){\small A3395 temperature}
\rput[lc]{0}(11.1,22.5){\small A3395 entropy}

\rput[tl]{0}(0,11.8){
\begin{minipage}{18cm}
\small\parindent=3.5mm
{\sc Fig.}~3.---Temperature and specific entropy maps of A3395 (see text).
Left panels are similar to Fig.\ 2. Lower right panel shows approximate
specific entropy in each region with respect to an arbitrary zero value;
horizontal line denotes a weighted mean value. Grayscale in the upper right
panel shows these specific entropy values overlaid on the image.
\end{minipage}
}
\endpspicture
\end{figure*}

\begin{figure*}[t]
\pspicture(0,2.6)(18.5,23.3)

\rput[tl]{0}(-0.1,23.5){\epsfxsize=6.5cm
\epsffile{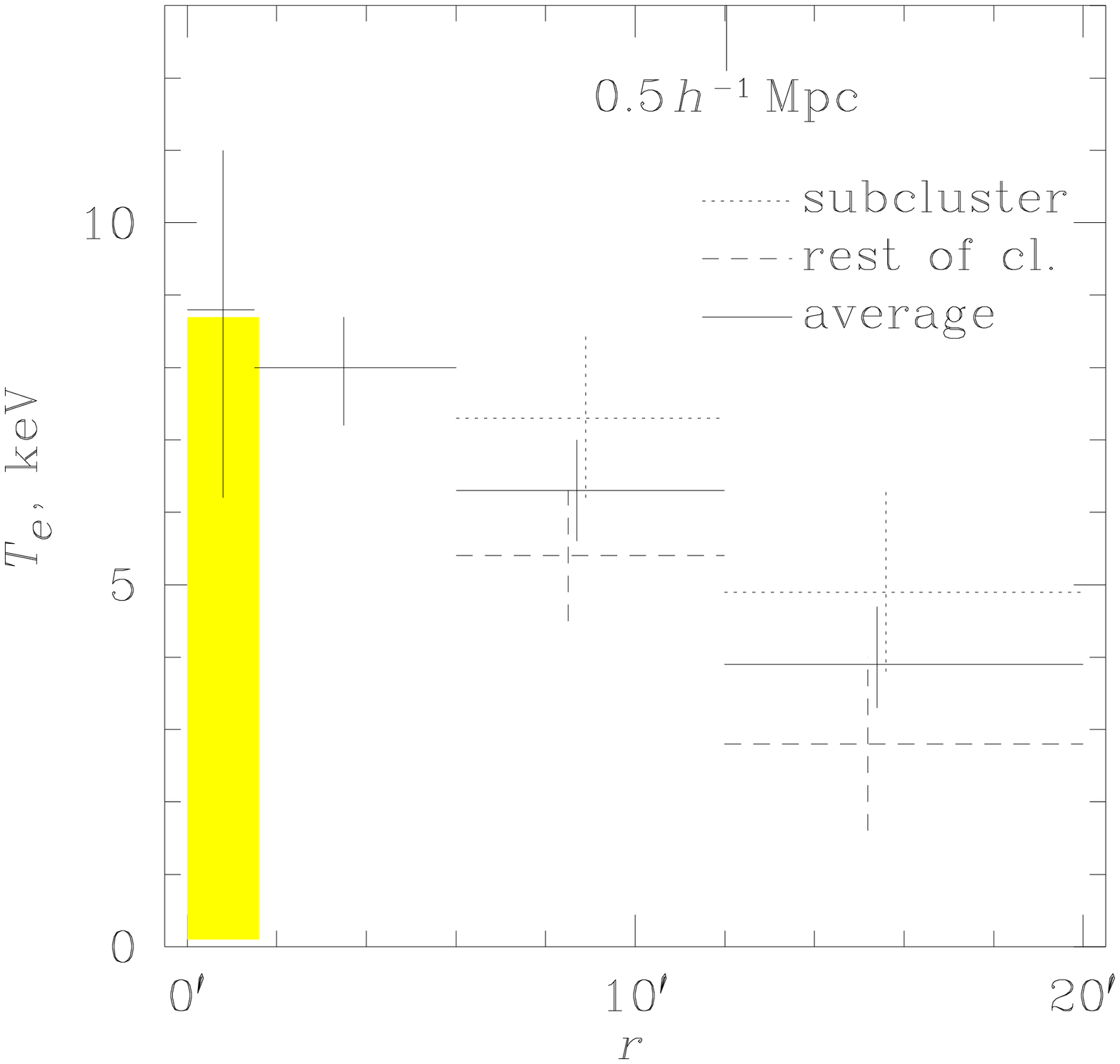}}

\rput[tl]{0}(6.1,23.5){\epsfxsize=6.5cm
\epsffile{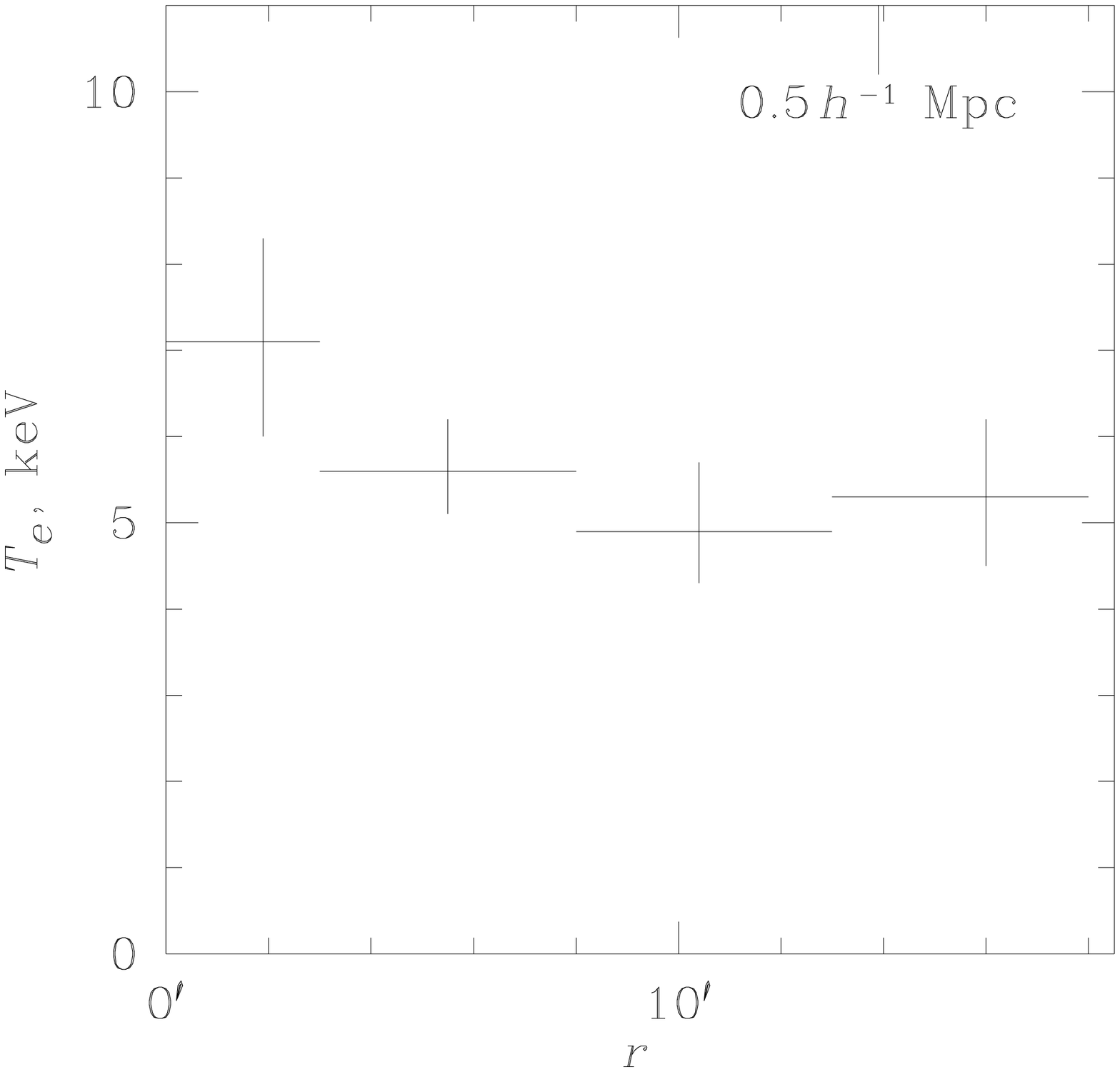}}

\rput[tl]{0}(12.4,23.5){\epsfxsize=6.5cm
\epsffile{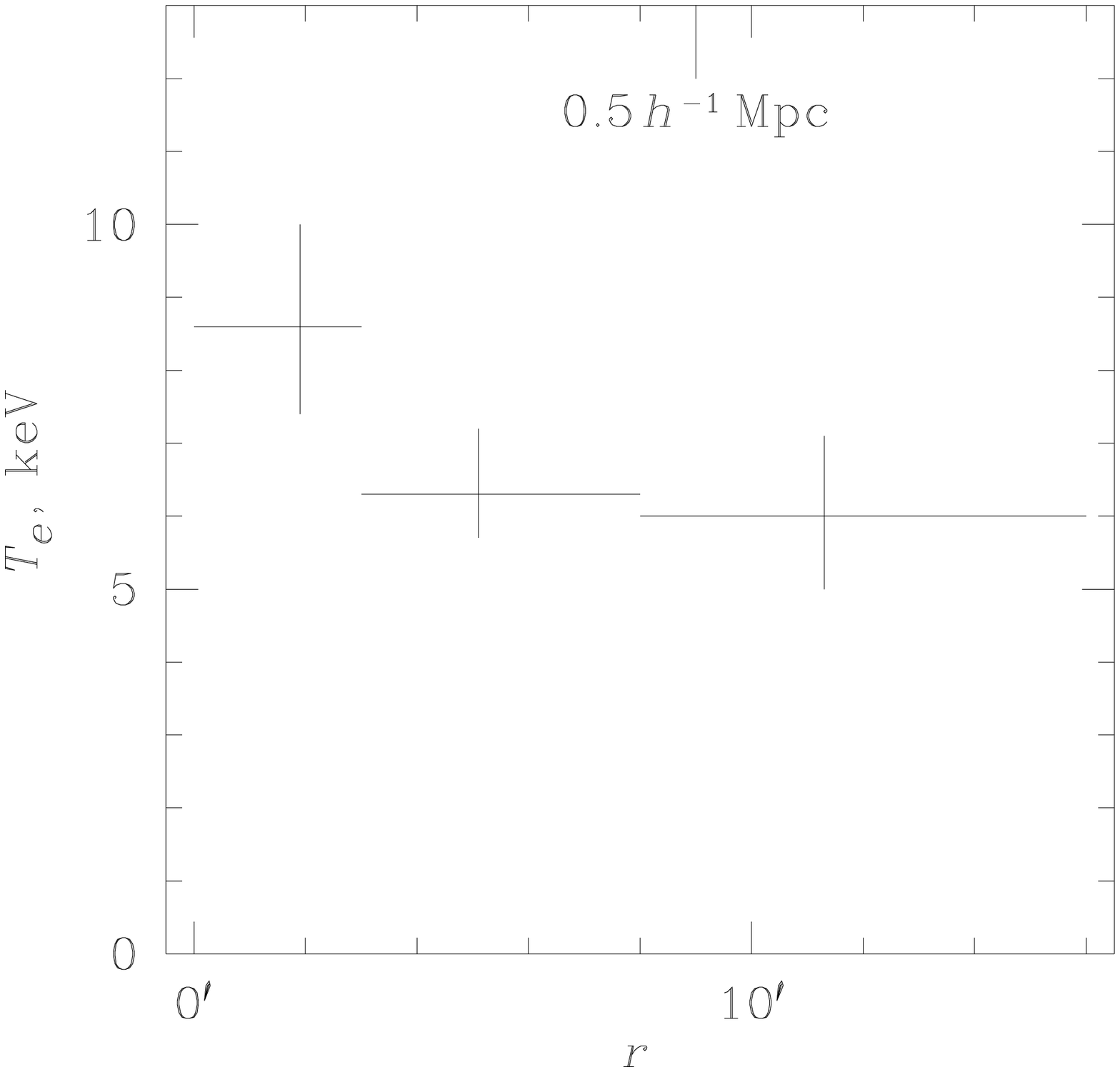}}

\rput[tl]{0}(-0.1,17.2){\epsfxsize=6.5cm
\epsffile{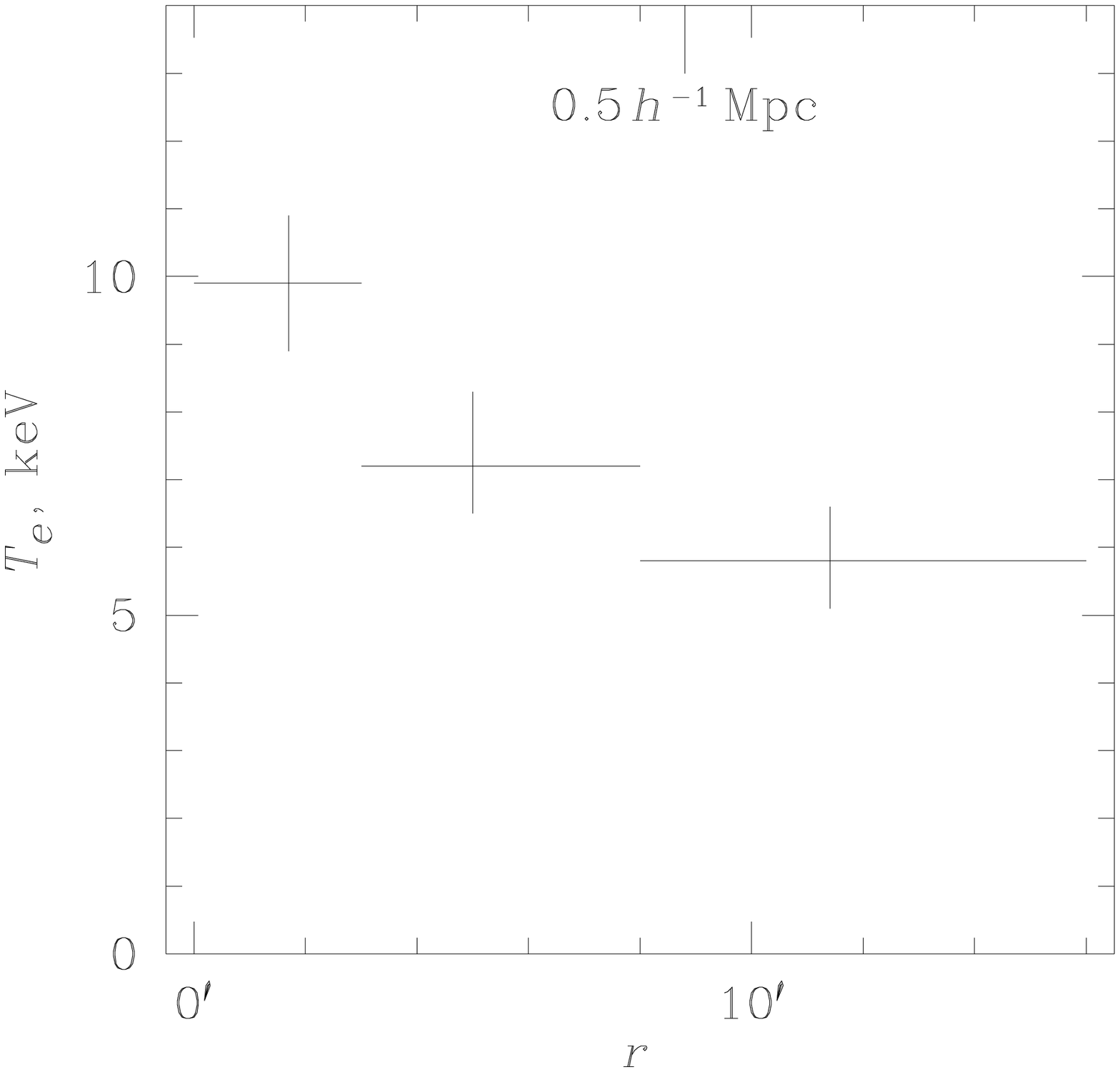}}

\rput[tl]{0}(6.1,17.2){\epsfxsize=6.5cm
\epsffile{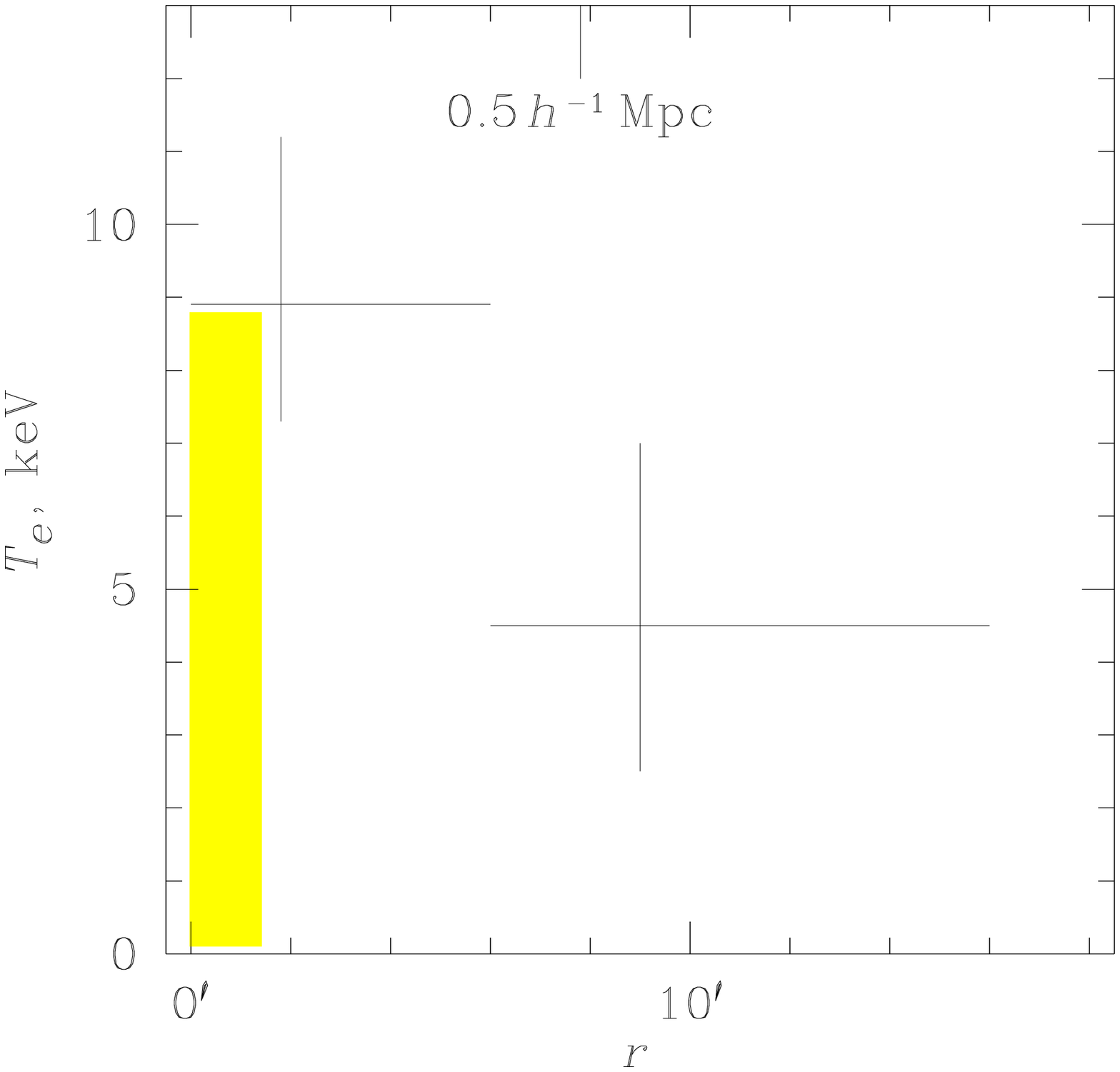}}

\rput[tl]{0}(12.4,17.2){\epsfxsize=6.5cm
\epsffile{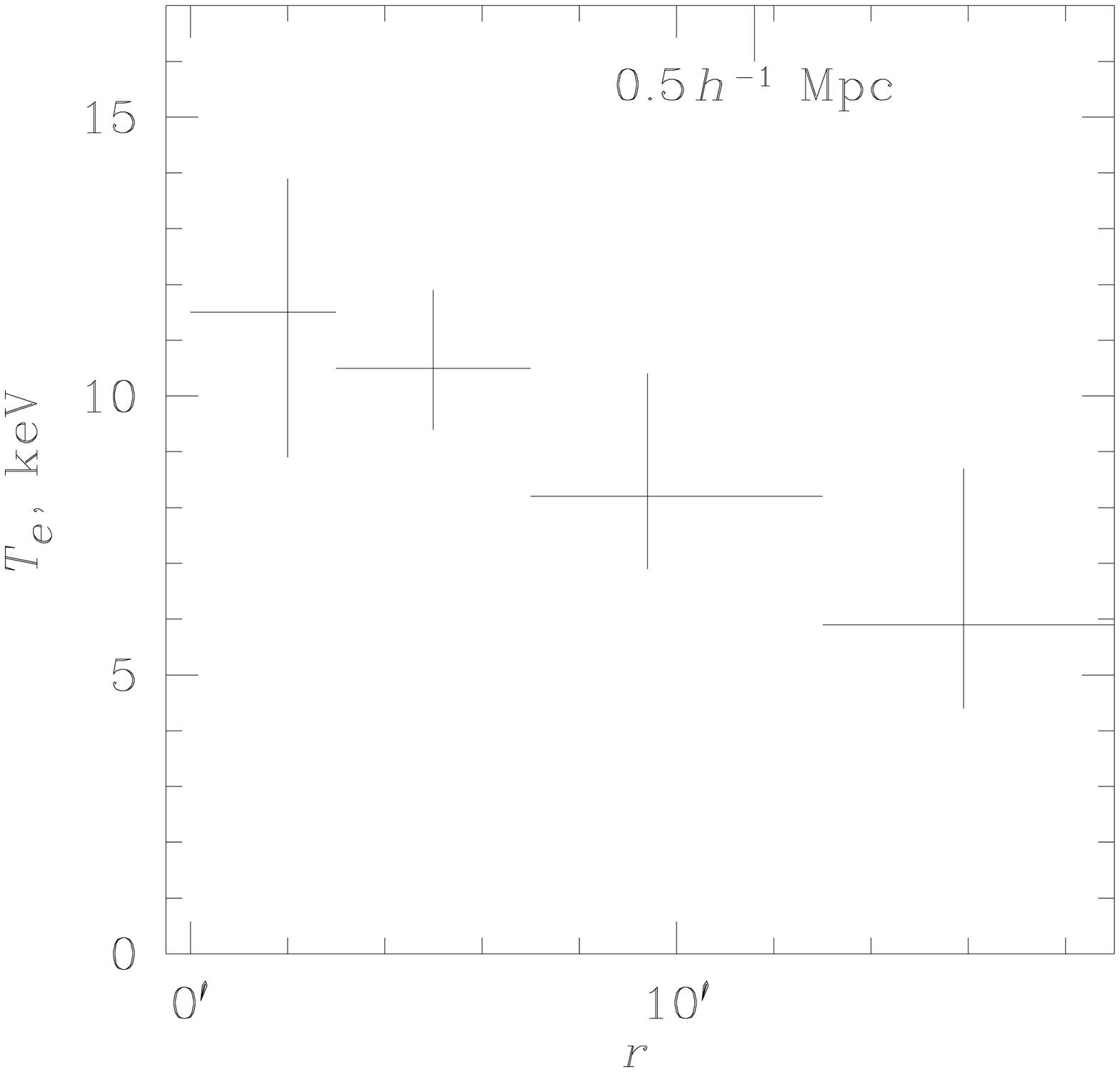}}

\rput[tl]{0}(-0.1,10.9){\epsfxsize=6.5cm
\epsffile{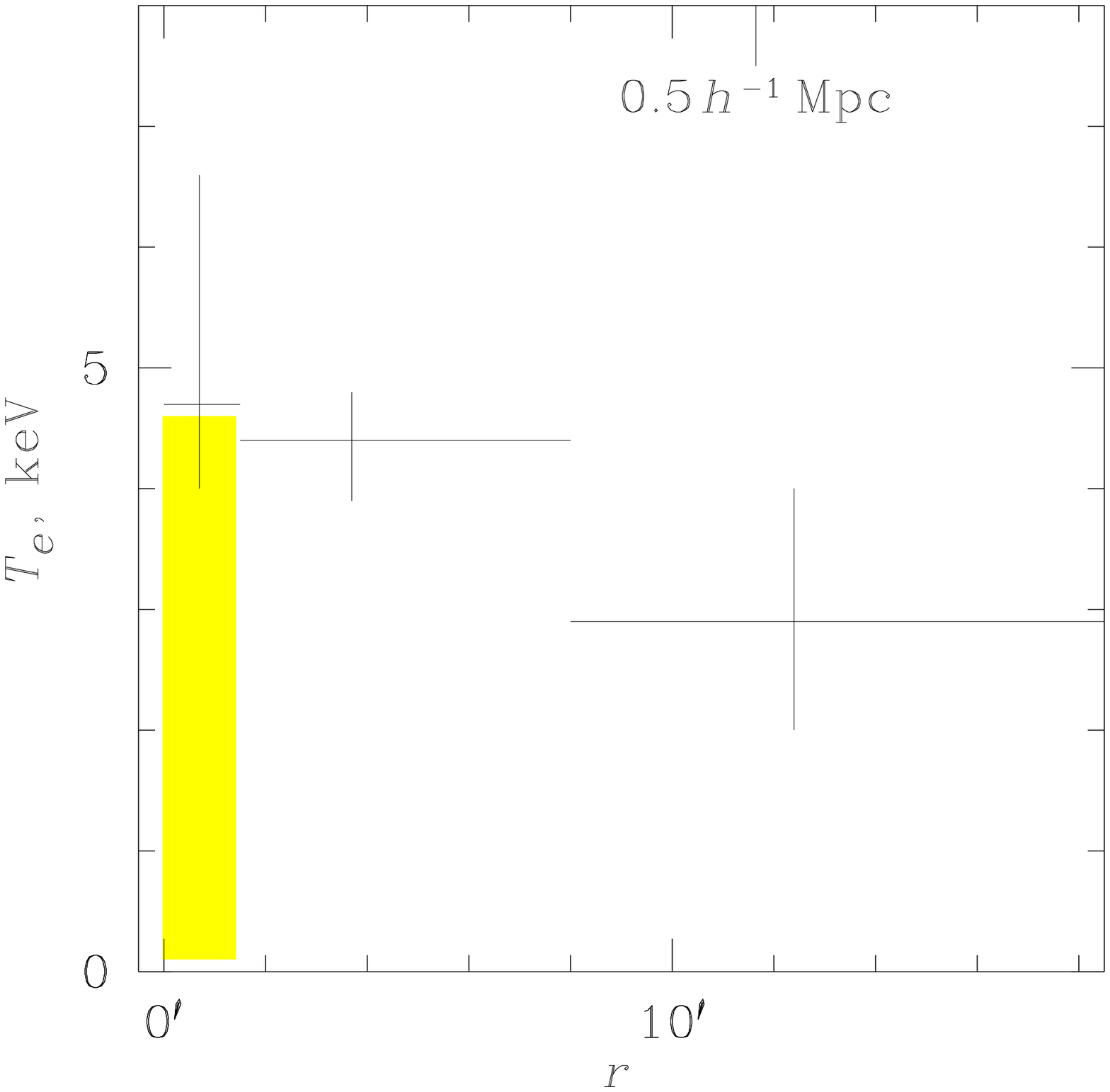}}

\rput[tl]{0}(6.1,10.9){\epsfxsize=6.5cm
\epsffile{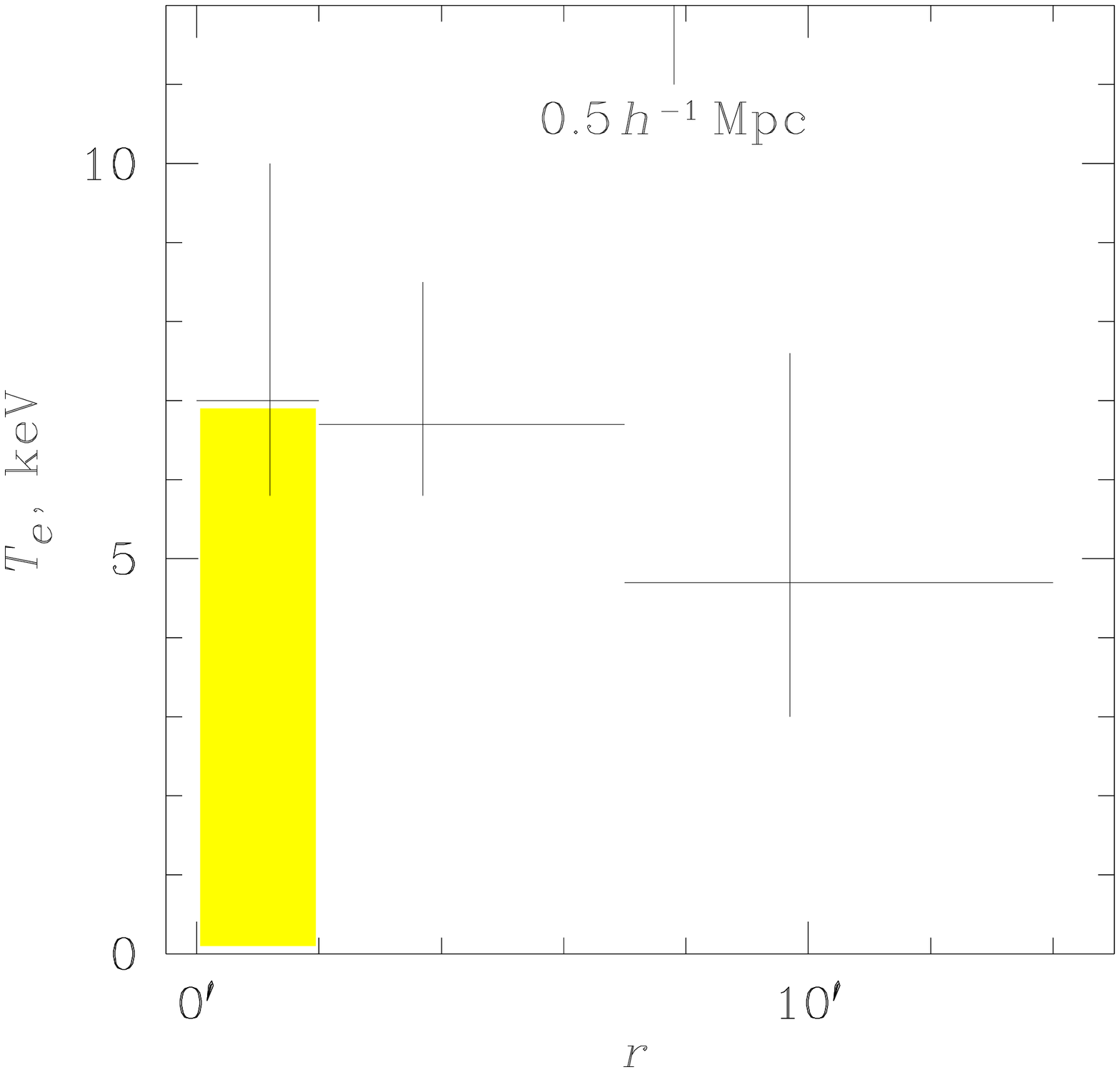}}

\rput[tl]{0}(12.4,10.9){\epsfxsize=6.5cm
\epsffile{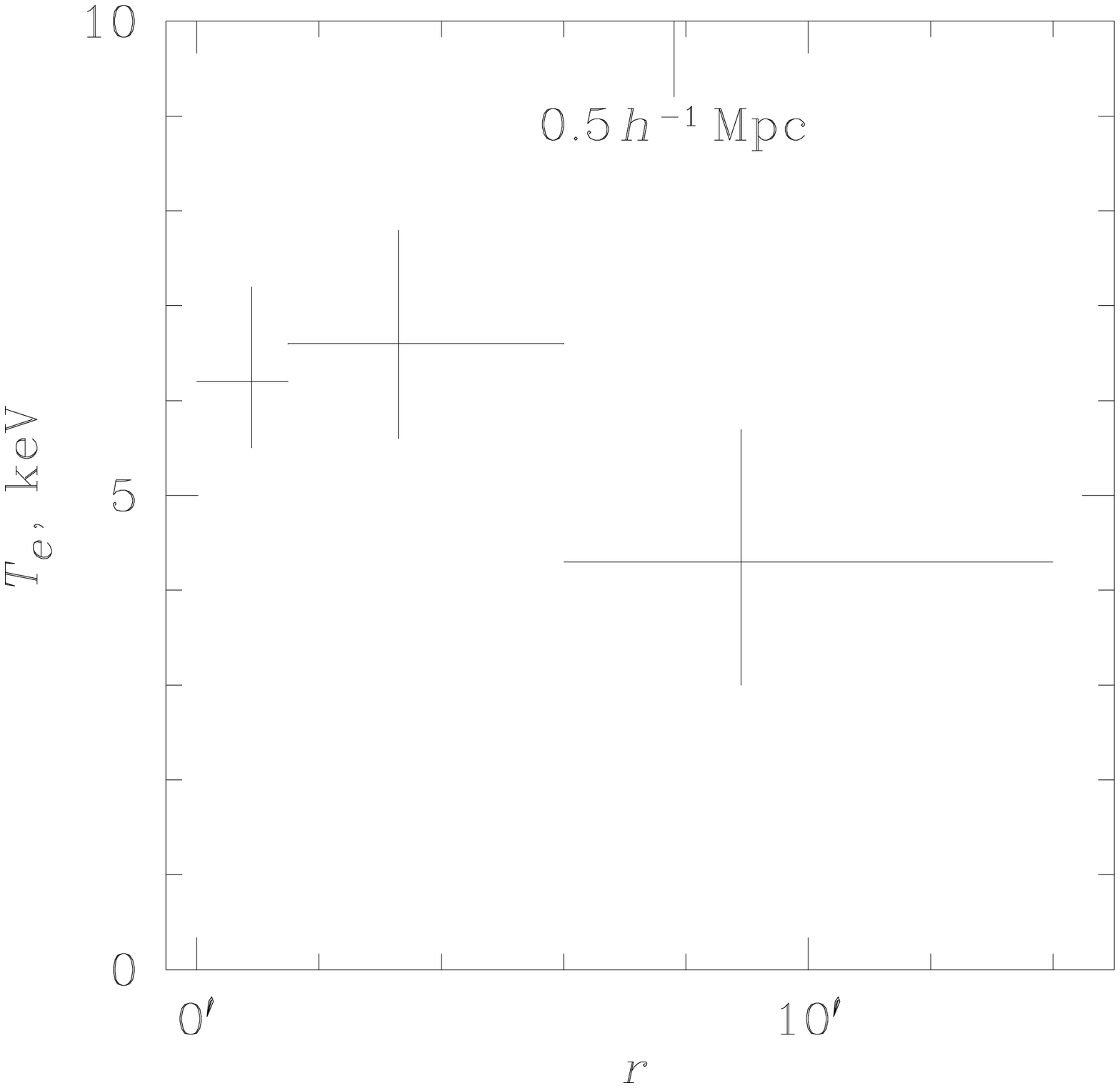}}

\rput[lc]{0}( 1.3,22.6){\small A85}
\rput[lc]{0}( 7.6,22.6){\small A119}
\rput[lc]{0}(13.8,22.6){\small A399}

\rput[lc]{0}( 1.3,16.3){\small A401}
\rput[lc]{0}( 7.6,16.3){\small A478}
\rput[lc]{0}(13.8,16.3){\small A754}

\rput[lc]{0}( 1.3,10.0){\small A780}
\rput[lc]{0}( 7.6,10.0){\small A1650}
\rput[lc]{0}(13.8,10.0){\small A1651}

\rput[tl]{0}(0,4.1){
\begin{minipage}{18.5cm}
\small\parindent=3.5mm
{\sc Fig.}~4.---Radial projected temperature profiles. Crosses are centered
on the emission-weighted radii. Vertical errors are 90\% and include
systematic uncertainties; horizontal error bars show the boundaries of the
annulus. Gray bands denote a continuous range of temperatures in a cooling
flow, or a power law component, in those clusters where these spectral
components are significantly detected by our analysis. For such clusters,
the central cross corresponds to the upper (ambient) temperature of the
cooling flow. For illustration, a central single-temperature fit is also
shown for A2065 as dotted cross.

\end{minipage}
}
\endpspicture
\end{figure*}

\begin{figure*}[t]
\pspicture(0,4.0)(18.5,23.3)

\rput[tl]{0}(-0.1,23.5){\epsfxsize=6.5cm
\epsffile{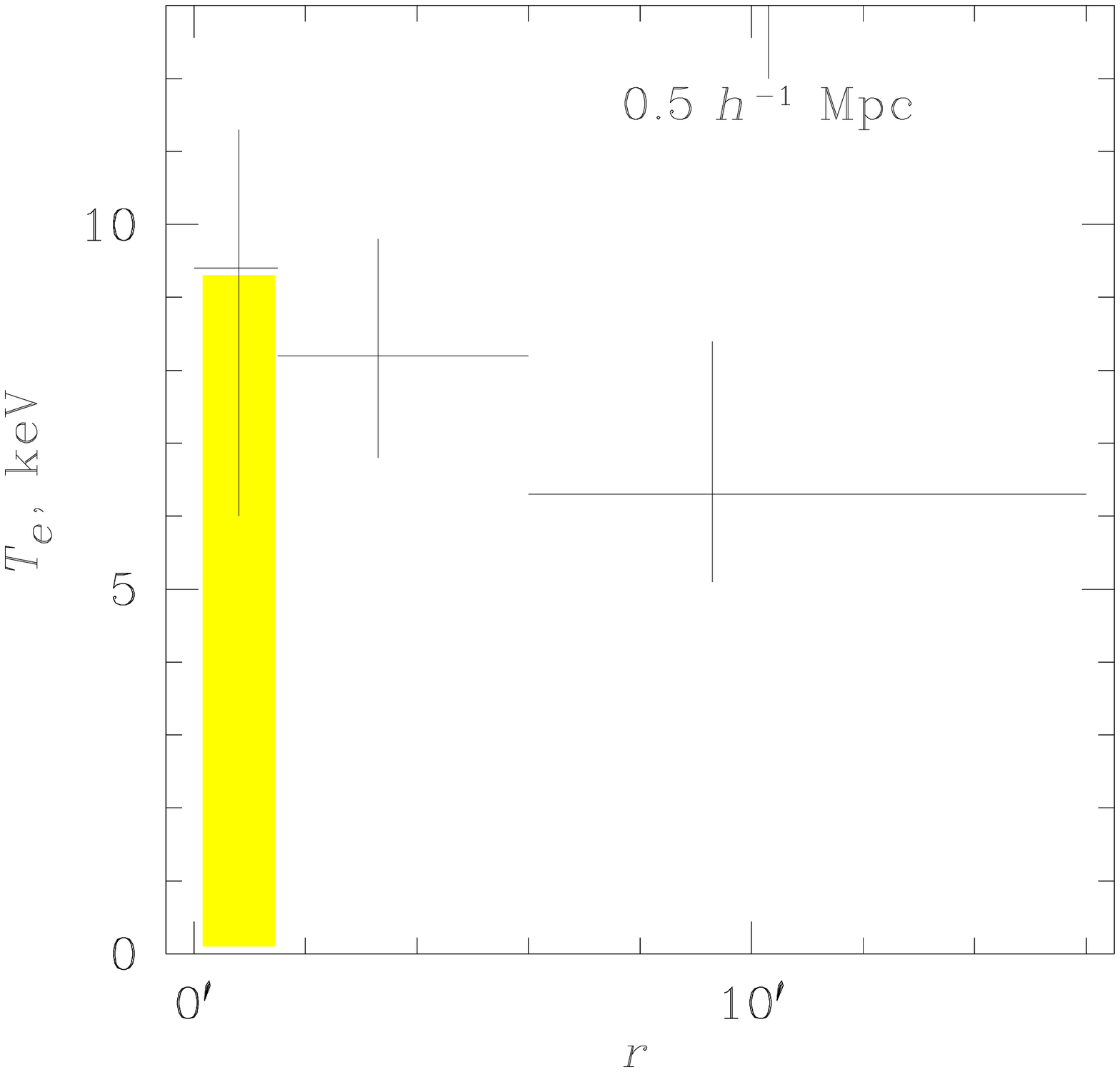}}

\rput[tl]{0}(6.1,23.5){\epsfxsize=6.5cm
\epsffile{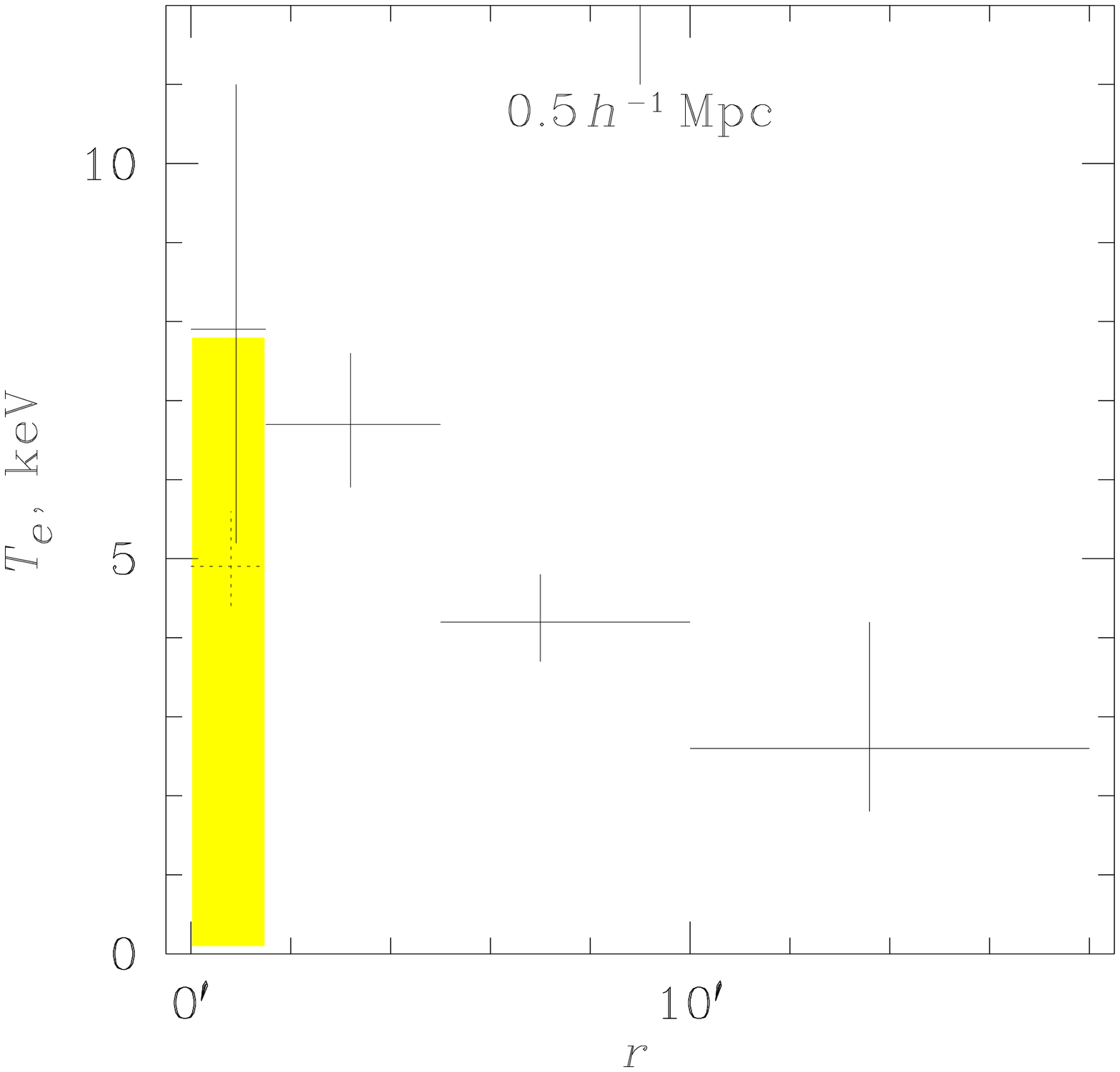}}

\rput[tl]{0}(12.4,23.5){\epsfxsize=6.5cm
\epsffile{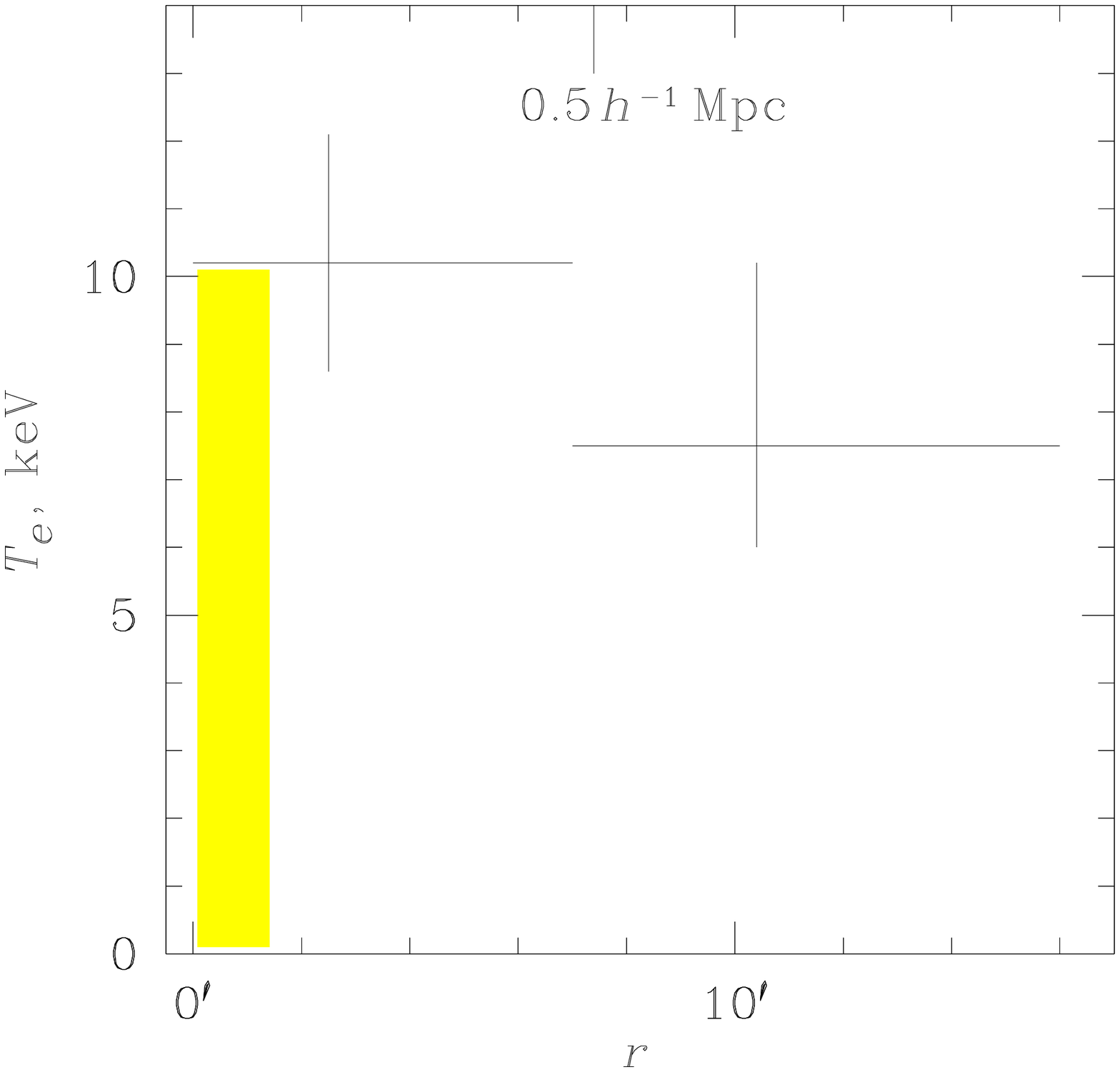}}

\rput[tl]{0}(-0.1,17.2){\epsfxsize=6.5cm
\epsffile{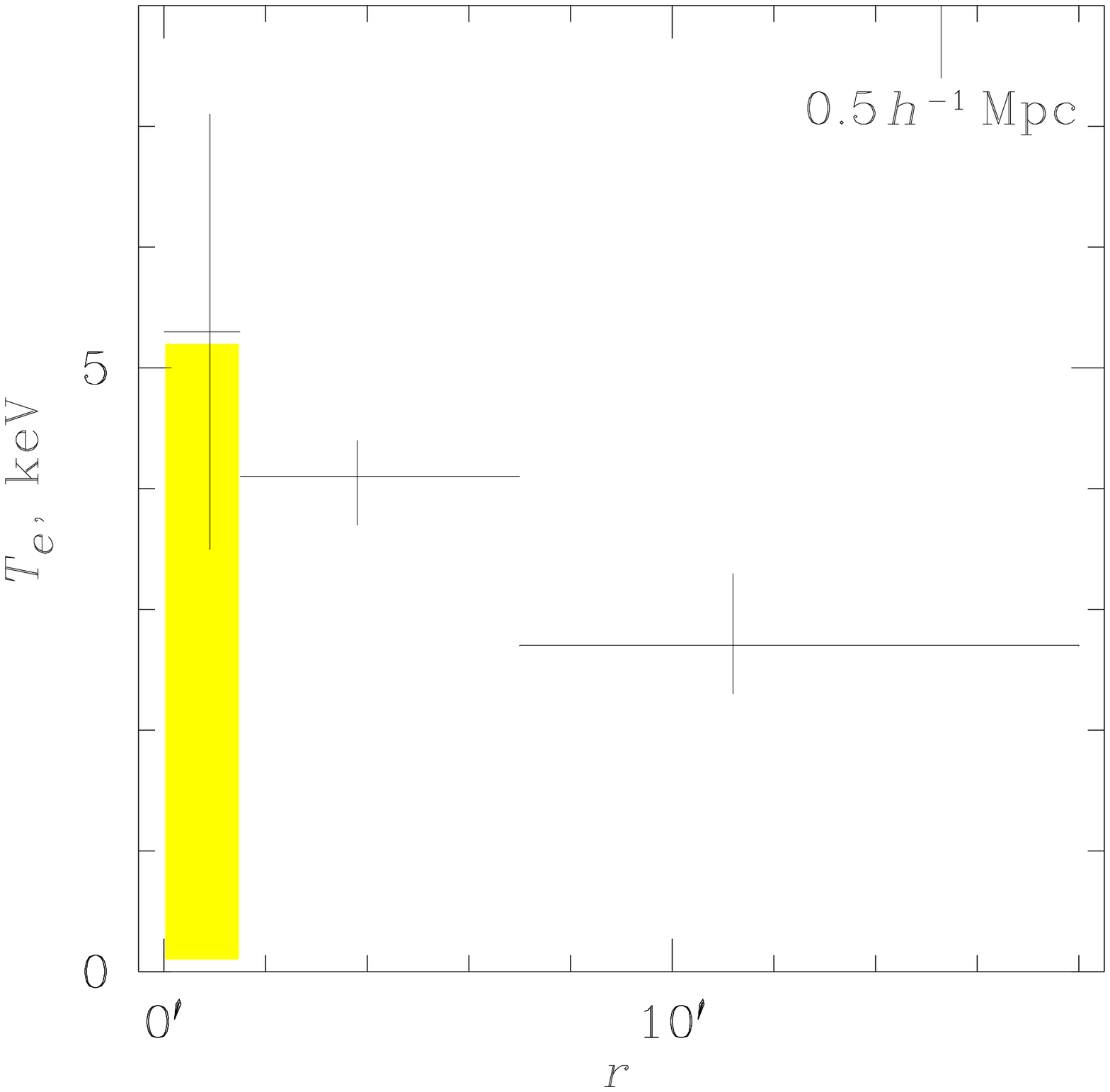}}

\rput[tl]{0}(6.1,17.2){\epsfxsize=6.5cm
\epsffile{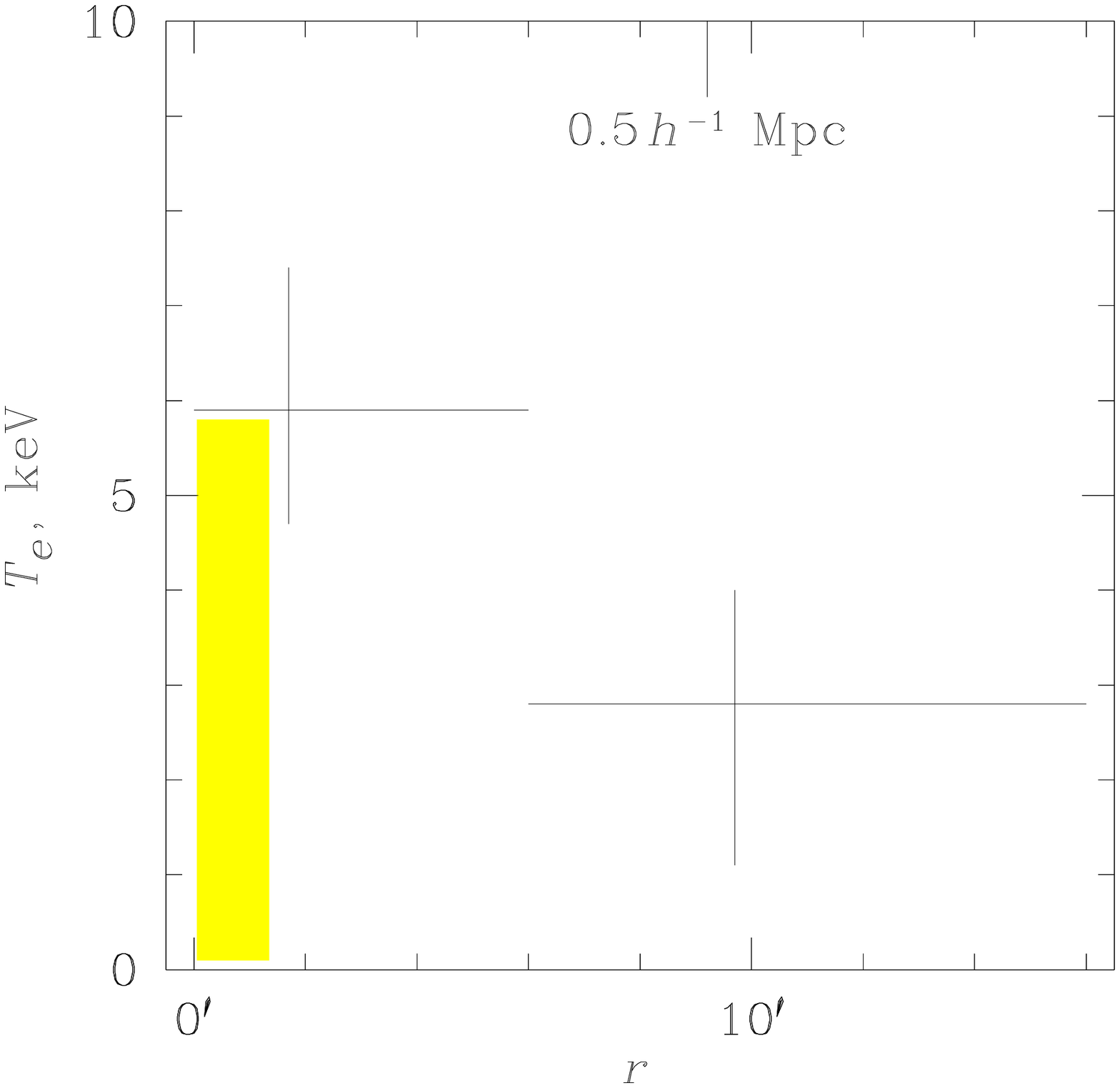}}

\rput[tl]{0}(12.4,17.2){\epsfxsize=6.5cm
\epsffile{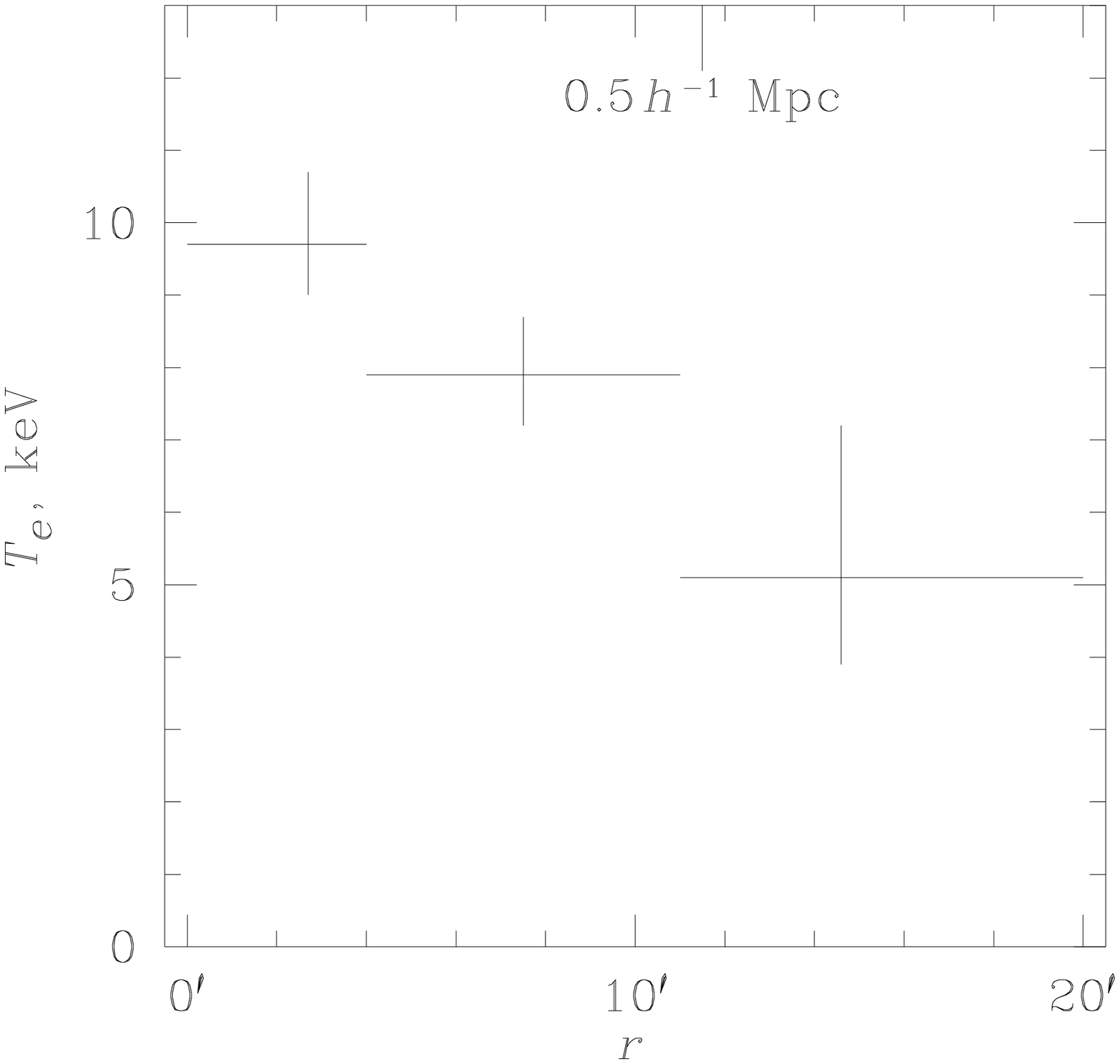}}

\rput[tl]{0}(-0.1,10.9){\epsfxsize=6.5cm
\epsffile{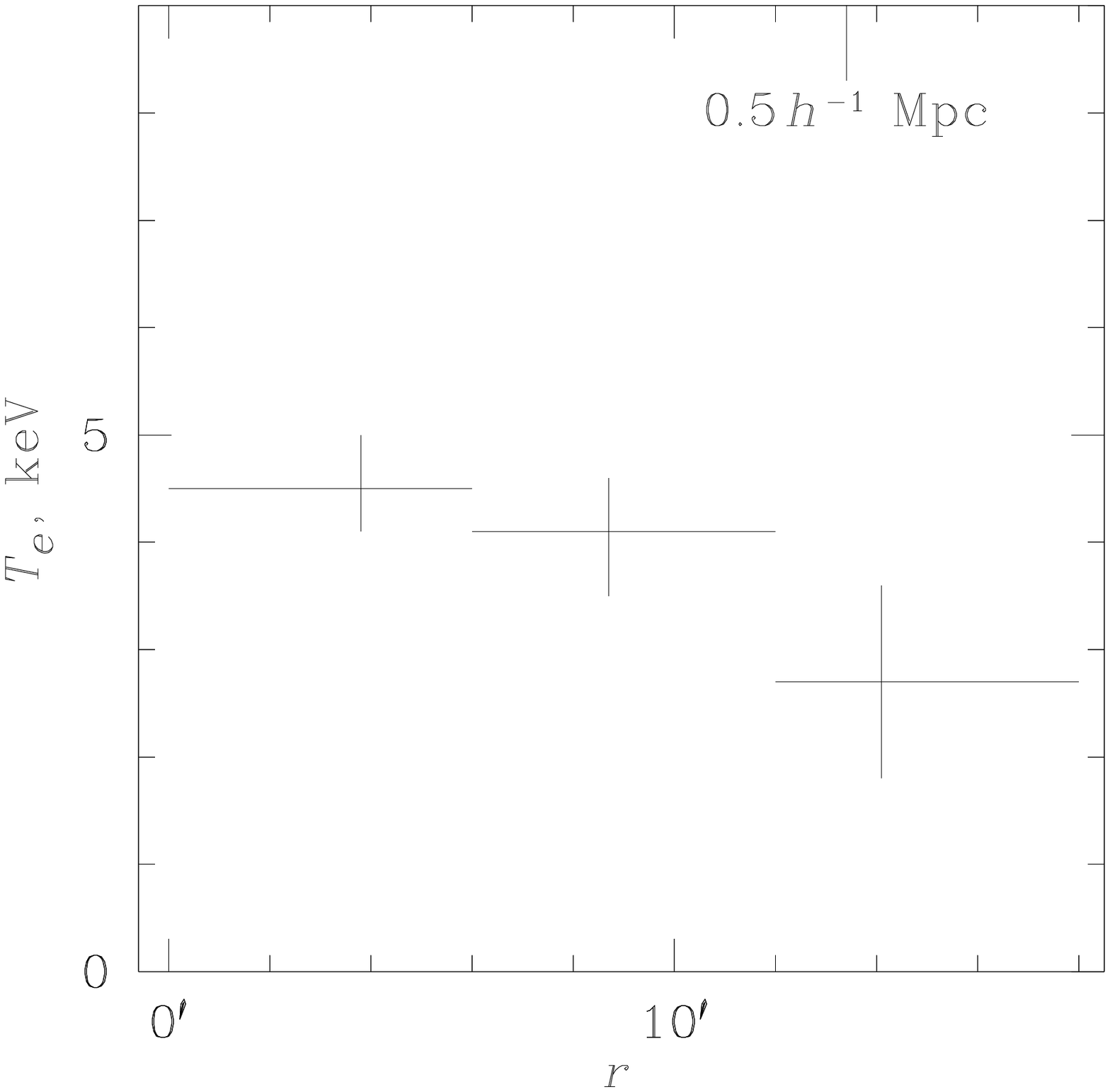}}

\rput[tl]{0}(6.1,10.9){\epsfxsize=6.5cm
\epsffile{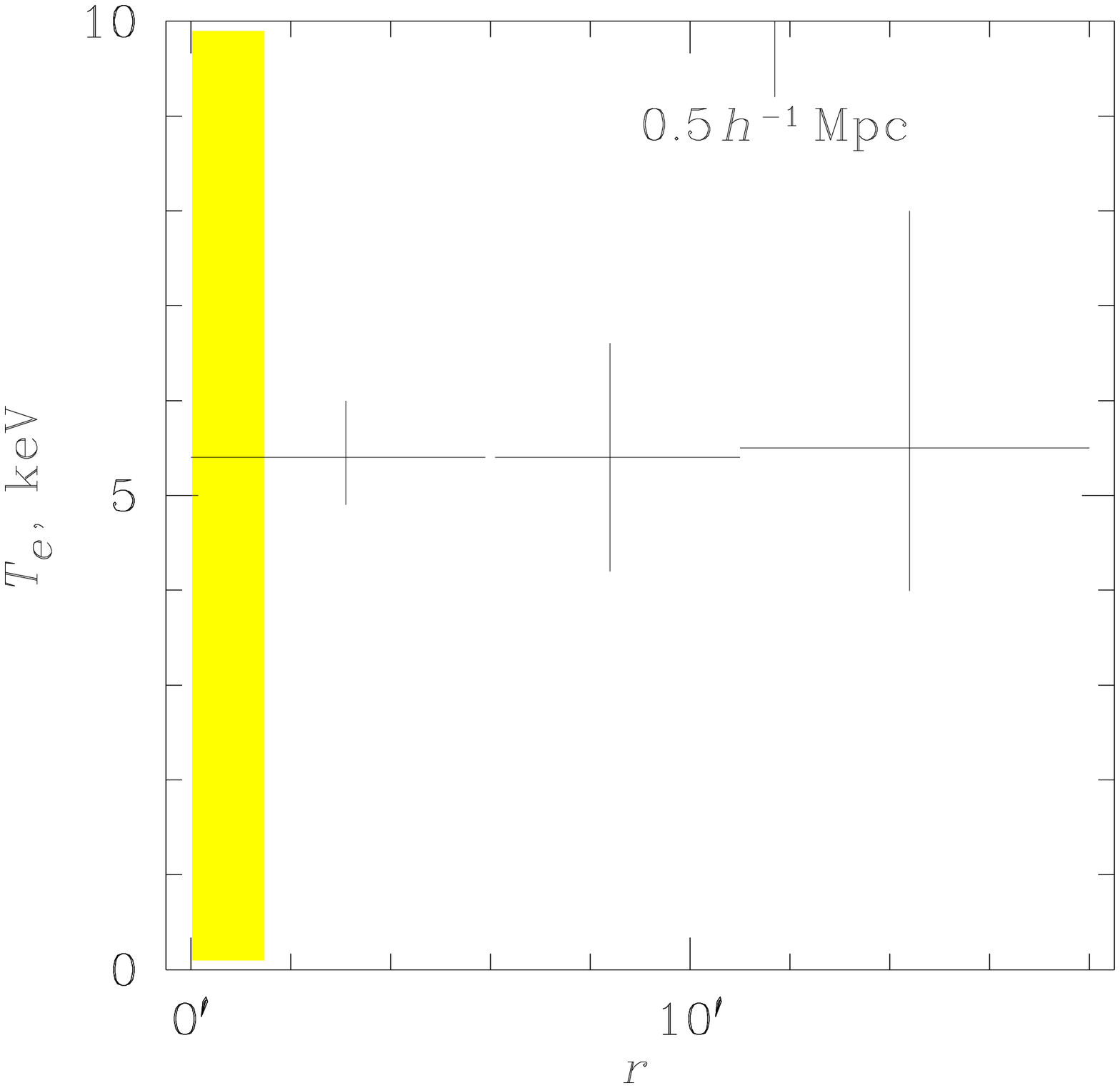}}

\rput[tl]{0}(12.4,10.9){\epsfxsize=6.5cm
\epsffile{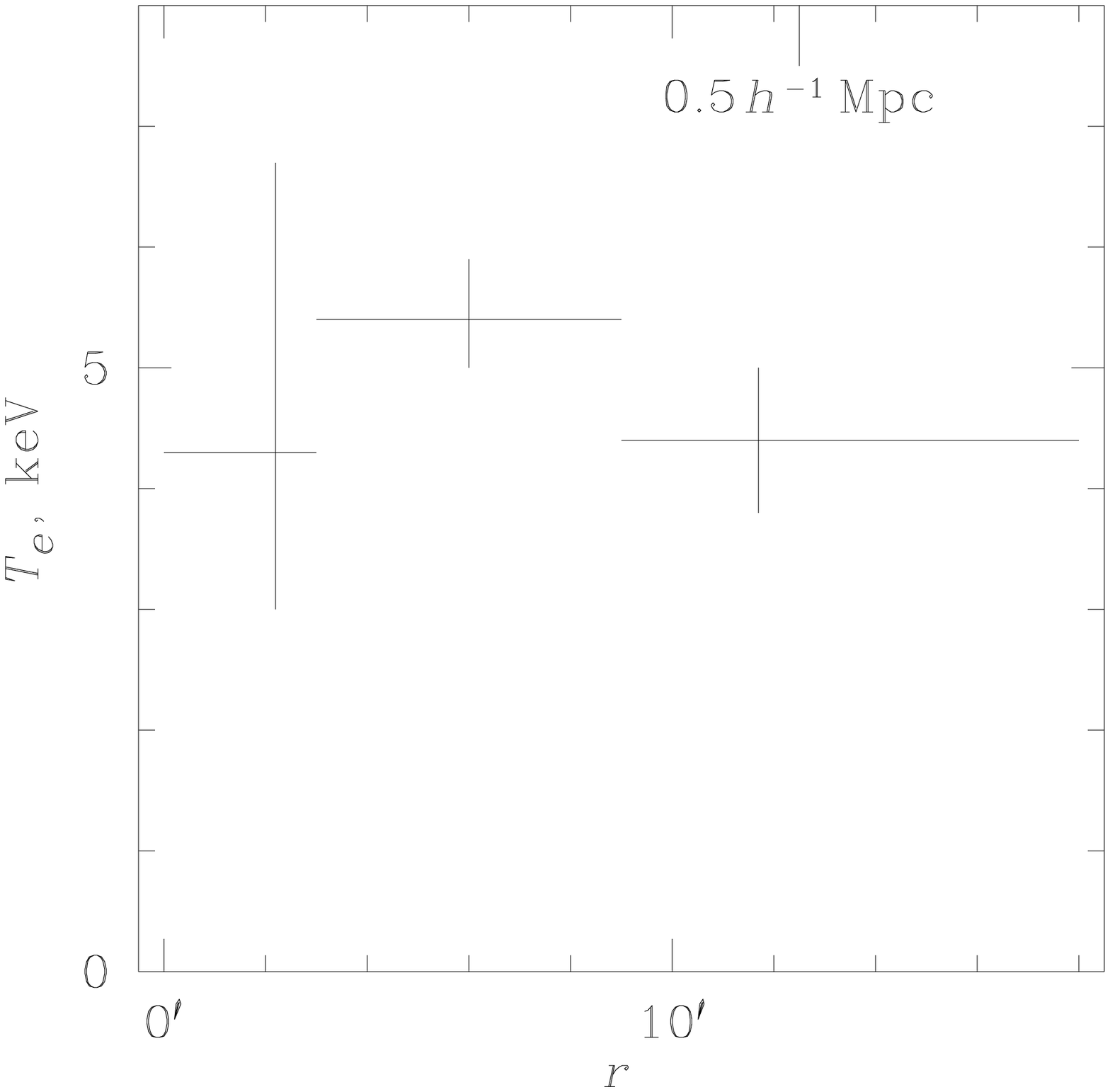}}

\rput[lc]{0}( 1.3,22.6){\small A1795}
\rput[lc]{0}( 7.6,22.6){\small A2065}
\rput[lc]{0}(13.8,22.6){\small A2142}

\rput[lc]{0}( 1.3,16.3){\small A2657}
\rput[lc]{0}( 7.6,16.3){\small A3112}
\rput[lc]{0}(13.8,16.3){\small A3266}

\rput[lc]{0}( 1.3,10.0){\small A3376}
\rput[lc]{0}( 7.6,10.0){\small A3391}
\rput[lc]{0}(13.8,10.0){\small A3395}

\rput[tl]{0}(0,4.1){
\begin{minipage}{18.5cm}
\small\parindent=3.5mm
{\sc Fig.}~4.---Continued
\end{minipage}
}
\endpspicture
\end{figure*}

\begin{figure*}[t]
\pspicture(0,10.4)(18.5,23.3)

\rput[tl]{0}(-0.1,23.5){\epsfxsize=6.5cm
\epsffile{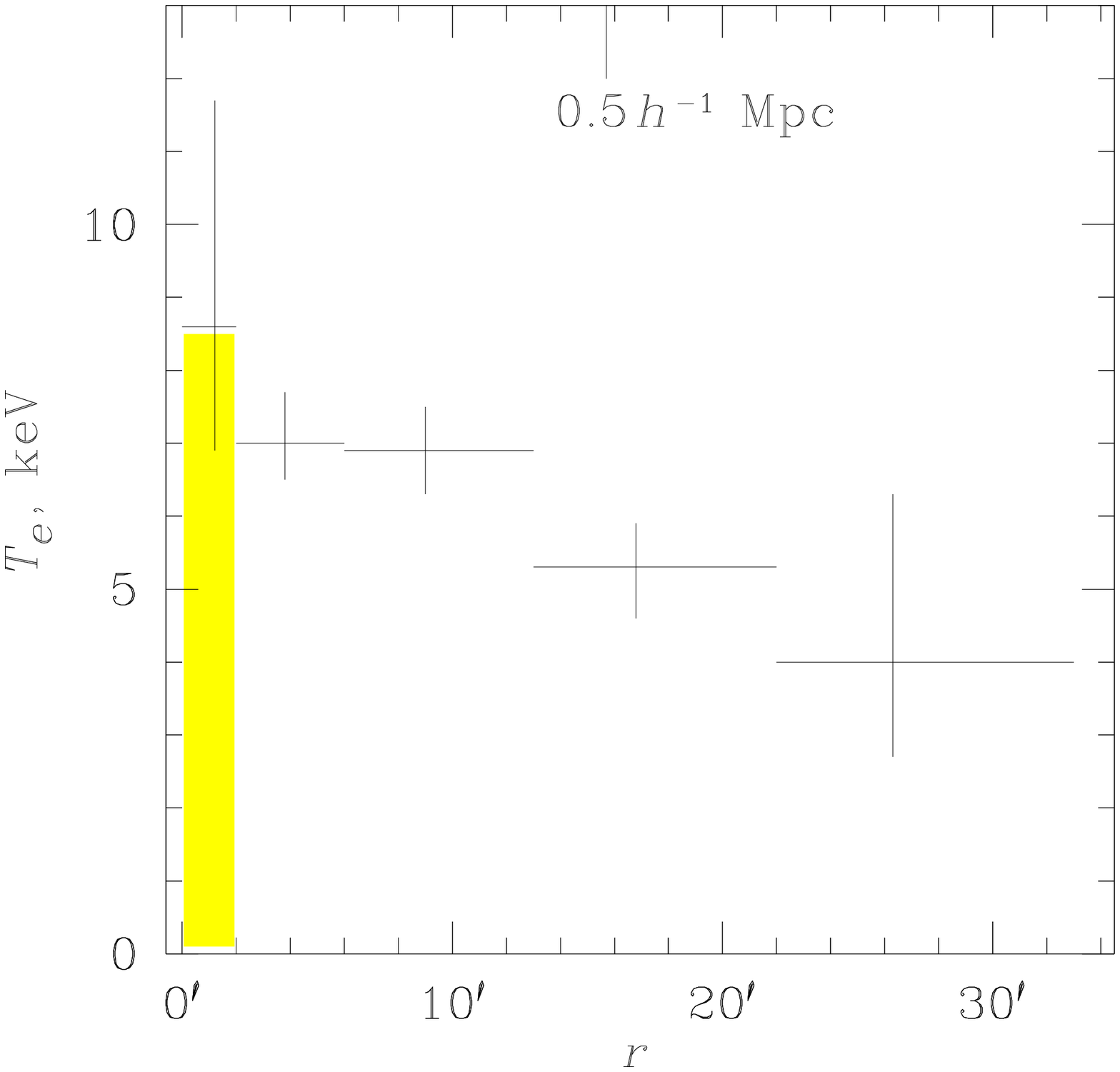}}

\rput[tl]{0}(6.1,23.5){\epsfxsize=6.5cm
\epsffile{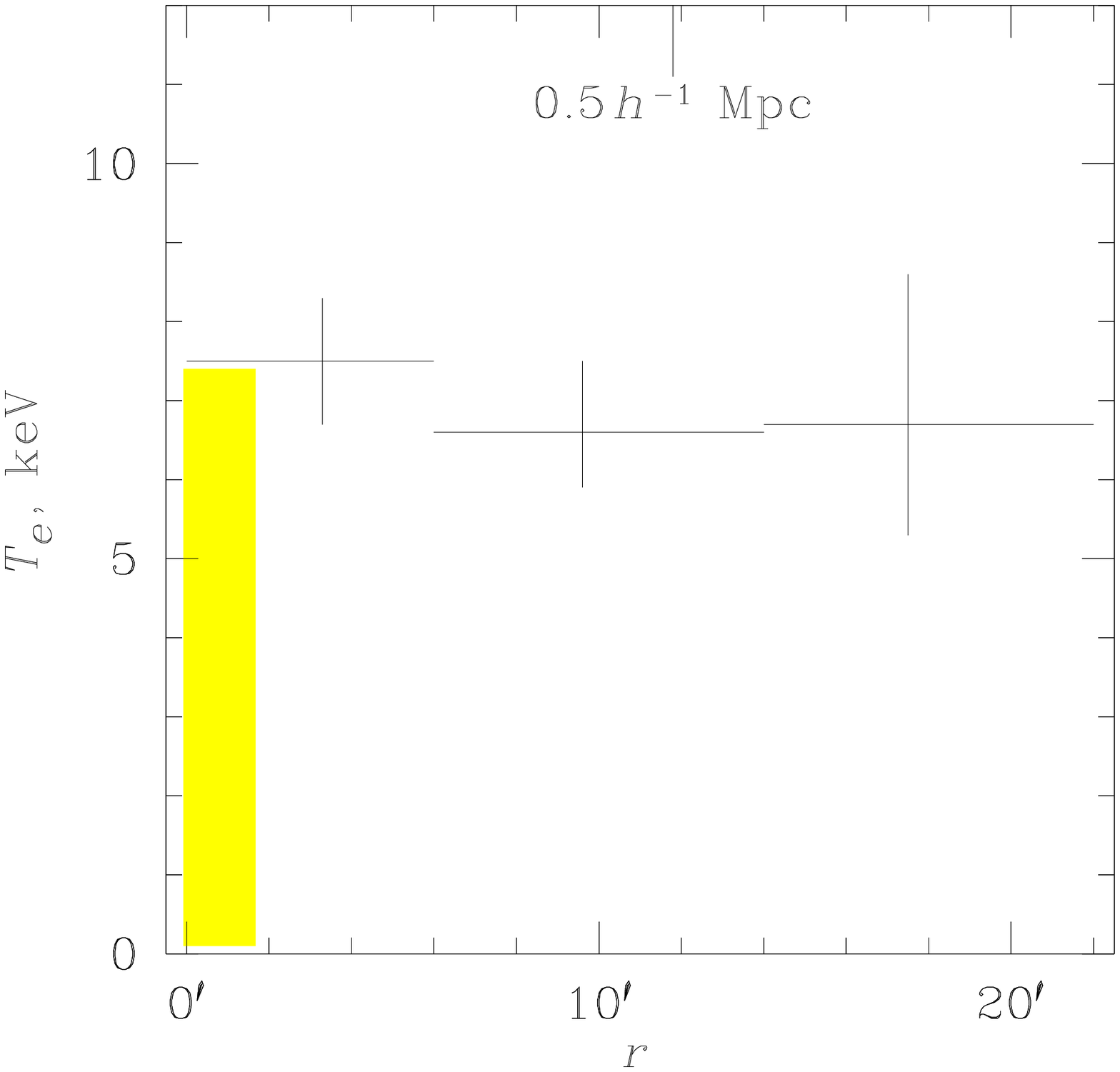}}

\rput[tl]{0}(12.4,23.5){\epsfxsize=6.5cm
\epsffile{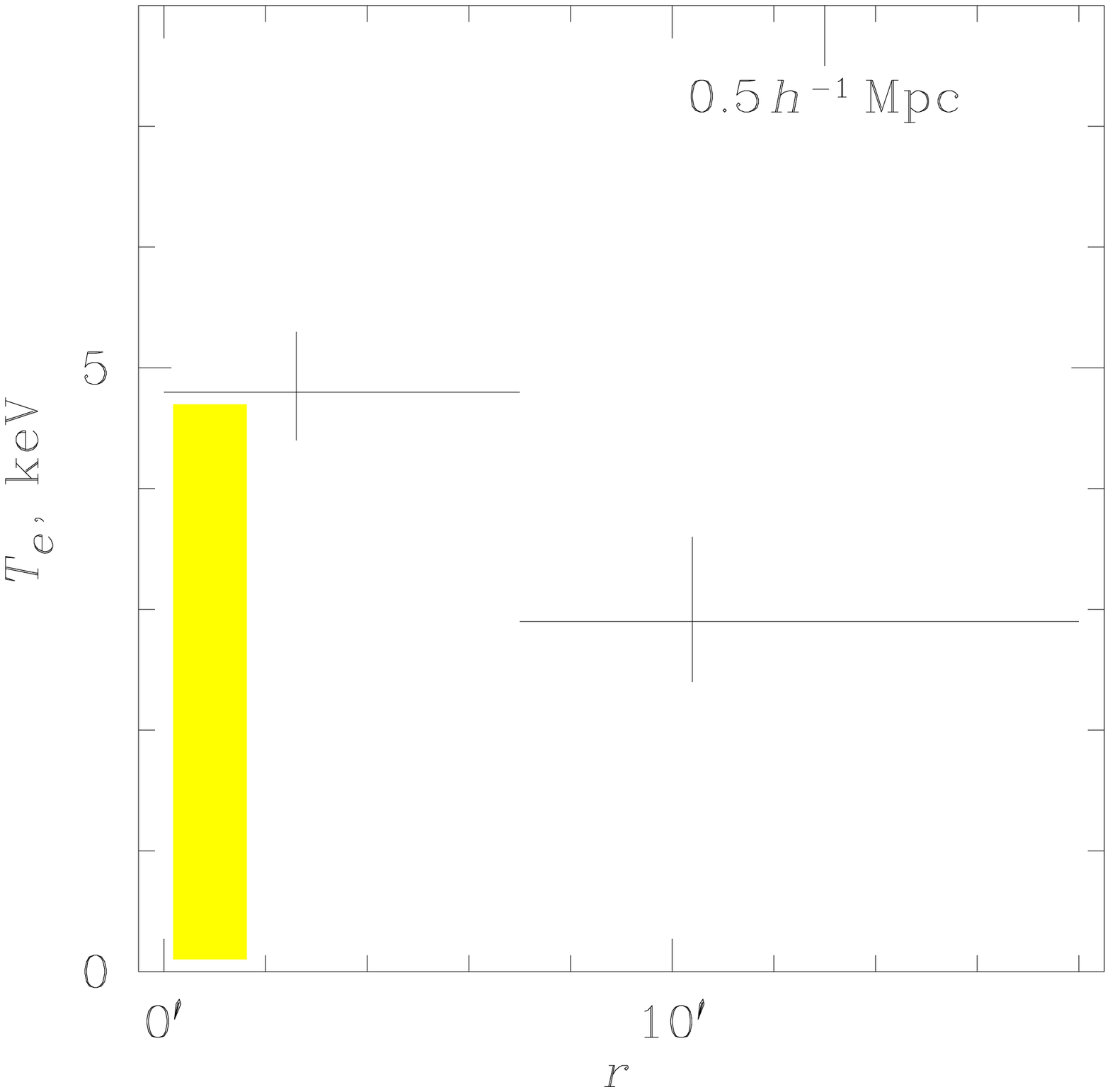}}

\rput[tl]{0}(-0.1,17.2){\epsfxsize=6.5cm
\epsffile{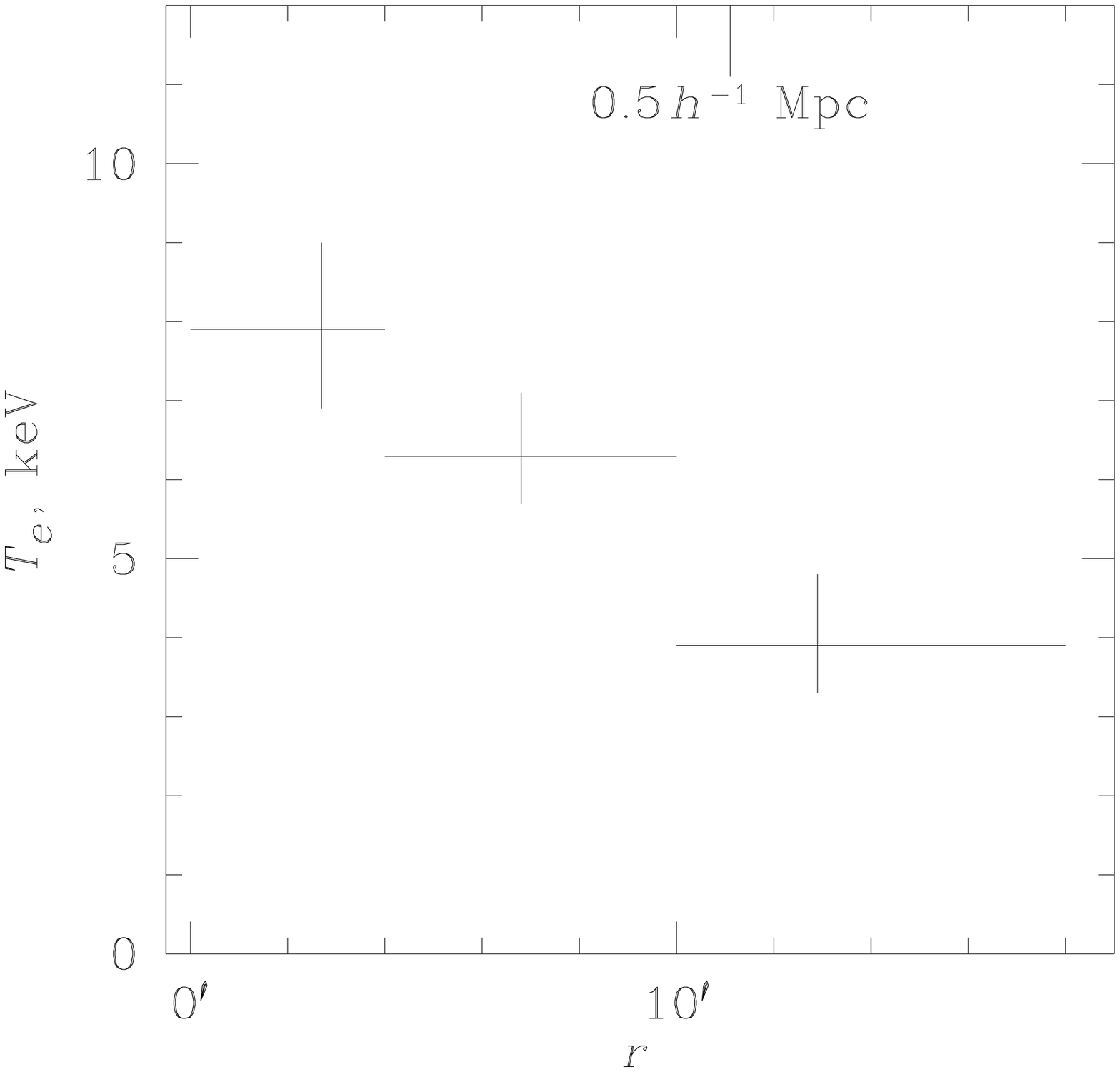}}

\rput[tl]{0}(6.1,17.2){\epsfxsize=6.5cm
\epsffile{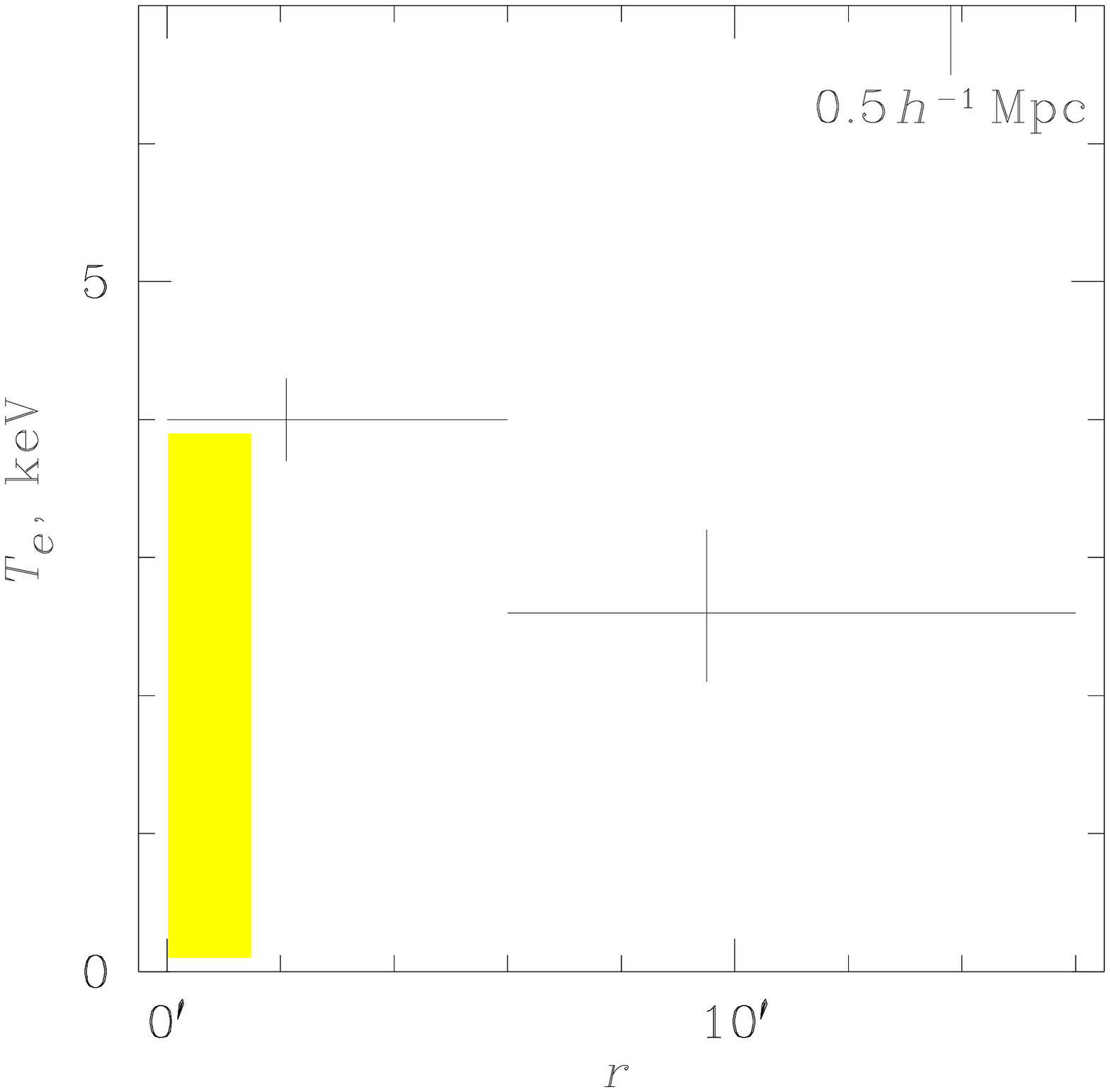}}

\rput[tl]{0}(12.4,17.2){\epsfxsize=6.5cm
\epsffile{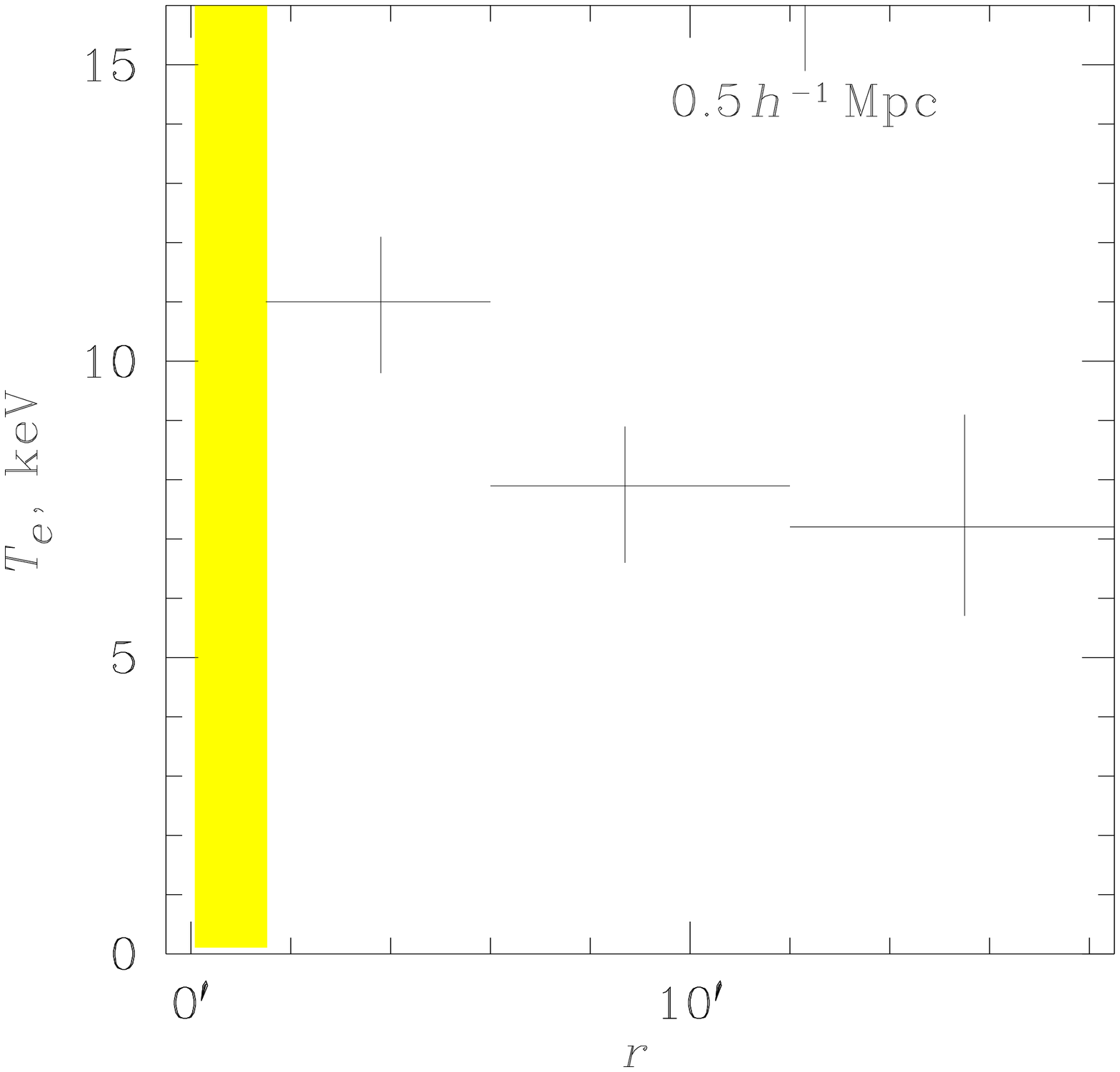}}

\rput[lc]{0}( 1.3,22.6){\small A3571}
\rput[lc]{0}( 7.6,22.6){\small A3667}
\rput[lc]{0}(13.8,22.6){\small A4059}

\rput[lc]{0}( 1.3,16.3){\small Cyg A}
\rput[lc]{0}( 7.6,16.3){\small MKW3S}
\rput[lc]{0}(13.8,16.3){\small Triangulum}

\rput[tl]{0}(0,10.5){
\begin{minipage}{18.5cm}
\small\parindent=3.5mm
{\sc Fig.}~4.---Continued
\end{minipage}
}
\endpspicture
\end{figure*}

\subsection{Temperature Maps and Radial Profiles}

Figures 2 and 3 present two-dimensional temperature maps for 9 clusters with
superposed \rosat\ PSPC brightness contour maps. We show only clusters for
which the temperature maps have useful accuracy. Temperature maps obtained
for A644, A754, A2029, A2065, A2256, A2319, A3558, A3667, Cygnus A, and
Triangulum Australis are published separately (see references in Table 1).
In Figs.\ 2 and 3, we tried to assign different shades of gray to
significantly different temperatures, but it was not always possible due to
the very different confidence intervals. Individual clusters are discussed
in the section below.

Figure 4 shows radial projected temperature profiles for all clusters in our
sample except for A644, A2029, A2256, A2319, and A3558 published elsewhere
(see references in Table 1), A2597 which is dominated by a cooling flow (see
below), and A1736 for which we do not have a profile (see \S\ref{fitsec}).
The profiles were obtained in the annuli around the centroid of the
large-scale emission, which did not always coincide with the brightness peak
(an extreme example is the double cluster A3395, see Fig.\ 3).  Generally,
we did not exclude from the radial profiles any substructures (except for
the contaminating point sources) even when they were obvious, in order to be
able to compare these results to the future low statistics data on distant
clusters.

\subsection{Results for Individual Clusters}

{\em A85.}---A temperature map is reconstructed in the annuli
$r=0-1.5'-6'-12'-20'$, with the second and third annuli divided into 4 equal
sectors and the outermost annulus in two unequal sectors (Fig.\ 2).  Sectors
6 and 10 coincide with an infalling subcluster (e.g., Durret et al.\ 1997).
In the center, we detect a strong cooling flow. The normalization of this
spectral component is greater than zero at greater than 99.9\% confidence. A
single-temperature fit is shown in the map and the upper temperature of a
cooling flow is shown in the radial profile in Fig.\ 4. The temperature
declines with radius, and the map shows no highly significant deviations
from azimuthal symmetry.  However, if the less massive infalling subcluster
were only a projection, the correspondent sectors would appear cooler than
adjacent regions (as, for example, in A2256 or A2319, M96).  Instead, the
map and the profile (Fig.\ 4) show that the subcluster regions are probably
hotter than the adjacent ones, suggesting the presence of shock-heated gas
arising from an ongoing collision. A more detailed temperature map, e.g.,
from \axaf, should detect such shocks.

\vspace{1mm}
{\em A119.}---The temperature map is reconstructed on a grid of $6'\times
6'$ boxes, with regions 1 and 2 composed of 2 boxes and region 15 of 4 boxes
for better accuracy. Region 10, coincident with one of the infalling or
projected subclusters seen both in the X-ray image and in the optical data
(Way, Quintana, \& Infante 1997), is significantly cooler than the cluster
average. Region 13, located apparently ``in front'' of this subcluster, is
marginally hotter than the cluster average, suggesting that there indeed is
a mild merger shock and therefore simple projection is unlikely. Our map
excludes the possibility of a major merger comparable to the ones seen in
A754, A3667 or Cygnus A. The radial temperature profile of this cluster is
nearly constant with a probable mild central peak.

\vspace{1mm}
{\em A399--A401.}---\asca\ observed this pair with 3 pointings centered at
each cluster and between them. Our regions for A401 are sectors of
concentric annuli with $r=0-2'-7'-14'$ (Fig.~2). In the A399 map, the
regions are the same except that the innermost one is centered on the
brightness peak while others are centered on the cluster centroid. The
overall temperatures for A401, A399, and the link region are in agreement
with the earlier \asca\ analysis of Fujita et al.\ (1996) that did not take
into account the PSF effects. Our analysis ignores the effects of \asca\
stray light (Ishisaki 1996) that may be important when a bright cluster is
just outside the mirror field of view, as in two of these three pointings.
However, an estimate shows that such contamination is comparatively small
for these observations.

The temperature map shows a centrally symmetric temperature structure in
both clusters, with sectors 6 and 16, each pointing toward the opposite
cluster, being hotter, with marginal significance, than the corresponding
sectors at the same radius. The temperature declines with radius, with peaks
in the centers of both of these cD clusters (Fig.~4). Despite the large
uncertainty, the gas in region 10 between the clusters does not seem to
follow the temperature decline, arguing against projection and possibly
suggesting the beginning of a collision of these clusters or a massive dark
matter filament between them. Our analysis requires no cooling flow
components in these clusters, in agreement with Edge, Stewart, \& Fabian
(1992). This makes A401 a rather unusual cD cluster since neither the
temperature map nor the X-ray image indicate recent merger activity in the
central regions of this cluster, and yet it has no cooling flow.

\vspace{1mm}
{\em A478.}---This distant cluster has a strong cooling flow (e.g., White et
al.\ 1994) and is at the limit of the \asca's angular resolution.  For this
reason, reconstruction of its two-dimensional map is currently impossible
and even a radial profile (for which we use radii $r=0-1.5'-6'-16'$) has
very large uncertainties (Fig.\ 4). We fixed the central temperature, which
is also a cooling flow upper temperature, to be equal to that in the
surrounding annulus. A central cool component is required at 94\%
confidence and there is a marginally significant temperature decline with
radius.

\vspace{1mm}
{\em A644.}---This cluster is analyzed in detail by Bauer \& Sarazin (1998),
who find a merger signature in the temperature map and marginal evidence for
a central cooling flow. The temperature declines with radius.  Note that our
wide-beam temperature in Table~1 differs slightly from their value due to
our wider integration region.

\vspace{1mm}
{\em A754.}---A temperature map of this merging cluster was discussed in
detail in Henriksen \& Markevitch (1996). Here we present its radial
temperature profile (Fig.~4), although for such a highly asymmetric cluster
with complex temperature structure, it does not have much meaning. The
profile shows a temperature decline.

\vspace{1mm}
{\em A780 (Hydra A).}---Ikebe et al.\ (1997) previously analyzed the same
data on this cooling flow cluster with their independent code. We find a
temperature decline with radius (Fig.~4) outside the central cluster region
similar to that in Ikebe et al., and also detect the presence of cooler gas
in the center at greater than 99.9\% confidence. A two-dimensional
temperature map of this cluster is not given since it would be inaccurate
because of the strong cooling flow.

\vspace{1mm}
{\em A1650.}---This is another example of a distant cooling flow. Our
analysis of this cluster with a large cD was complicated by the absence of a
PSPC image. Instead, we used the lower-resolution \einstein\ IPC image.  As
a result, the accuracy of the temperature profile is poor (Fig.~4). A cool
central component is detected at 99\% confidence.

\vspace{1mm}
{\em A1651.}---This is a distant cD cluster, for which our data do not
require a central cool component. Its PSPC image is rather regular and our
temperature map ($r=0-2'-8'-14'$ annuli with the second annulus divided into
4 sectors, Fig.~2) shows no interesting structures, although its accuracy is
poor. A radial temperature decline is suggested (Fig.~4).

\vspace{1mm}
{\em A1736.}---For this cluster, included here for completeness, we obtain
only a wide-beam temperature (see \S\ref{fitsec}). \einstein\ IPC and \asca\
images indicate that A1736 does not exhibit a strong cooling flow or any
bright contaminating sources. Therefore, its weighted temperature $T_X$
should be close to the wide-beam temperature, as for other such clusters (see
\S\ref{tempsec} below); this will be assumed in Markevitch (1998).

\vspace{1mm}
{\em A1795.}---This cluster has one of the strongest known cooling flows
(Edge et al.\ 1992; Fabian et al.\ 1994a [an earlier \asca\ analysis]; Briel
\& Henry 1996). The cool component is required at greater than 99.9\%
confidence in the central $r=1.5'$ region. This small central region
contains almost half of the total cluster emission in the \rosat\ band,
which makes our \asca\ spatially resolved analysis very uncertain. We detect
some indication of a radial temperature decline. A bright point source 6\am\
from the center was fit separately and found to be equally well described by
either a thermal or a soft power law spectrum; this uncertainty does not
affect the temperature values in other cluster regions.

Note that our temperatures at all radii are higher than the
single-temperature fit to the overall spectrum (Fig.~4 and Table 1); this
cluster is the most prominent example of how the presence of a strong
cooling flow results in an underestimate of the cluster temperature. This
issue will be discussed below.

\vspace{1mm}
{\em A2029.}---This strong cooling flow and well relaxed cluster is
discussed in detail in Sarazin et al.\ (1997), who detect a significant
radial temperature decline in it.

\vspace{1mm}
{\em A2065.}---A temperature map for this cluster is presented and discussed
in M98. A prominent asymmetric temperature pattern is
detected, as well as a central cool component (at 99\% confidence) which
must have survived the ongoing major merger. In Fig.\ 4, we present its
radial temperature profile, which is declining with radius.  Dashed cross
shows a single-temperature fit in the central bin.

\vspace{1mm}
{\em A2142.}---This is the most distant cluster in our sample. Its \rosat\
PSPC and HRI images suggest an ongoing merger. However, we cannot
reconstruct its temperature map due to the presence of a cooling flow (Edge
et al.\ 1992; Henry \& Briel 1996) which we detect with greater than 99\%
confidence. An AGN 4\am\ from the cluster center is fitted separately by a
power law with the best-fit index of --1.9. The temperature profile has
large uncertainties and is consistent with a constant value, although a
temperature decline with radius is suggested (Fig.~4).

\vspace{1mm}
{\em A2256}---A detailed discussion of this cluster, including its total
mass derivation, is presented in M96 and Markevitch \& Vikhlinin (1997b,
hereafter MV97b). Those works inferred that, most probably, the two large
subclusters of A2256 have not yet started interacting. An ongoing collision
between them is excluded, since the temperature map does not exhibit
structures characteristic of shock heating, predicted by simulations and
indeed observed in several merging clusters such as A754 and Cygnus A.  In
the discussion below, we use the projected temperature profile for the main
component of A2256 obtained in MV97b.

\vspace{1mm}
{\em A2319.}---GIS results for this cluster were presented in M96. The
GIS+SIS results are similar with a small revision of the overall temperature
(Table 1). M96 found that there is a cool region near the center coincident
with a subcluster seen in the X-ray image. As for A2256, the temperature map
does not suggest any major merger, although it has a poorer accuracy than
the A2256 map.

\vspace{1mm}
{\em A2597.}---This is a distant cluster dominated by a cooling flow (e.g.,
Edge et al.\ 1992; Sarazin \& McNamara 1997). Almost half of the total
emission in the \rosat\ band originates from the central $r=50\,h^{-1}$ Mpc
(0.8\am) region. For this reason, we were only able to fit spectra in two
radial bins, $r=0-1.5'-15'$, to separate the cooling flow from the rest of
the cluster.  The cooling flow sufficiently dominated the central bin that
we could not determine an independent temperature there, so we assumed that
the ambient temperature was constant at all radii. A cool component is
required at greater than 99\% confidence in the central region.

\vspace{1mm}
{\em A2657.}---A slight westward elongation apparent in the image coincides
with significantly hotter sectors in the temperature map (consisting of
sectors of $r=0-1.5'-7'-18'$ annuli, Fig.~2), indicating a merger rather
similar to that in A85. We detect a moderate central cooling flow with 94\%
confidence (the map shows a single-temperature fit). The azimuthally
averaged temperature declines with radius (Fig.~4). A point source south of
center (white circle in region 5) is an AGN for which we obtain
$\gamma=-1.60\pm0.16$.

\vspace{1mm}
{\em A3112.}---This cluster possesses a strong cooling flow (Edge et al.\
1992) which we detect at 98\% significance. Again, for this reason, the
radial temperature profile has a large uncertainty. Still, there is a
significant indication of a radial temperature decline (Fig.~4).

\vspace{1mm}
{\em A3266.}---A \rosat\ PSPC image of this cluster indicates ongoing
merger. Our temperature map (which consists of sectors of annuli with
$r=0-2.5'-8'-18'$ centered on the brightness peak, Fig.\ 2) does not have
sufficient resolution, but may indeed suggest an asymmetric temperature
pattern expected for a merger.  Sector 4, which is coincident with an
infalling subcluster, is cooler than the annulus average, while the
neighboring sector 3 is hotter, although with only marginal significance.
The cluster brightness peak is hot; on average, the temperature declines
with radius (Fig.~4).

\vspace{1mm}
{\em A3376.}---A temperature map of this cluster (made of $8'\times 8'$
boxes) is shown in Fig.\ 2. A comet-like shape of the \rosat\ image
indicates a merger. However, our temperature map probably excludes any
strong shocks. A radial profile (Fig.~4) shows that the outer cluster
regions are significantly cooler than the center, although a radial profile
has little meaning for this asymmetric cluster. This object is at the lower
flux limit of our sample and only has a short GIS exposure, so our results
are limited by statistics.

\vspace{1mm}
{\em A3391.}---This relatively regular-looking cluster and A3395 form a pair
separated by 48\am\ or $2\,h^{-1}$ Mpc in projection, similar to the
A399-A401 pair. Stray light contamination from the neighboring cluster is
estimated to be small. Our temperature map consists of sectors of annuli
with $r=0-2'-8'-18'$ (Fig.~2); the cluster northern region does not have
sufficient \asca\ coverage and is not shown (but was included in the fit).
As in the A399-A401 pair, the sector pointing toward A3395 is marginally
hotter, suggesting that the two clusters are beginning to collide or are
connected by a massive filament. There is some temperature asymmetry in the
cluster's inner $0.7\,h^{-1}$ Mpc as well, possibly a remnant of earlier
mergers. The azimuthally averaged temperature is constant with radius
(Fig.~4). In the central radial bin, we detect an additional spectral
component best described by an absorbed power law. If its slope is fixed at
$-1.7$, the redshifted absorbing column is between $0.3-80\times
10^{21}$\cmsq. The normalization of this component is greater than zero at
92\% confidence. In Fig.~2, a single-temperature fit is shown for the
central region, while in Fig.~4, the result of including the additional
spectral component (denoted by gray band) is shown.

\vspace{1mm}
{\em A3395.}---A temperature map of this double cluster is shown separately
from others in Fig.\ 3. Regions of the map are nine $5'\times 5'$ boxes,
with boxes 4 and 6 coincident with the peaks of the two subclusters, and
four sectors of an $r=16'$ circle outside the boxes. The temperature map
excludes the presence of a very hot gas between the cluster peaks (region 5)
which one would expect if the two clusters were colliding head-on. The peaks
(regions 4 and 6) are marginally hotter than the cluster average, but
temperatures in other regions are consistent with the average within their
large uncertainties. The uncertainties are large mostly because of the low
statistics of the \rosat\ PSPC image obtained from a very short (2 ks)
exposure.

A closer look at the elongated X-ray images of each of the two A3395
components suggests that they may in fact be in the course of an offset
merger, with the northeast and southwest peaks moving west and east,
respectively. Interestingly, the statistically insignificant positive
temperature deviations are observed in regions 1, 2, 8, and 9 ``in front''
of the subclusters if this speculation is correct. Entropy maps are a
sensitive indicator of shock heating (MSI) and may reveal additional detail.
We show in Fig.\ 3 (right panel) approximate specific entropies of the gas
in the inner 9 cluster regions.  The entropy per particle is defined here as
$\Delta s\equiv s-s_0=\frac{3}{2}k\;{\rm ln}\left[
(T/T_0)(\rho/\rho_0)^{-2/3}\right]$, where the subscript 0 refers to any
fiducial region in the cluster. For a qualitative estimate, we approximate
$\rho/\rho_0 \sim (S_x/S_{x0})^{1/2}$, where $S_x$ is the cluster X-ray
surface brightness in a given region. This approximation is adequate in the
central regions but is inaccurate in the outer image regions which are not
shown (the entropy values there are higher than in the center as in any
isothermal cluster). With marginal significance, regions 1, 2, 8, and 9 have
higher specific entropy than other central regions which appear to be
approximately adiabatic. Note that region 5, between the peaks, has a low
density similar to those in high-entropy regions, but has an average
entropy.  Such an entropy distribution suggests the presence of shock-heated
gas in front of each of the two subclusters, which should then indeed
collide as proposed above. We note that, of course, dividing temperatures by
a power of surface brightness to calculate entropy does not increase the
statistical significance of the observed temperature structure, but it helps
to present it in a more illuminating way. From the observed entropy
differences one can estimate, for example, the relative velocity of the
colliding subclusters; however, any quantitative analysis must wait for more
accurate measurements.

\vspace{1mm}
{\em A3558.}---This cluster is analyzed in MV97a, who detected temperature
structure indicating a major merger, and a slow radial temperature decline
The region around the central galaxy requires a cool and a hot or nonthermal
component.

\vspace{1mm}
{\em A3571.}---This cluster has the best-quality data in our sample.
Temperatures are obtained in 90\deg\ sectors for annuli with
$r=0-2'-6'-13'-22'-35'$. In the last annulus, only two sectors are covered
by \asca; in one of them, the individual temperature is not usefully
constrained and is not shown in Fig.\ 2, but it is included in the
calculation of the radial profile. The map in Fig.\ 2 shows no significant
deviations from azimuthal symmetry, except for the possibility that the
southern half of the cluster may be slightly hotter than the northern half.
Together with the symmetry of the cluster image, the featureless temperature
map suggests that A3571 is a well-relaxed cluster. This cluster is known to
have a small cooling flow (Edge et al.\ 1992); a central cool component is
detected with 92\% confidence in our data. The radial temperature profile,
shown in Fig.\ 4, will be used for a derivation of the total mass profile in
a separate paper.

\vspace{1mm}
{\em A3667.}---A temperature map of this cluster shows an ongoing major
merger and is presented and discussed in detail in M98. A trace of cooler
gas, which must have survived a merger, is detected in the vicinity of the
central galaxy with 95\% confidence. The radially-averaged temperature
profile (Fig.\ 4) is constant; however, a radial profile has little meaning
for this highly asymmetric cluster.

\vspace{1mm}
{\em A4059.}---This is a cooling flow cluster (e.g., Huang \& Sarazin 1998);
we detect a central cool component with 99\% confidence. A crude temperature
map (which consists of two annuli with $r=0-3'-16'$ with the outer one
divided into 4 sectors, Fig.~2) shows no significant detail. A
single-temperature fit for the central region is shown in the map. A radial
profile shows a significant temperature decline with radius (Fig.~4).

\vspace{1mm}
{\em Cygnus A.}---A temperature map of this cluster is presented in M98.
The map reveals temperature structure indicating an ongoing merger of two
large subclusters seen in the X-ray image (and perhaps in the optical as
well; Owens et al.\ 1997), one of which harbors a strong radio galaxy Cygnus
A. There is a strong point source at the position of Cygnus A in the
hardest-band \asca\ images where the surrounding cluster emission is faint,
which shows that \asca\ can confidently separate the AGN component from the
cluster. Fitting the cluster temperatures simultaneously with a
self-absorbed power law in a $r=1.5'$ region centered on the AGN, we obtain
for the AGN component a slope of $-2.0^{+0.3}_{-0.4}$ and a column density
of $4.8^{+1.2}_{-1.1}\times 10^{23}$\cmsq, in agreement with the \ginga\
wide-angle fit (Ueno et al.\ 1994). The temperature of the surrounding
cluster component is also in agreement with Ueno et al.  Reynolds \& Fabian
(1996) detect a cooling flow around Cygnus A in the \rosat\ data. A cooling
flow component is not required in our fit, nor is it excluded, due to the
spectral complexity of the AGN region. A radial profile, centered on the
cluster large-scale centroid between the colliding subclusters, is shown in
Fig.\ 4.

\vspace{1mm}
{\em MKW3S.}---This apparently relaxed, slightly elliptical cluster has a
strong cooling flow which we detect with 96\% confidence. Because of this,
the temperature map does not have useful accuracy and is not presented. The
teprerature declines with radius (Fig.~4).

\vspace{1mm}
{\em Triangulum Australis.}---Temperature, pressure, and entropy maps of
this hot cluster are discussed in MSI, who find an indication for
nonadiabatic heating of the intracluster gas. Here we present the cluster's
radial temperature profile. MSI noted a problem with the fit to the central
$r=3'$ region of this cluster. A smaller, $r=1.5'$ central bin which we use
here (in addition to including lower-energy \asca\ data) requires an
additional spectral component with 99\% confidence. From \asca\ data alone,
we cannot distinguish between a power-law and a cooling flow component with
a high upper temperature. \rosat\ PSPC data which are sensitive to the
presence of cool gas, do not require a cooling flow (MSI), therefore, a
nonthermal contribution is more plausible. A complex spectrum in the central
radial bin, denoted by the gray band in Fig.\ 4, does not significantly
affect measurements in the outer annuli. The profile shows the temperature
rising toward the cluster center.

\begin{figure*}[th]
\pspicture(0,11.8)(18.5,20.3)

\rput[tl]{0}(12.1,20.4){\epsfxsize=6.9cm
\epsffile{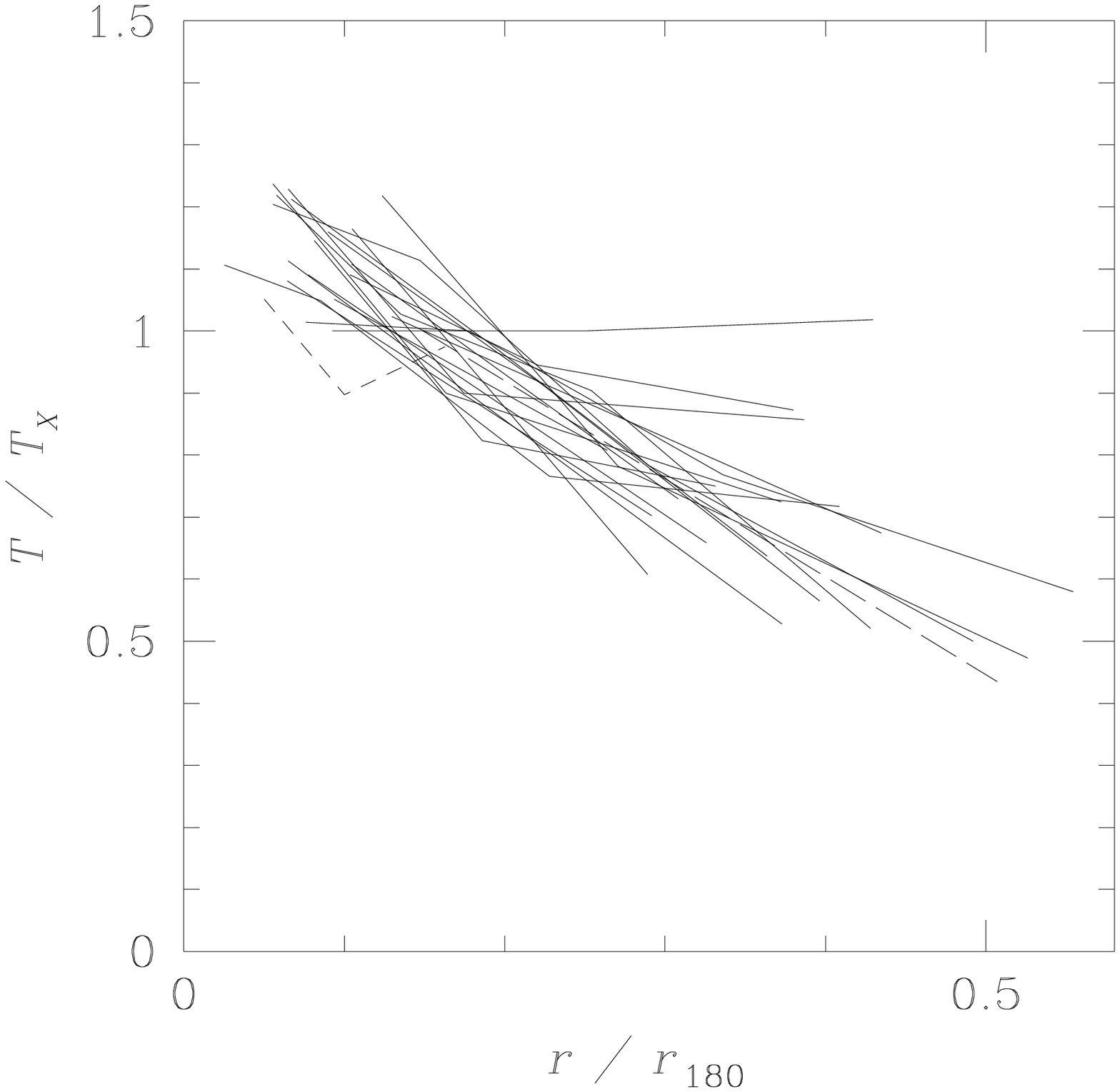}}

\rput[cc]{0}(12.2,17.3){\psframebox*{\white \rule{5mm}{2cm}}}

\rput[tl]{0}(6.0,20.4){\epsfxsize=6.9cm
\epsffile{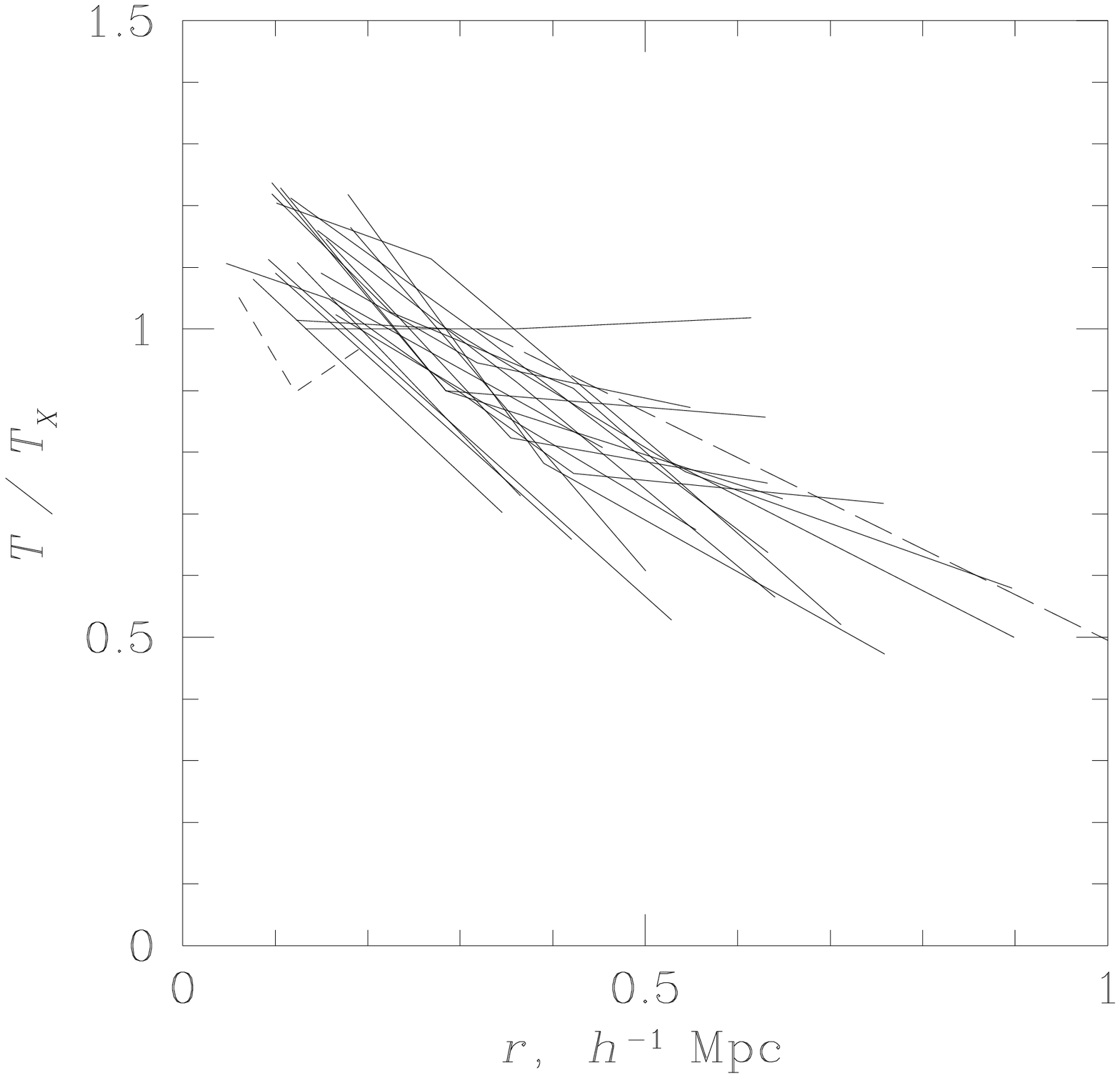}}

\rput[tl]{0}(-0.1,20.4){\epsfxsize=6.9cm
\epsffile{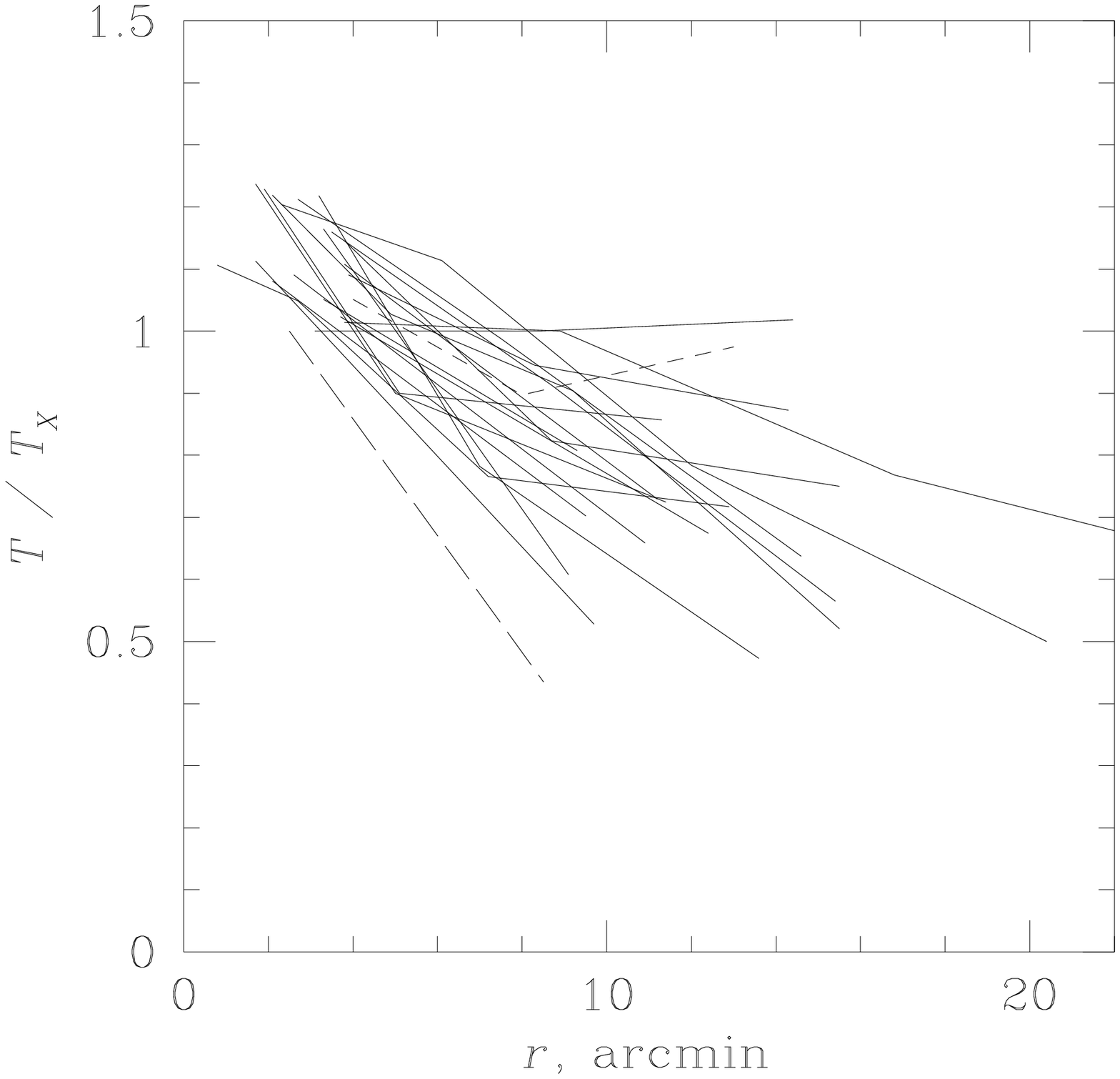}}

\rput[lc]{0}( 5.7,19.3){\it a}
\rput[lc]{0}(11.8,19.3){\it b}
\rput[lc]{0}(17.9,19.3){\it c}

\rput[tl]{0}(0,13.2){
\begin{minipage}{18.5cm}
\small\parindent=3.5mm
{\sc Fig.}~5.---Normalized temperature profiles of 19 relatively symmetric
clusters, plotted against radii in ({\em a}) arcminutes, ({\em b})
megaparsecs, and ({\em c}) in units of $r_{180}$. Cooling flow components
are removed. Error bars and central temperatures with large uncertainties
are not shown for clarity. Also shown for comparison are nearby AWM7
(short-dashed line) and distant A2163 (long-dashed line), which are not part
of the sample. Cluster profiles are quite different in detector units ({\em
a}), but appear rather similar in physical units ({\em b} and especially
{\em c}). The biggest deviation in panel ({\em c}) is A3391, which is still
consistent with others within its errors (see Fig.~7 and discussion in
\S\ref{outsec}).
\end{minipage}
}
\endpspicture
\end{figure*}

\begin{figure*}[tbh]
\pspicture(0,11.1)(18.5,27)

\rput[tl]{0}(2.0,27.2){\epsfxsize=6.9cm
\epsffile{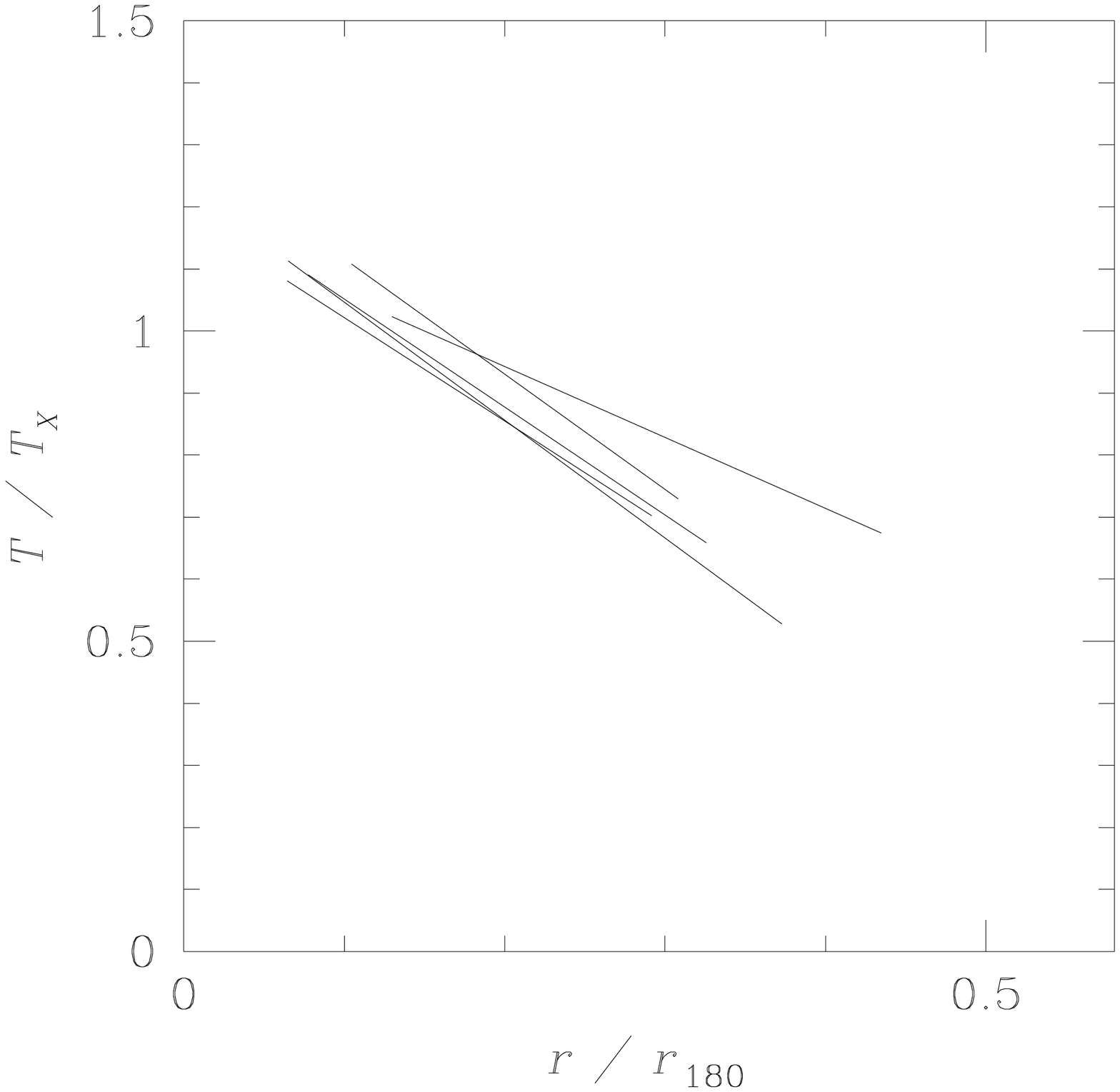}}

\rput[tl]{0}(10.0,27.2){\epsfxsize=6.9cm
\epsffile{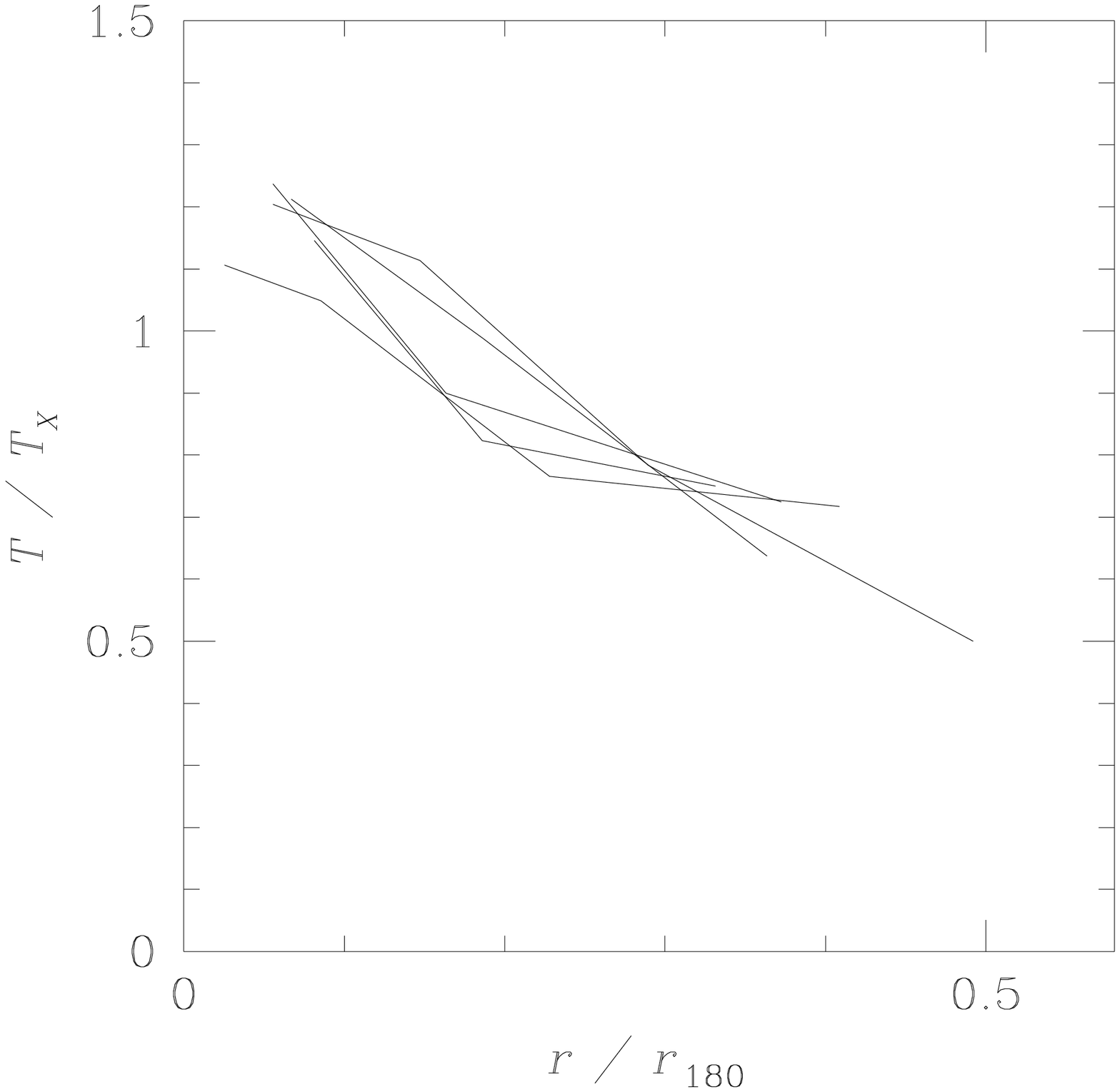}}

\rput[lb]{0}( 7.8,26.1){\it a}
\rput[lb]{0}(15.8,26.1){\it b}

\rput[tl]{0}(12.1,20.4){\epsfxsize=6.9cm
\epsffile{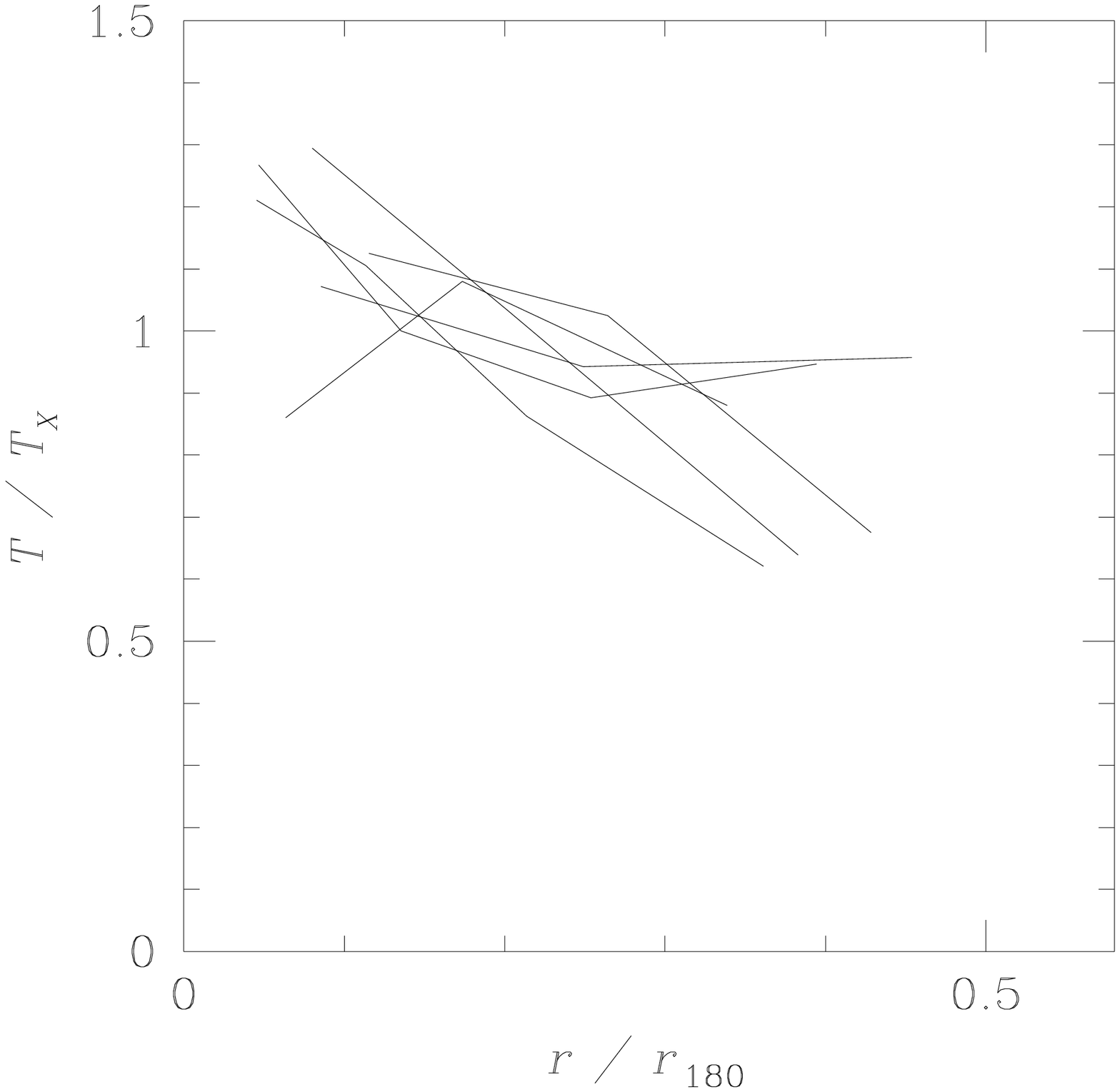}}

\rput[cc]{0}(12.2,17.3){\psframebox*{\white \rule{5mm}{2cm}}}

\rput[tl]{0}(6.0,20.4){\epsfxsize=6.9cm
\epsffile{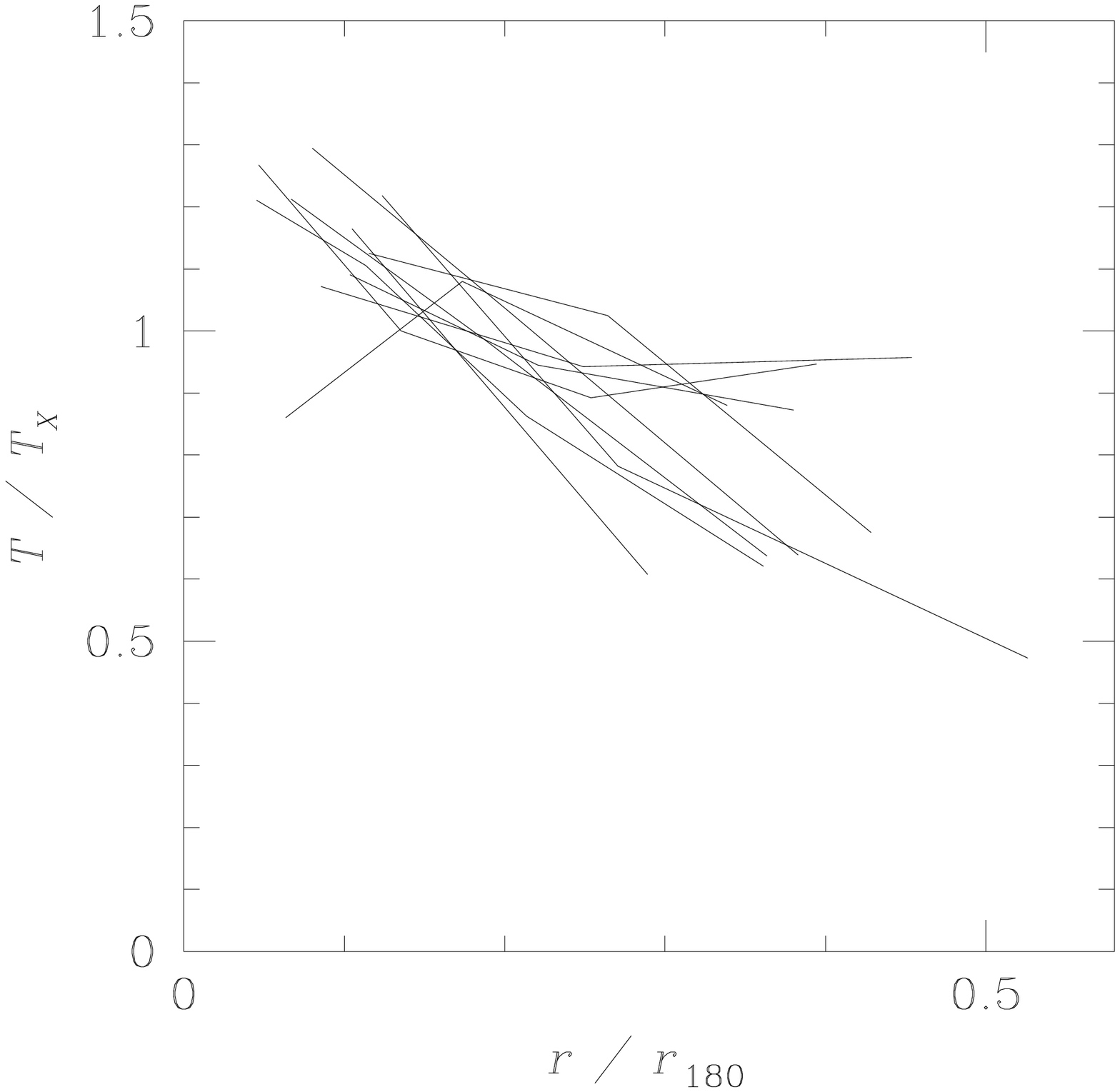}}

\rput[cc]{0}(6.1,17.3){\psframebox*{\white \rule{5mm}{2cm}}}

\rput[tl]{0}(-0.1,20.4){\epsfxsize=6.9cm
\epsffile{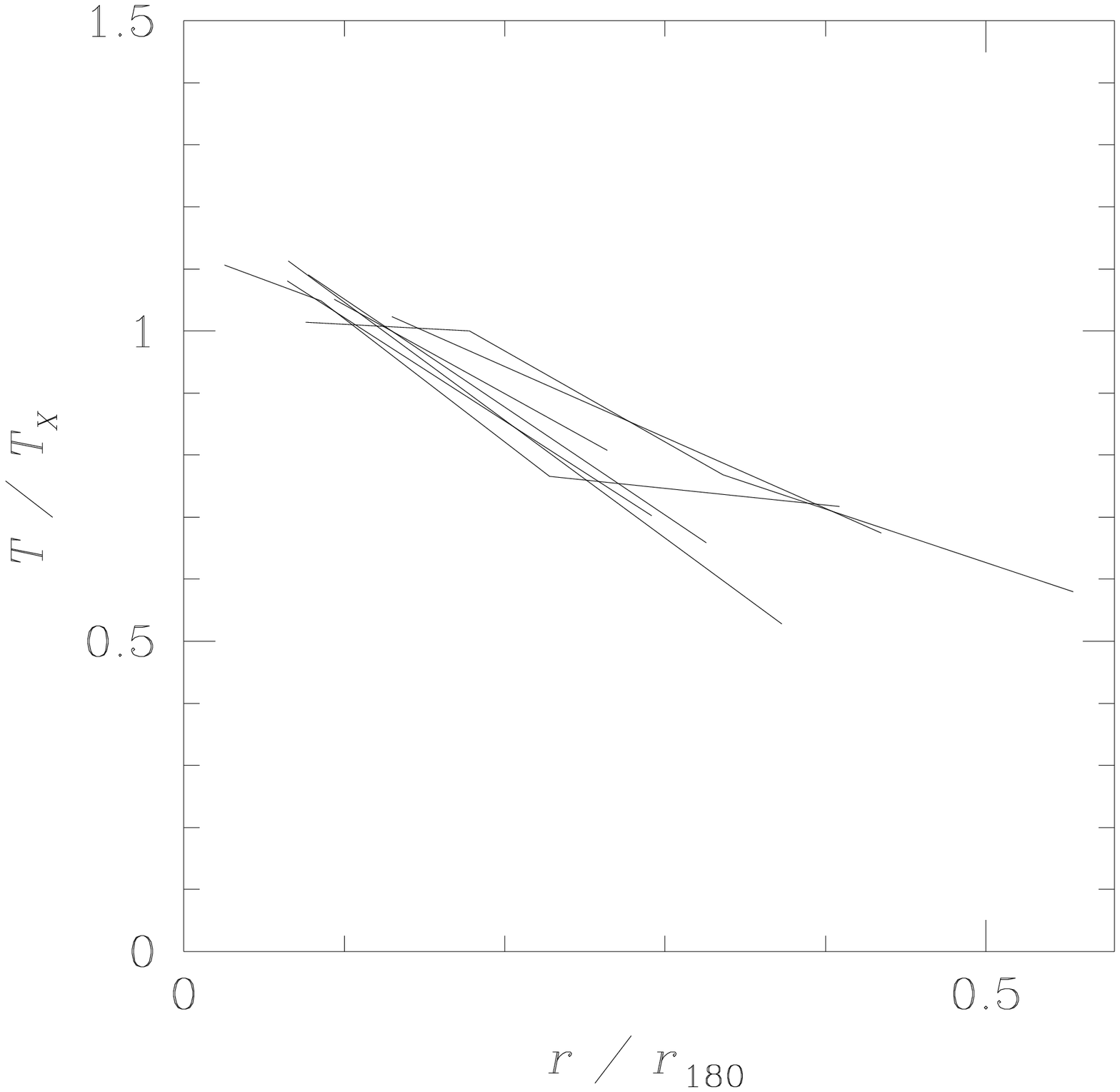}}

\rput[lc]{0}( 5.7,19.3){\it c}
\rput[lc]{0}(11.8,19.3){\it d}
\rput[lc]{0}(17.9,19.3){\it e}

\rput[tl]{0}(0,13.2){
\begin{minipage}{18.5cm}
\small\parindent=3.5mm
{\sc Fig.}~6.---Normalized temperature profiles of several cluster subsets.
Error bars are omitted for clarity. Panels ({\em a}) and ({\em b}) show five
coolest ($T_X\leq 5.3$ keV) and five hottest ($T_X\geq 8$ keV) clusters of
those shown in Fig.\ 5 (excluding AWM7 and A2163), respectively. Panel ({\em
c}) shows relaxed clusters with cooling flows (A780, A1795, A2029, A3112,
A3571, A4059, MKW3S), panel ({\em d}) shows asymmetric and strong merger
clusters (A119, A644, A754, A2065, A3266, A3376, A3395, A3558, A3667, Cygnus
A), and panel ({\em e}) shows only the asymmetric clusters excluded from
Fig.\ 5 (A119, A754, A3376, A3395, A3667, Cygnus A). There is no apparent
systematic difference between the subsamples, except for a greater scatter
and slightly less steep decline of profiles for irregular clusters (panels
{\em d} and {\em e}).
\end{minipage}
}
\endpspicture
\end{figure*}

\section{CORRECTING AVERAGE TEMPERATURES FOR COOLING FLOWS}
\label{tempsec}

We noted above that the presence of a strong cooling flow results in a
significant underestimate of the average temperature for some clusters. The
possibility of such underestimates was suggested earlier by Fabian et al.\
(1994b) as one explanation for the observed difference in the $L_X-T$
relations for clusters with and without cooling flows. However, because
spatially resolved spectral data for clusters were unavailable, those
authors did not foresee the amplitude of the underestimate that we observe
(for even more extreme examples see, e.g., Allen 1998). Since cooling flows
occupy a small fraction of the cluster volume and are governed by different
physics than the rest of cluster gas (for a review see, e.g., Fabian 1994),
their contribution should be excluded from the temperature estimates, if one
intends to compare the global cluster properties with theoretical and
numerical predictions that at present cannot model radiative cooling in
detail.

The temperature maps and profiles presented above provide the necessary data
to calculate average, emission-weighted gas temperatures corrected for
cooling flows and any unrelated sources that contaminated earlier,
unresolved data. To do this, we co-add the fitted temperatures, $T_i$, for
all regions of a cluster map or a profile with weights proportional to the
\rosat\ flux, $f_i$, from the respective $i$-th region (which for the
temperature range of our sample essentially means weighting with the
emission measure). For regions with cooling flows or other contaminating
components (assigned $i=1$ below), only the main thermal (ambient)
component, $T_1$, was included in the calculation of the weighted
temperature $T_X$:
\begin{equation}
T_X=\frac{T_1 f_1 \eta_1 (1-\eta_{\rm cf}) + \sum_{i=2}^n T_i f_i}
	 {f_1 \eta_1 (1-\eta_{\rm cf}) + \sum_{i=2}^n f_i}.
\end{equation}
Here, $\eta_1$ denotes the overall normalization of region 1 relative to that
given by its \rosat\ flux and $\eta_{\rm cf}$ denotes the fraction of the
excluded spectral component in the projected emission measure of region 1,
both fitted as free parameters (see \S\ref{cfsec}). Confidence intervals on
$T_X$ were calculated by Monte-Carlo simulations to take into account
correlations between the temperature measurements in different regions.
Using a map or a profile for this procedure did not result in significant
differences in the weighted temperatures. Although in reality the
temperature changes continuously rather than abruptly as in our maps, each
$T_i$ in turn is, essentially, the emission-weighted temperature over the
respective region, thus the approximation (1) is sufficient for our level of
accuracy. Despite its simplicity, our treatment of central regions is
adequate even for strong cooling flows, because for those clusters $\eta_{\rm
cf}$ is near 1 and the contribution of the central region to $T_X$ is small
anyway.

The resulting weighted temperatures are given in Table 1.  The table shows
that, as expected, the weighted temperatures are consistent with the
single-temperature fits for clusters without cooling flows or other obvious
spectral complications. Strong cooling flow clusters have significantly
higher mean temperatures than their single-temperature fits imply, with the
greatest difference, by a factor of 1.3, obtained for A1795.  A separate
paper (Markevitch 1998) shows that the difference between the $L_X-T$
relations for clusters with and without cooling flows essentially disappears
when these weighted temperatures and the luminosities with excised cooling
flows are used. Such an improvement strongly suggests that our measurements
indeed result in more physically meaningful temperatures. We will use these
temperatures below to normalize the radial temperature profiles and estimate
cluster virial radii.

\section{THE COMPOSITE TEMPERATURE PROFILE}
\label{profsec}

Figure 4 shows that, with a few exceptions, cluster temperatures
significantly decline with radius. Those few clusters that do not show a
general temperature decline are mostly ongoing mergers (e.g., A3667) or
highly asymmetric clusters (e.g., A3395) for which radial profiles do not
have much physical meaning. Below we show that temperature profiles of
almost all azimuthally symmetric clusters are similar when compared in
physical units. 

We normalize the radial temperature profile for each cluster by the weighted
average temperature $T_X$ obtained in \S\ref{tempsec}, and plot it against
the radius in units of $r_{180}$, within which the mean density (total mass
divided by volume) is 180 times the critical density. This radius is
approximately the radius of the virialized region for clusters (the virial
radius) in an $\Omega=1$ universe (e.g., Lacey \& Cole 1993). It is natural
to scale radii by the virial radius, because cluster total density profiles
are expected to be similar in these units (e.g., Bertschinger 1985). The
scaled radii are independent of the Hubble constant.

The values of $r_{180}$ can in principle be derived directly by
reconstructing the mass distribution using the gas temperature and density
profiles, if the cluster gas is in hydrostatic equilibrium. However, only
several clusters in our sample have sufficiently accurate data and symmetric
images for such a determination (MV97b; Sarazin et al.\ 1997; our ongoing
work). Alternatively, simulations of Evrard, Metzler, \& Navarro (1996,
hereafter EMN) predict that the average cluster temperature strongly
correlates with the cluster mass, even for moderately irregular clusters. We
take advantage of this correlation to calculate $r_{180}$ for all clusters
in a uniform manner, using our weighted temperatures and the EMN fit to the
simulations, $r_{180}=1.95\,h^{-1}\,{\rm Mpc}\; (T_X/10\;{\rm keV})^{1/2}$.
Although the simulations generally do not reproduce the observed temperature
profiles (see \S\ref{simsec} below) and, therefore, this relation may be
inaccurate, we are mostly interested in the scaling of $r_{180}$ with $T_X$
which is unlikely to be seriously wrong. We exclude from this exercise all 5
clusters with $z>0.08$ (A478, A1650, A1651, A2142, and A2597) because of
large errors in their profiles, and 6 clusters with strongly asymmetric
images (A119, A754, A3376, A3395, A3667, and Cygnus A) because their radial
profiles have little meaning. Note that we do not specifically exclude
mergers but only those of them with asymmetric images, due to the above
consideration. For comparison, we include projected temperature profiles for
two clusters that are not part of the sample due to their redshifts, A2163
at $z=0.201$ from M96, and AWM7 at $z=0.018$ from MV97a.

Figure 5 shows normalized temperature profiles (without error bars for
clarity) plotted together against radius in arcminutes, megaparsecs, and in
units of $r_{180}$. Those central temperature values with large
uncertainties (resulting from the removal of the cooling flow component) are
not shown. It is clear from the figure that in angular (detector) units,
clusters have quite different temperature profiles. However, the profiles
become rather similar in linear distance units and still more similar, with
just one exception, when plotted against radius in units of $r_{180}$. This
comparison strongly suggests that we are observing a physical temperature
decline rather than some unaccounted for instrumental effect.

In Fig.\ 6({\em a, b}), we plot two subsets of the clusters shown in Fig.\
5, the five coolest ($T_X\leq 5.3$ keV) and hottest ($T_X\geq 8$ keV)
clusters.  There is no apparent systematic difference between them.  Note
that if we were accounting for the \asca\ PSF incorrectly, one would expect
systematic differences between cooler and hotter clusters, because the
effects of PSF scattering are greater at higher energies. Panels {\em c, d,
e} in Fig.\ 6 show, respectively, temperature profiles for strong cooling
flow clusters that are thought to be the most relaxed, strong mergers and
asymmetric clusters (including those not shown in Fig.\ 5), and only the
asymmetric clusters excluded from Fig.\ 5. There is no strong qualitative
difference between the profiles in these subsets except for the large
scatter for asymmetric clusters, and a possibly shallower median slope for
the excluded clusters. The shallower slope is in part due to the fact that
our profiles are not centered on the brightness peaks but centered on the
large-scale emission centroids between the subclusters.  Normalized profiles
of the five high-redshift clusters in our sample that we do not use because
of large errors are consistent with the others.

We now try to quantify the slope of our composite radial temperature profile
using a polytropic relation. Figure 7 shows profiles of symmetric clusters
(those in Fig.\ 5 excluding AWM7 and A2163). Also plotted is an approximate
band that encloses most profiles and their error bars. We assume a gas
distribution with the core radius $a_x=0.15\,h^{-1}$ Mpc and $\beta=0.67$
typical for our cluster sample median temperature of 7 keV (Jones \& Forman
1997), and $r_{180}=1.63\,h^{-1}$ Mpc which corresponds to this temperature.
The observed temperature decline between 1 and 6 $a_x$ (0.09 and 0.55
$r_{180}$) in Fig.\ 7 corresponds to a polytropic index $\gamma\simeq
1.24^{+0.20}_{-0.12}$. The error on $\gamma$ corresponds to the width of the
90\% error band at large radii as shown in Fig.\ 7. A simple polytropic
dependence is not a particularly good description of the composite profile;
the ratio of temperatures between 1 and 3 $a_x$ corresponds to $\gamma\simeq
1.19$ while the ratio between 3 and 6 $a_x$ corresponds to $\gamma\simeq
1.29$. For individual clusters, this equivalent polytropic index would of
course depend on the individual gas density profile and the appropriate
scaling of $r_{180}$, and in fact may be quite different from these values.
We did not attempt to derive a more physically motivated functional form for
the composite temperature profile. It should be searched for among the
solutions of the hydrostatic equilibrium equation for various forms of the
dark matter distribution; this issue will be addressed in a separate paper.

\section{DISCUSSION}

\subsection{Merger and Cooling Flow Fractions}

The last column in Table 1 marks those clusters in which our temperature
maps (or entropy maps for A3395 and Triangulum Australis) indicate
significant merging. In addition, \rosat\ images of A2142, A3266, and A3376
undoubtedly indicate mergers in progress; for these clusters, we were unable
to obtain sufficiently accurate spatially resolved temperatures due to
limited angular resolution or statistics. Thus, in total, about half of the
sample shows signs of ongoing mergers. The frequency of mergers is an
indicator of the cluster formation rate which depends on cosmology (e.g.,
Richstone, Loeb, \& Turner 1992). Temperature maps of mergers provide
information complementary to that contained in the X-ray image, such as
evidence for physical interaction, the merger direction, the collision
velocity, etc.  This makes temerature maps potentially more discriminating
between cosmological models than tests based on the frequency of
substructure in images alone (e.g., Tsai \& Buote 1996; Thomas et al.\
1997). Therefore, a detailed comparison of the merger fraction in our sample
with simulations capable of modeling shocks may yield
interesting cosmological constraints.

Table 1 also shows that we detect cooling flow spectral components in about
60\% of the sample. We may have missed a few cooling flows (e.g., Cygnus A)
due to the \asca\ limited angular resolution. The fraction of clusters with
cooling flows is in general agreement with predictions of Edge et al.\
(1992) from their \exosat\ image analysis.

\begin{figure*}[thb]
\pspicture(0,-2.9)(18.5,9.8)

\rput[tl]{0}(0.,9.7){\epsfxsize=9cm
\epsffile{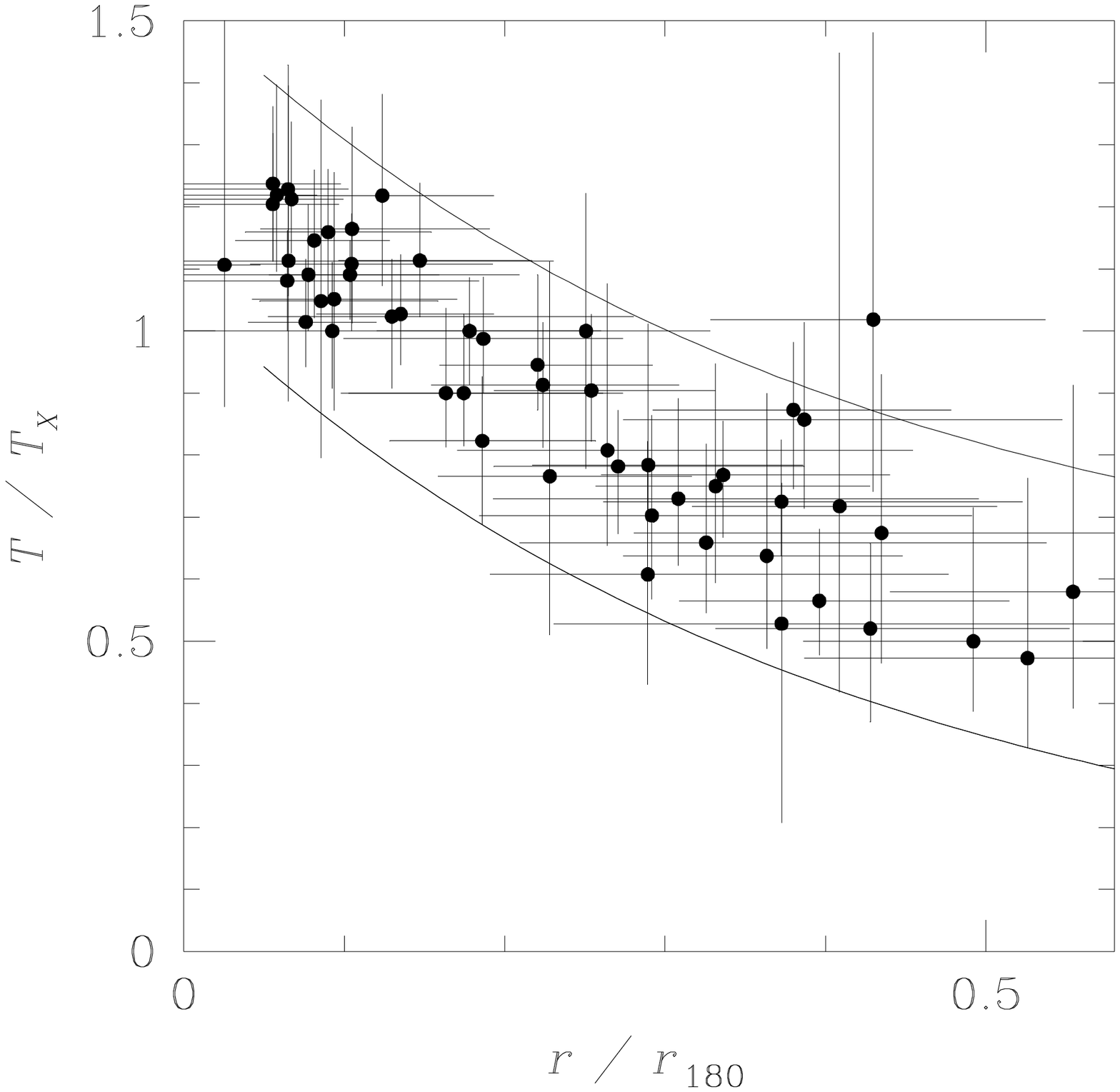}}

\rput[tl]{0}(-0.1,0.8){
\begin{minipage}{8.85cm}
\small\parindent=3.5mm
{\sc Fig.}~7.---Normalized profiles of symmetric clusters (those shown in
Fig.\ 5 without AWM7 and A2163) with their 90\% errors. Again, cooling flow
components are excluded. Smooth curves show an approximate band that
encloses profiles and most of their errors. The outlier is A3391.
\end{minipage}
}

\rput[tl]{0}(9.6,9.7){\epsfxsize=9cm
\epsffile{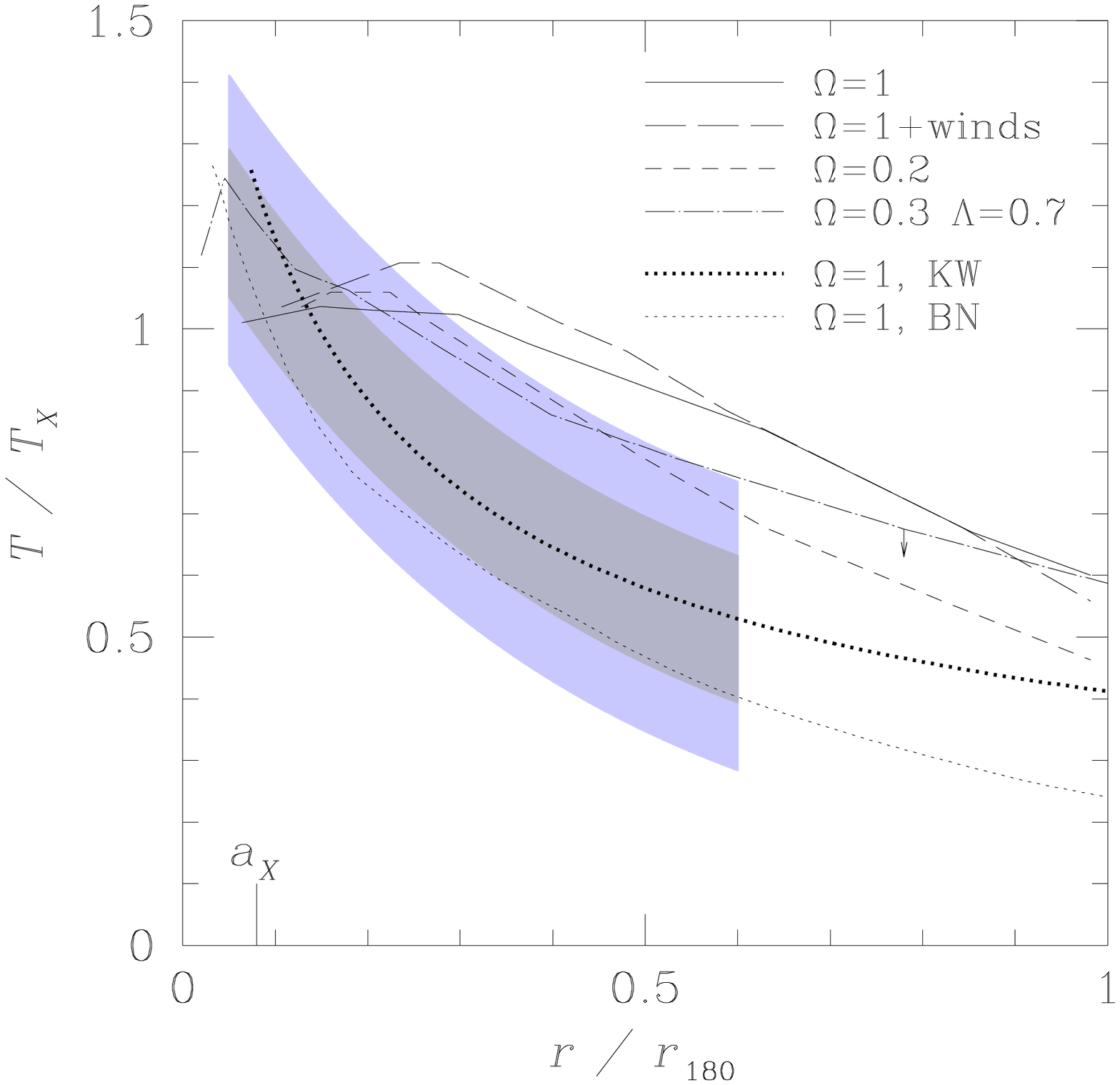}}

\rput[tl]{0}(9.6,0.8){
\begin{minipage}{8.85cm}
\small\parindent=3.5mm
{\sc Fig.}~8.---Temperature profiles from cluster simulations overlaid on
our results. Light gray band is the same as that in Fig.\ 7; dark gray band
encloses only the scatter of the best-fit values in Fig.\ 7.  An approximate
X-ray core radius for a 7 keV cluster is marked.  Three-dimensional average
profiles for $\Omega=1$, $\Omega=1$+winds, and $\Omega=0.2$ models are from
Evrard et al.\ (1996); $\Omega=0.3, \Lambda=0.7$ model is from Eke et al.\
(1997). Projection effect is shown by an arrow. Also shown are projected
profiles from simulations of single clusters of Katz \& White (1993) and
Bryan \& Norman (1997), both for $\Omega=1$. The latter simulation is
Eulerian, all others are Lagrangian.
\end{minipage}
}
\endpspicture
\end{figure*}

\subsection{The Composite Temperature Profile}

The main result of our paper is the observed similarity of most cluster
temperature profiles in units of average temperatures and virial radii. We
discuss some implications of this finding below, after commenting on
relevant earlier work.

\subsubsection{Comparison with Earlier Work}

Since \asca\ is the first instrument offering the possibility of (almost)
direct spatially resolved cluster temperature measurements, it is difficult
to find accurate earlier data with which to compare our findings. Pre-\asca\
spatially resolved temperature measurements were mostly limited to the
nearby clusters Coma and Perseus. The Coma radial temperature profile
obtained by Hughes et al.\ (1988) from \exosat\ is in rather good
qualitative agreement with our composite profile. The \spacelab\ profile for
Perseus (Eyles et al.\ 1991) only covers relatively small radii, but it
agrees with our composite profile when the cooling flow is taken into
%
%
account. Modeling data from \einstein\ SSS and \heao\ that have different
fields of view, Henriksen \& White (1996) find that several cooling flow
clusters, including A85 and A1795, have components outside the cooling flow
regions that are hotter and cooler than the single-temperature fit. This is
qualitatively similar to what we find in our direct measurements. MV97a
obtained a temperature profile for A2256 using \rosat\ PSPC which is in
excellent agreement with the \asca\ profile, although with large errors due
to the limited PSPC energy band.

Other spatially-resolved \asca\ temperature measurements are available; we
discuss only those which properly include the effects of the PSF. The Ikebe
et al.\ (1997) results for the Hydra A cluster are similar to our
independent analysis. The \asca\ analyses of the outer regions for nearby
relatively regular clusters such as Perseus, Coma, and AWM7 are complicated
by the presence of stray light; nevertheless, some preliminary results have
appeared. Honda et al.\ (1996) have derived a temperature map for Coma which
may indicate some radial temperature decline, although Coma appears to have
a complex temperature structure.  (T. Ohashi 1997 communicates that a more
sophisticated reanalysis results in even greater temperature variations.)  A
nearly isothermal temperature profile obtained by Ezawa et al.\ (1997) for
the symmetric cluster AWM7 apparently disagrees with our composite profile;
however, when a central cooling flow is taken into account (MV97a), the
temperature in the central region increases as in other cooling flow
clusters. The resulting AWM7 profile is consistent with the observed range
of profiles. A definitive check of our results should be possible in the
near future with \axaf.

\subsubsection{Effect on Mass Estimates}

The temperature profile is critical in deriving the gravitating mass under
the assumption of hydrostatic equilibrium (e.g., Sarazin 1988). To evaluate
the effect of the observed temperature decline on the mass estimates, we
take a polytropic temperature profile with $\gamma=1.24$ which approximately
represents our data for a typical 7 keV cluster (see \S\ref{profsec}). For
the gas density parameters used in \S\ref{profsec}, the mass estimates
within 1 and 6 core radii (0.15 and 0.9$\,h^{-1}$ Mpc) are approximately
factors of 1.35 and 0.7 of the isothermal $\beta$-model estimates,
respectively. This is similar to the results of a more detailed modeling of
the A2256 \asca\ data presented in MV97b (who find that the dark matter
profile of the Navarro, Frenk, \& White 1997 form was marginally allowed by
the data but a steeper profile was preferred). It is also qualitatively
similar to the A2029 \asca\ analysis in Sarazin et al.\ (1997). As discussed
in MV97b, the increased mass at small radii and decreased mass at large
radii have several important implications. One of them is the convergence of
the X-ray and lensing mass estimates at the cluster central regions;
Miralda-Escud\'e \& Babul (1995) first noted that a temperature decline such
as those we observe would be sufficient to explain most of the mass
discrepancy in A2218. Another implication is a steep rise of cluster baryon
fraction with radius, which leads to a more pronounced ``baryon
catastrophe'' (e.g., White et al.\ 1993) and probably indicates a presence
of sources of gas thermal energy other than gravity and merger shocks (e.g.,
David et al. 1995). A model-independent estimate of the average cluster
total mass distribution using our composite temperature profile will be
attempted in a later paper.

\subsubsection{Comparison with Cluster Simulations}
\label{simsec}

In Fig.\ 8, our composite temperature profile is shown by the light gray
band which corresponds to the error band in Fig.\ 7, and by a less
conservative dark gray band which approximates the scatter of the best-fit
profile points. Overlaid on the data are radial temperature profiles from
several published hydrodynamic cluster simulations with spatial resolution
comparable to that of our measurements. We plot median temperature profiles
for the simulated cluster samples from EMN (see also references therein),
and the average profile from Eke, Navarro, \& Frenk (1997; hereafter ENF).
We also show profiles of single clusters whose growth was simulated by Katz
\& White (1993, hereafter KW; further details given in Tsai, Katz, \&
Bertschinger 1994) and Bryan \& Norman (1997, hereafter BN). The smallest
radii that we show represent the claimed spatial resolution in the
simulations.

For the two steep KW and BN three-dimensional (i.e., not projected)
temperature profiles, we performed the emission-weighted projection using
the gas density distributions presented in the respective papers. Profiles
from EMN and ENF are much shallower and the projection effect (illustrated
approximately by an arrow in Fig.\ 8) is small; these profiles are shown
without projection. The EMN models are appropriately normalized by the
authors using the X-ray emission weighted temperature. ENF report that X-ray
temperatures of their simulated clusters are on average equal to the virial
temperatures which they use to normalize the profiles. To normalize the KW
profile, we use an average X-ray temperature given in Tsai et al. For BN
cluster, we calculated the average temperature using the temperature and
density profiles.

EMN simulated clusters in $\Omega=1$ cosmological models with and without
inclusion of galactic winds, and in an open model with $\Omega=0.2$. They
also present a flat model (cosmological constant $\Lambda = 1 - \Omega$),
which is in general agreement (except
in the center) with the better-resolution ENF $\Omega=0.3$, $\Lambda=0.7$
simulation, and we show the latter. All these model profiles are in apparent
disagreement with our results, being less steep than the observed profiles.
The profile for the flat $\Lambda$ model of ENF only marginally resembles
our observations. The EMN $\Omega=0.2$ model and possibly the
($\Omega=1$ + winds) model may have the correct slope at $r>0.2 r_{180}$, but
they, too, disagree within the overall range covered by the data. Note,
however, that the central regions and the normalizations of the EMN profiles
may be considerably incorrect, because their simulations did not resolve
cluster cores where a large fraction of the X-ray emission originates. If
their simulations underestimate temperatures in the cluster cores because of
resolution effects (as is suggested by comparison of their and ENF
results for flat $\Lambda$ models), then an open model profile may succeed
in describing the data.

The no-winds $\Omega=1$ model of EMN is in disagreement with the
observations. However, an independent simulation assuming an $\Omega=1$
cosmology by KW, who employed the same Lagrangian approach with resolution
comparable to EMN, but including radiative cooling, produced a very
different temperature profile. The KW profile is in agreement with our
observations. A high-resolution Eulerian simulation of an $\Omega=1$ cluster
by BN (which did not include cooling) also produced a steeper temperature
profile. KW and BN followed the evolution of only one cluster in each case,
and it is unclear to what extent their particular cluster realizations are
representative. However, BN report that the dark matter profile of their
cluster is well fit by the ``universal'' profile of Navarro et al.\ (1997)
that fits most simulations with comparable resolution, suggesting that their
cluster realization may be close to a typical cluster.

Because the EMN simulations do not reproduce the observed temperature
profiles, the $T_X-r_{180}$ relation from EMN that we used to scale the
observed profiles may be incorrect. Indeed, the MV97b mass profile measured
for A2256 corresponds to an $r_{180}$ a factor of $\sim1.2$ smaller than
predicted. This means that our temperature data may in fact subtend a larger
fraction of $r_{180}$ than shown in Fig.\ 8. This difference is small and
will not qualitatively alter the comparison with models.

We excluded very asymmetric clusters (6 out of 21) in order to obtain a
meaningful composite radial profile. In principle, this can bias our
comparison with simulations that apparently do not make such a selection.
However, as we noted in \S\ref{profsec}, the excluded cluster profiles are
not qualitatively different and would only add scatter to the composite
profile. A possibly shallower median slope of the 6 excluded clusters (Fig.\
6{\em e}) is within the scatter of the profiles shown in Figs.\ 7 and 8 and
therefore will not change any of our conclusions.

The comparison presented above shows that our temperature profile can
potentially provide a useful constraint for cluster formation models and,
possibly, for the underlying cosmology. At present, however, the
disagreement among different simulation techniques is greater than the
uncertainty of our measurements. It is also noteworthy that none of the
simulations discussed above reproduces the observed shallow gas density
profiles (see, e.g., Jones \& Forman 1984 for a large sample; Briel, Henry,
\& B\"ohringer 1992 and Elbaz, Arnaud, \& B\"ohringer 1995 for data at large
cluster radii). A possible exception is the model with galactic winds
(Metzler \& Evrard 1997, presented in EMN) that predicts a shallow gas
profile for cool but not for hot clusters. The temperature and density
distributions are related through the hydrostatic equilibrium equation, and
if one is not predicted correctly then the other would also be in error, for
a given dark matter profile.  Therefore, in the short term, our
temperature profiles underscore the need to improve the cluster simulations.

\subsubsection{The Outliers}
\label{outsec}

Although all but one of the 19 symmetric clusters shown in Figs.\ 5{\em c}
and 7 exhibit remarkably similar temperature profiles and the remaining one
is still consistent within errors with the common profile, it is interesting
to identify the most prominent outliers. The isothermal profile corresponds
to A3391 and two other shallow profiles correspond to A399 and A3558.
Curiously, all three clusters are located in the regions of high local
matter density --- A3391 and A399 are members of close pairs and A3558 is in
the center of the dense Shapley Supercluster. If the simulations discussed
above capture the qualitative dependence of the profiles on $\Omega_0$
correctly, one would indeed expect to find less steep average temperature
profiles in clusters located in relatively overdense regions of the Universe
(that is, regions with a high local value of $\Omega_0$).

\subsection{A Note of Caution}
\label{caut}

Finally, we would like to emphasize that the temperature profiles presented
here are not direct measurements but result from considerable corrections
for the complex scattering of the \asca\ mirrors. Therefore, for almost all
clusters in our sample, the errors are dominated by the systematic
component, mainly due to the uncertainties in the \asca\ PSF and effective
area. If our current understanding of the \asca\ instruments is
significantly flawed, the likely result would be that all our temperature
profiles become systematically steeper or shallower. Therefore, even though
the true individual temperature values should be within our confidence
intervals that include conservative estimates for all current uncertainties,
the individual profiles shown in Fig.\ 7 cannot be averaged in any sense,
because their errors are not independent.

\section{SUMMARY}

We systematically analyzed \asca\ data for 30 nearby bright clusters and
found that none of them is isothermal, excluding those few for which our
accuracy is insufficient. Apart from cooling flows, the gas temperature
varies with position within each cluster by a factor of 1.3--2 and sometimes
stronger. For most clusters, we were able to reconstruct crude
two-dimensional gas temperature maps. These maps (together with the images
for three clusters without accurate maps) show that half of the clusters in
our sample exhibit signs of ongoing merging. In about 60\% of the sample, we
detect a central cooling flow component.

For all clusters, we obtained radial temperature profiles. Almost all
clusters show a temperature decrease with radius (in addition to the central
cool components found in many clusters). We excluded the most asymmetric
clusters for which a radial profile has no meaning, and compared the
normalized temperature profiles for the remaining clusters. While the
profiles are different in angular (or \asca\ detector) units, when plotted
in radial units normalized to the estimated virial radius for each cluster,
they are remarkably similar. For a 7 keV cluster with a typical gas density
profile, the observed temperature decline can be characterized by a
polytropic index of 1.2--1.3, although this is not a particularly good
description over the range of measured radii.

The observed temperature decline implies that, at small radii, an analysis
that assumes the hydrostatic equilibrium and a constant temperature
underestimates the cluster mass, while at large radii, the gravitating mass
falls below the isothermal estimate. In particular, for a 7 keV cluster with
our median temperature profile and a typical gas density distribution, the
total mass estimates within 0.15 and 0.9$\,h^{-1}$ Mpc are approximately
factors of 1.35 and 0.7 above and below the isothermal $\beta$-model
estimates, respectively. As discussed in the study of A2256 (MV97b), this
general result strengthens the argument for a low-$\Omega_0$ cosmology based
on the high baryon fraction in clusters (e.g., White et al.\ 1993). It also
implies a strong segregation of gas and dark matter, possibly indicating
that sources other than gravity have produced a significant fraction of the
gas thermal energy (e.g., David et al.\ 1995).

Finally, we compared our composite temperature profile to the results of
cluster hydrodynamic simulations. We find that most simulations predict a
considerably shallower average radial temperature decline, with the possible
exception of those for low-$\Omega$ cosmologies. This comparison suggests
that, potentially, the \asca\ temperature profiles can constrain cluster
formation models. However, at present, there is a discrepancy between
different simulations that is greater than the uncertainty in our
measurements. This underscores the need for further theoretical and
numerical work before conclusions can be drawn regarding which cosmological
parameters best describe the observations.

\acknowledgments

The results reported here would not be possible without the dedicated work
of the entire \asca\ team on building, calibration and operation of the
observatory. We thank NASA/GSFC for maintaining the HEASARC online data
archive which we used extensively. We are grateful to Franz Bauer for
sharing his results prior to publication, and to Ue-Li Pen, Nancy Brickhouse
and the anonymous referee for useful comments. M. M. and W. R. F.
acknowledge support from Smithsonian Institution, NASA grant NAG5-2611, and
NASA contract NAS8-39073.  C. L. S.  was supported in part by NASA grants
NAG5-2526 and NAG5-4516. A.  V.  was supported by CfA postdoctoral
fellowship.

\end{document}